%% file: coronet.tex
\documentclass[emulateapj]{emulateapj}
\usepackage{natbib}
\begin{document}
%\title{Protoplanetary Disks in the Coronet Cluster: 
%\\ The Transitional Disk Phase as a Function of Age and Stellar Mass}
\title{The Transitional Protoplanetary Disk Frequency as a Function of Age: 
 Disk Evolution in the Coronet Cluster, Taurus, and Other 1--8 Myr-old Regions}
\author{Thayne Currie\altaffilmark{1} and Aurora Sicilia-Aguilar\altaffilmark{2}}
\altaffiltext{1}{NASA-Goddard Space Flight Center, Greenbelt, MD}
\altaffiltext{2}{Max-Planck-Institute for Astronomy, K\"{o}nigstuhl 17, 
69117 Heidelberg, Germany}
\begin{abstract}
We present Spitzer 3.6--24 $\mu m$ photometry and spectroscopy for stars in the 
1--3 Myr-old Coronet Cluster, expanding upon the survey of Sicilia-Aguilar et al. (2008).
  Using sophisticated radiative transfer models, we analyze these new data and those 
from Sicilia-Aguilar et al. (2008) to identify disks with evidence 
for substantial dust evolution consistent with disk clearing: transitional disks.
We then analyze data in Taurus and others young clusters -- IC 348, NGC 2362, and 
$\eta$ Cha -- to constrain the transitional disk frequency as a function of time. 
Our analysis confirms previous results
finding evidence for two types of transitional disks -- those with inner holes and those 
that are homologously depleted.  The percentage of disks in the transitional phase 
increases from $\sim$ 15--20\% at 1--2 Myr to $\ge$ 50\% at 5--8 Myr; the 
mean transitional disk lifetime is closer to $\sim$ 1 Myr than 0.1--0.5 Myr, 
consistent with previous studies by Currie et al. (2009) and Sicilia-Aguilar et al. (2009).  
In the Coronet Cluster and IC 348, transitional disks are more numerous for very low-mass 
M3--M6 stars than for more massive K5--M2 stars, while Taurus lacks a strong 
spectral type-dependent frequency.  Assuming standard values for the gas-to-dust ratio and 
other disk properties, the lower limit for the masses of optically-thick primordial disks is 
M$_{disk}$ $\approx$ 0.001--0.003 M$_{\star}$.  We find that single color-color diagrams do not 
by themselves uniquely identify 
transitional disks or primordial disks.  Full SED modeling is required to accurately assess 
disk evolution for individual sources and inform statistical estimates of the transitional disk 
population in large samples using mid-IR colors.
\end{abstract}
\section{Introduction}
\textit{Transitional} protoplanetary disks identify 
important stages in planet formation, linking epochs when disks are optically-thick \textit{primordial} 
protoplanetary disks, rich in gas and containing massive amounts of small dust,
 and when they are gas-poor/free, optically-thin \textit{debris} disks and 
thus are in the post gas planet-building stage \citep[][]{CurrieLada2009,Strom1989}.  
Compared to primordial disks, transitional disks have weaker near-to-mid IR emission 
indicating that they are depleted in warm, micron-sized dust within several AU of the host star
 \citep[e.g.][]{Strom1989, Calvet2002}.  If this dust traces the bulk mass of solids, 
transitional disks then show evidence for the removal of planet-building material.
  
Though transitional disks are not primarily identified by their levels of gas, they 
typically have a lower rate/frequency of stellar accretion 
\citep{SiciliaAguilar2006a,Muzerolle2010, Cieza2010}.  
Transitional disks may also be losing much of their gas from
a photoevaporative wind driven by stellar EUV photons \citep{Pascucci2009}, 
may lack gas-rich inner disks \citep{Salyk2009},  or may have a low density of cold outer disk gas \citep{Hughes2010}.
Thus, transitional disks are also 
plausibly depleted of circumstellar gas responsible for the envelopes of 
gas/ice giant planets and may be signposts for the last phases of protoplanetary 
disk evolution and gas giant planet formation.

 Spitzer studies have better constrained the morphologies and 
lifetimes of transitional disks around solar to slightly subsolar-mass stars 
(M$_{\star}$ $\sim$ 0.5--1.4 M$_{\odot}$, see \citealt{Currie2010p} and \citealt{Muzerolle2010r} for reviews).
Transitional disks exhibit one of two general morphologies:
1) disks with weak or negligible levels of near-to-mid IR emission indicative of inner dust holes 
 but more strongly emitting, typically optically-thick outer disks consistent with an inside-out clearing 
of dust or 2) disks lacking evidence for inner holes but having weaker emission at all IR 
wavelengths consistent with a more homologous depletion of dust \citep[][]{CurrieLada2009}.
Transitional disks comprise $\lesssim$ 10--20\% of the disk population in $\approx$ 1 Myr-old 
clusters \citep[e.g.][]{Muzerolle2010,CurrieKenyon2009} but 
are relatively more frequent by $\sim$ 5 Myr, consistent with a transitional 
disk phase comprising a substantial fraction of the total protoplanetary disk lifetime 
\citep[e.g. $\sim$ 1 Myr out of 3--5 Myr,][]{CurrieLada2009,SiciliaAguilar2009}.

The 1--3 Myr-old Coronet Cluster \citep[][and references therein]{MeyerWilking2009} may provide a particularly useful probe of 
transitional disk properties in the youngest clusters and as a function of stellar mass. 
\citet{SiciliaAguilar2008} find that transitional disks comprise nearly half of the 
disk population, much larger than frequencies derived for other clusters of comparable 
age \citep[e.g.][]{CurrieKenyon2009, Muzerolle2010}.  At first glance, these results 
appear to be in conflict, but \citet{SiciliaAguilar2008} note that the members in their 
sample have systematically low masses \citep[see also][]{LopezMarti2010}.
Compared to solar and slightly subsolar-mass stars, disks around very low-mass stars 
and brown dwarfs may have very different structures and different dispersal timescales
 \citep[e.g.][]{Lada2006,Hartmann2006}.
Thus, it is unclear whether the high reported transitional disk frequency in the Coronet Cluster 
is anomalous or identifies a stellar mass-dependent transitional disk frequency.  
Addressing this issue requires 
analyzing disks around higher-mass Coronet stars and assessing the transitional disk frequency 
as a function of stellar mass for many clusters.

In this paper, we investigate the transitional disk population
in the Coronet Cluster as a function of spectral type/stellar mass by presenting Spitzer photometry and spectroscopy for additional, predominantly 
solar-mass members.  By comparing results for solar mass and subsolar-mass members, we determine whether the Coronet Cluster 
disk population shows evidence for a stellar mass-dependent frequency of 
transitional disks.   \S 2 describes our new observations, data reduction, and photometry/spectroscopy.  
In \S 3, we model the SEDs of disk-bearing stars to determine plausible disk evolutionary states following 
previous methods \citep[e.g.][]{CurrieLada2009}.  Finally, in \S 4 we compare these results 
to those for other clusters to investigate the duration of the transitional disk phase.

\section{Data}
\subsection{Spitzer IRAC and MIPS Photometry}
To the sample of stars analyzed by \citet{SiciliaAguilar2008}, we add
  IRAC and MIPS photometry of other Coronet Cluster members 
 located on the processed IRAC and MIPS mosaics obtained from Program IDs 6 (AORs 3650816 and 
3664640), 248 (AOR 13469696), and 30784 (AOR 17672960). 
Image processing and photometry for these stars was performed 
following methods described in \citet{SiciliaAguilar2008}.
Briefly, the basic calibration data (BCD) frames were mosaiced together using MOPEX using 
standard input parameters for the overlap correction, pixel interpolation, and outlier rejection.  
Aperture photometry was performed with APEX using a 3 pixel aperture, a 12--20 pixel background 
radius, and aperture corrections listed in the IRAC data handbook (1.112, 1.113, 1.125, and 1.218 
for the four channels, respectively).  For MIPS, we used 
a 5 pixel aperture with a 8--13 pixel background annulus, and the aperture correction 
listed in the MIPS data handbook (1.167).

We select cluster stars from the catalog of \citet{Forbrich2007}, which 
identifies x-ray bright sources with near-IR/optical counterparts as Coronet Cluster members.  
We add 32 cluster members from \citet{Forbrich2007} to the list of stars analyzed in \citet{SiciliaAguilar2008}.
The IRAC and MIPS coverage areas do not completely overlap,
and many cluster members lack photometry in two of the four IRAC channels ([3.6] and [5.8] or [4.5] and [8]) 
or lack MIPS data.  Other sources, particularly Class I protostars and 
bright, higher-mass stars (e.g. R CrA) saturate the IRAC and/or MIPS detector.  
Table \ref{newsources} lists the names and coordinates for these targets.
Table \ref{newphotometry} lists their photometry.  Sources with an "FP" prefix for their 
names were classified as members only by \citet{Forbrich2007}: the numbers 
following this prefix correspond to the row number in Table 2 of \citet{Forbrich2007}.

\begin{figure}
\centering
\epsscale{0.8}
\plotone{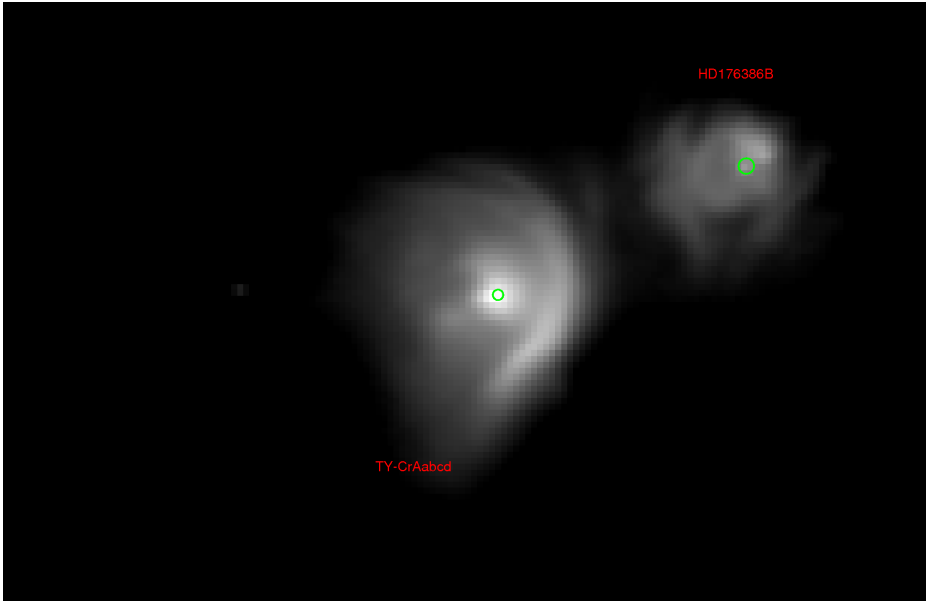}
\caption{Region of the IRAC [5.8] mosaic showing nebular emission 
surrounding TY CrA and HD 176386b.  The positions of both 
sources are shown as green circles.  Nebular emission contaminates both stars 
and is responsible for positional offsets between the 2.2--4.5 $\mu m$ 
centroids and 5.8--8 $\mu m$ centroids.} 
\label{contamination}
\end{figure}

To further assess photometric quality, we inspected the processsed 
IRAC and MIPS mosaics and compared the known target 
positions to the computed target centroids.
Most sources lie in regions of low background and are thus uncontaminated 
by nebulousity.  Most sources located in high-background regions are 
much brighter than the background, which is typically uniform.  
However, two sources -- TY CrA and HD 176386B -- are located in regions 
of high and highly-variable background.  In the longer wavelength 5.8 $\mu m$ 
and 8 $\mu m$ channels, the computed centroid positions are well offset from 
the apparent star positions, especially for HD 176386B.  
Figure \ref{contamination} illustrates this offset for the 5.8 $\mu m$ channel.  

To investigate how the mid-IR fluxes from TY CrA and HD 176386 
are affected by nebular emission, we recomputed the flux for these sources using much 
smaller apertures and background annuli, ranging from our default assumptions to 
a small 2 pixel aperture radius and 2--6 pixel background annulus.  
Depending on the choice of aperture radii and background annuli, their 
fluxes at 5.8 $\mu m$ and 8 $\mu m$ vary by $\sim$ 50\%--200\%. 
Therefore, we consider the 5.8 $\mu m$ and 8 $\mu m$ photometry for these sources 
to be unreliable.  Since the nebular background emission increases 
with wavelength, it also renders fluxes for these sources uncertain 
at 24--100 $\mu m$.  

\subsection{Spitzer IRS Spectroscopy}
To supplement our photometric data, we add spectroscopic 
data from the Infrared Spectrograph \citep{Houck2004} spectroscopy 
for several intermediate-mass members: S CrA, V709, T CrA, 
IRS7w, and IRS5.  
The spectra were processed starting from the BCD data and
using the spectral extraction tools developed for the
FEPS Spitzer science legacy team \citep{Bouwman2008},
based on the SMART software package \citep{Higdon2004}.
The spectra were extracted using 6 and 5 pixel wide
apertures in the spatial dimension for short-
(7.5 - 14 $\mu$m) and the long-wavelength (14 - 35 $\mu$m)
modules, respectively. The background was subtracted using
associated pairs of imaged spectra from the two nodded
positions along the slit, also eliminating stray light
contamination and anomalous dark currents. The long-wavelength
modules were not available for S CrA, and in the cases of
IRS7w, IRS5, and T CrA, strong background contamination
and/or extended emission makes the spectral extraction 
uncertain at $\lambda$ $>$ 14$\mu$m.  Therefore, we
only include in the analysis the short-wavelength region
for these sources.

 \subsection{Ancillary Data}
To compare the stellar and circumstellar properties of 
all Coronet Cluster members, we add IRAC/MIPS photometry and 
IRS spectroscopy presented in \citet{SiciliaAguilar2008}; 
optical photometry from \citet{LopezMarti2005}, \citet{Torres2006} and from 
the \textit{SIMBAD} Astronomical Database (various sources); near-IR 
photometry from 2MASS \citet{Skrutskie2006} and \citet{MeyerWilking2009}.  
Spectral types for most 
cluster stars with new Spitzer data derive from \citet{MeyerWilking2009} 
or \citet{Nisini2005}.  Most stars studied in \citet{SiciliaAguilar2008}, 
have spectral types derived from VLT/FLAMES.

We also consider new, higher signal-to-noise VLT/FLAMES spectra 
of Coronet Cluster members, which yield new spectral types for several 
low-mass cluster stars -- G-1, CrA-4107, CrA-4109, CrA-432, 
CrA-468, CrA-452, G-85, and G-87.  These data will be discussed and 
analyzed in detail in a later contribution (Sicilia-Aguilar et al., 2011, in preparation).
  Briefly, the medium resolution (R $\sim$ 6000) 
spectra were centered on three bands covering 0.6--0.9 $\mu m$ ($\lambda_{c}$ = 
0.682, 0.773, and 0.881 $\mu m$).  
For spectral classification, we first determined 
whether the stars were likely earlier or later than M3 using standard spectral indices.  
For later stars, we derived spectral types using spectral indices of TiO bands following 
\citet{SiciliaAguilar2008}.  
Earlier stars were classified by comparing their spectra to Cep OB2 spectra 
described in \citet{SiciliaAguilar2005}.  

We estimate the spectral types of stars without spectra by deriving 
the best-fit effective temperature from model the optical to near-IR SED 
using the \citet{Robitaille2006} radiative transfer grid and then 
using the \citet{Currie2010b} effective temperature scale for stars M2 
or earlier and \citet{Luhman2003} for later stars.
For some sources, we cannot derive a spectral type because 
the source lacks optical data or is a protostar.
Tables \ref{diskstatenew} and \ref{diskstateold} list the 
spectral types for each star.

To probe the mid-IR fluxes of bright stars that saturate the IRAC and MIPS detectors, we 
combined our source list with IRAS data for R CrA, TY CrA, and S CrA from \citet{Wilking1985,Wilking1992}.
  Our IRAC and MIPS mosaics indicate that the
 emission from R CrA and S CrA clearly dominates over the background nebular emission and 
thus are only weakly contaminated.
% at least at wavelengths less than $\sim$ 24 $\mu m$.  
As mentioned previously, TY CrA is heavily contaminated by background emission.
  Given the large beam size of IRAS compared to Spitzer, the IRAS flux quoted for TY CrA is also 
likely unreliable.

%Figure xx, our LABOCA data clearly detect S CrA.  Two other members, 
%G-65 and G-85, have marginal, low signal-to-noise detections.
\subsection{IRAC and MIPS Colors and Observed SEDs of Coronet Cluster Stars}
\begin{figure}
\epsscale{0.9}
\centering
\plottwo{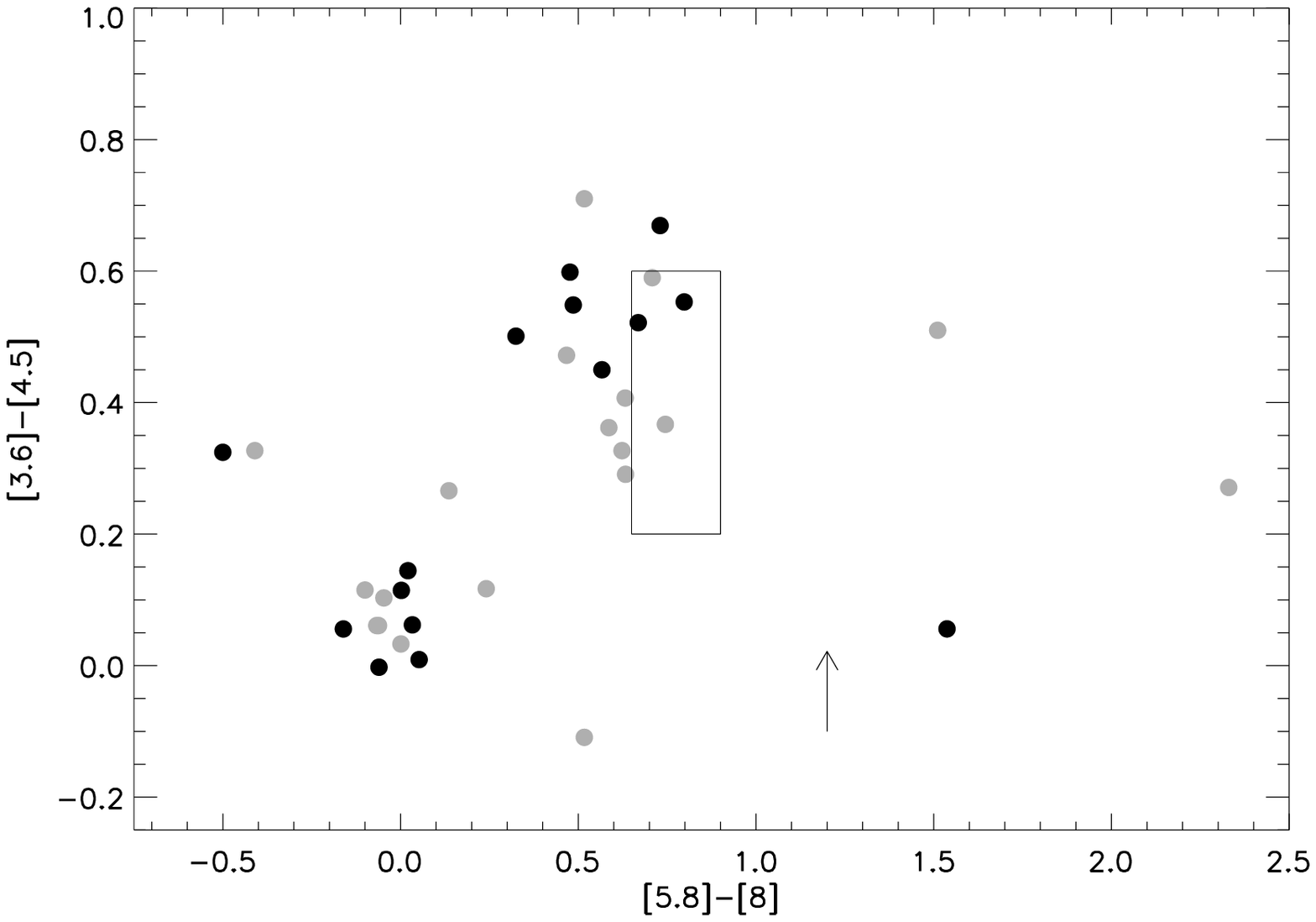}{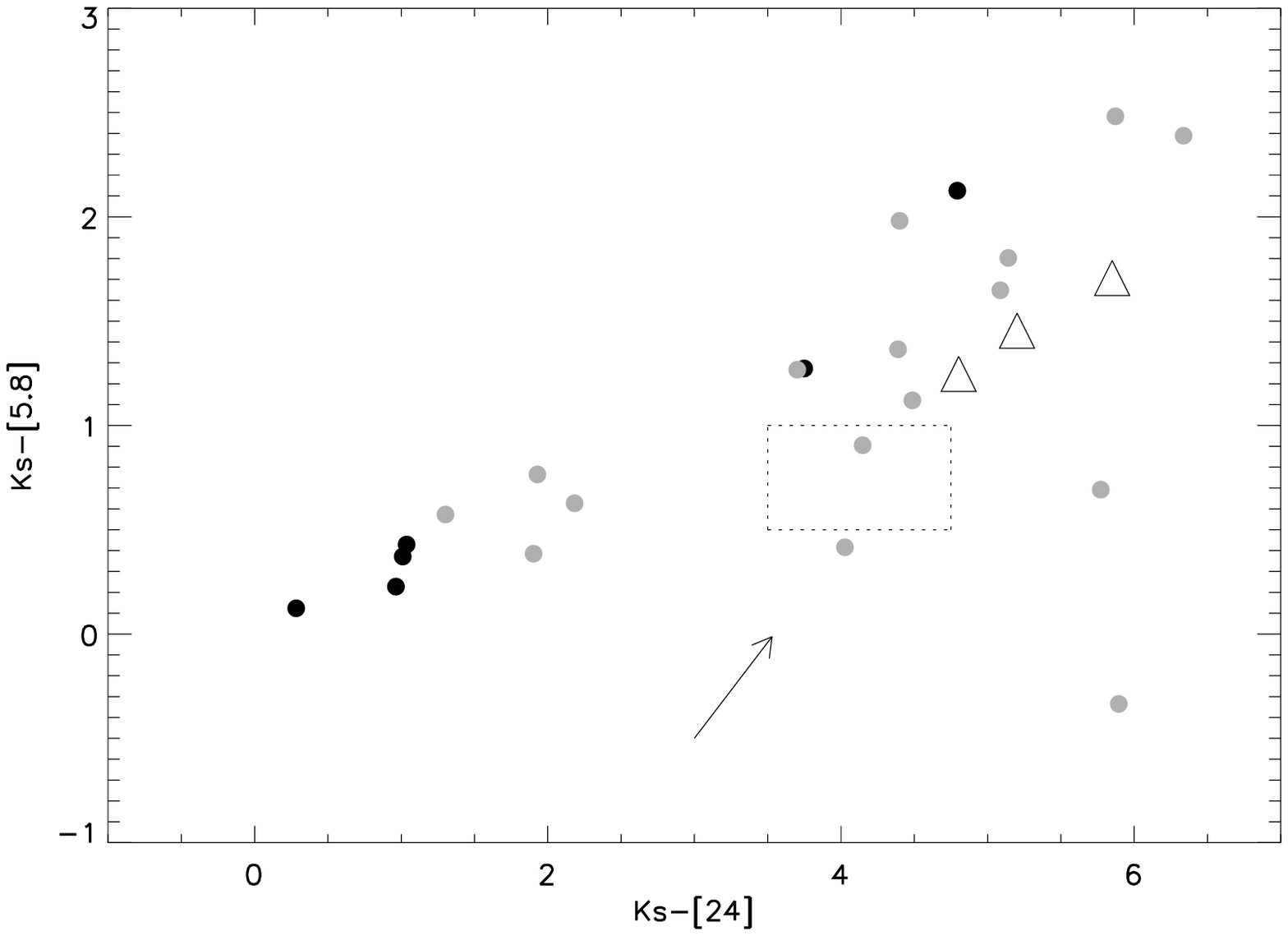}
\caption{(Left panel) IRAC color-color diagram of Coronet Cluster stars with new photometry (black circles) and those 
analyzed in \citet{SiciliaAguilar2008} (grey circles).  (Right panel) K$_{s}$-[5.8] vs. K$_{s}$-[24] color-color 
diagram for Coronet Cluster stars.  In each panel, we overplot a reddening vector of A$_{V}$ = 10, assuming 
the extinction relations of \citet{Indebetouw2005} for 2MASS and IRAC and \citet{Flaherty2007} for MIPS.  
In the left panel, the box corresponds to the range of colors for most stars in Taurus \citep{Hartmann2005}; the 
dotted box in the right-hand panel corresponds to the range of colors for most homologously depleted transitional disks 
in NGC 2362 identified by \citet{CurrieLada2009}.  The triangles in the right panel 
correspond to (from bottom to top) the lower-quartile Taurus SED, median Taurus SED, and upper-quartile 
Taurus SED for K5--M2 stars from \citet{Furlan2006}.}
\label{colorcolor}
\end{figure}

To provide a first-order investigation of the Coronet Cluster disk population, 
we analyze the distribution of near-to-mid IR colors and optical/IR spectral 
energy distributions (SEDs).  The left panel of Figure \ref{colorcolor} displays 
the observed IRAC colors of cluster stars with new photometry (black dots) and 
stars analyzed by \citet[][grey dots]{SiciliaAguilar2008}.  The distributions 
appear similar.  Many stars concentrate near zero color (as expected for bare stellar 
photospheres) and around [3.6] --[4.5] = 0.3--0.6 and [5.8]--[8] = 0.5--0.7.  Compared 
to the typical IRAC colors of primordial disks in 1--2 Myr-old clusters \citep[solid box,][]
{Hartmann2005}, colors for Coronet Cluster stars may be slightly bluer by 
$\sim$ 0.2 mag.  

The longer wavelength colors for cluster stars exhibit a wider diversity 
(Figure \ref{colorcolor}, right panel).  At least three sources, all from the \citet{SiciliaAguilar2008} 
sample, show evidence for weak/negligible 5.8 $\mu m$ excess but a $\gtrsim$ 4 magnitude excess at 24 $\mu m$ 
characteristic of transitional disks with inner holes.
  Others have 24 $\mu m$ and/or 5.8 $\mu m$ excess 
emission characteristic of more homologously depleted transitional disks, such as those found in NGC 2362 (dotted enclosed region, 
\citealt{CurrieLada2009}), which is weaker than emission for the lower-quartile median Taurus SED from \citet[][lower-left triangle]{Furlan2006}. 
The cluster also includes many stars with strong IRAC and MIPS excesses comparable to the median Taurus and upper-quartile Taurus SED.
Because only six stars in the new sample have K$_{s}$, [5.8] and [24] detections, it is less clear from the K$_{s}$-[5.8]/K$_{s}$-[24] 
diagram whether the distribution of colors from the new cluster sample displays the same diversity.

\begin{figure}
\epsscale{1.}
\centering
%\plotone{FP_seds_1.eps}
\plotone{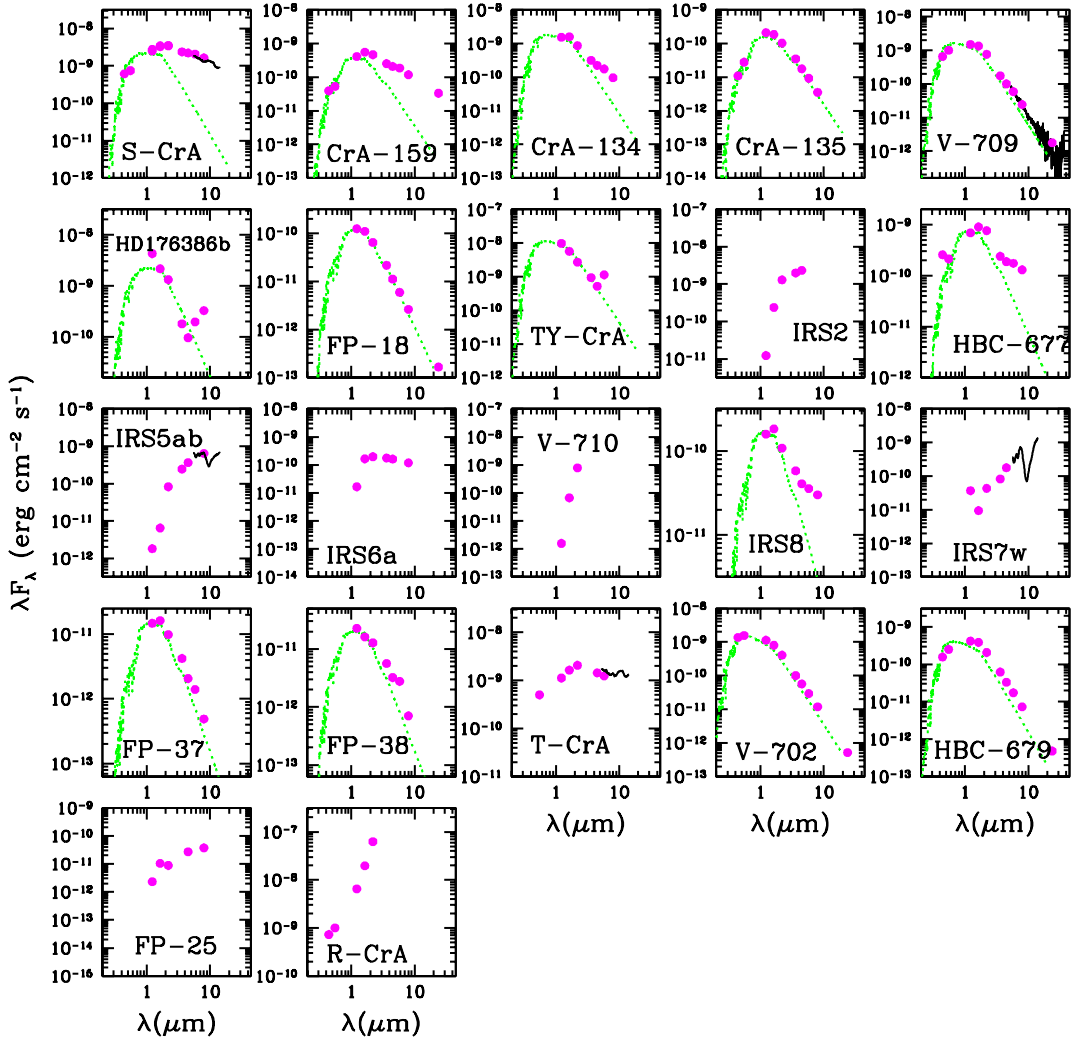}
\caption{Spectral energy distributions of Coronet Cluster stars with new Spitzer photometry.  In most cases,
 the SEDs can be modeled as a stellar photosphere (green) plus excess emission from a disk.  The stellar photospheres 
plotted come from the MARCS stellar atmosphere models \citep[][and references therein]{Gustafsson2008}.  
Sources not shown lack near-to-mid IR data required to constrain their SEDs.}
\label{sedsnew}
\end{figure}

Examining the SEDs of new sources clearly shows that they have a wide range of
 morphologies (Figure \ref{sedsnew}).  
Many sources (e.g. IRS2, IRS5ab, IRS7w) have rising near-to-mid IR SEDs indicative of 
 Class I protostars.  Well-known sources with optically-thick disks, such 
as the intermediate-mass stars S CrA and R CrA, have nearly flat SEDs from 1 $\mu m$ to 8 $\mu m$.

IRS spectra strengthens our identification of sources whose photometric data alone 
make them difficult to classify.  In particular, V709 shows 
evidence for a weak excess based on its MIPS-24 $\mu m$ flux and IRS spectra.  
Its SED gradually peels away from the stellar photosphere from $\sim$ 8 $\mu m$ 
to 30--35 $\mu m$.  Based on their 10 $\mu m$ absorption features,
IRS-5 and IRS-7w are likely protostars with cool envelopes \citep[e.g.][]{Willner1982,Andre1994}, 
consistent with previous classifications \citep{Henning1994,Chini2005,Groppi2007}. 
\section{Analysis}
To assess the evolutionary states of Coronet Cluster disks, we analyze source SEDs 
from the new sample and the \citet{SiciliaAguilar2008} sample using both simple 
theoretical comparisons and sophisticated radiative transfer modeling. 
We follow a slightly modified version of the analysis methods used in 
\citet{CurrieLada2009} to distinguish disks in different states.  
These classifications are based solely on 1--24 $\mu m$ data: 
since far-IR/submm data provides crucial constraints on disk properties, 
in particular the inferred disk mass, we consider these classifications to 
be provisional.  We will revisit these classifications later after 
analyzing stars in other clusters with longer wavelength data (\S 4).
%We later apply these same methods to analyze disks in other clusters.
%The Appendix discusses and fully justifies our adopted definitions for each disk state.  

In our classification, we consider protostars and disks with three main evolutionary states: primordial disks, 
transitional disks, and debris disk candidates.  We identify protostars as 
sources with rising SEDs from the near-IR to mid-IR consistent with emission 
predominantly from a cold dusty envelope (T $\le$ 200--300 K).  \textit{Primordial} disks 
are disks that are optically thick ($\tau_{IR}$ $>>$ 1) and thus lack evidence 
for dust evolution consistent with active disk clearing.  

\textit{Transitional disks} are disks that show substantial dust evolution consistent 
with active disk clearing.
%: a \textit{transition} from the primordial disk phase to the 
%debris disk phase.  
They can have one of two morphologies.  First, they can exhibit a 
drop in the optical depth ($\tau$ $\approx$ 1 or less) of emitting dust 
either at all IR wavelengths consistent with a 
reduced disk mass (a homologously depleted transitional disk).
Second, they can also have photospheric or weak, optically-thin emission 
at shorter wavelengths (e.g. 3.6--5.8 $\mu m$) 
and more optically thick emission at longer wavelengths consistent 
with inside-out disk clearing (a transitional disk with an inner hole).  In this latter group,
we include so-called 'pre-transitional disks' \citep[][]{Espaillat2007}.  These disks have 
a small amount of high optical depth material very close to the star but otherwise 
have large, dust-poor inner regions (e.g. large gaps) and optically-thick outer 
regions and thus also identify an inside-out disk clearing.  

  \textit{Debris disk} candidates lack evidence for accretion and have very optically-thin emission 
($\tau_{IR}$ $<<$ 1) from a low mass ($\lesssim$ 0.01 M$_{lunar}$, \citealt{Chen2005}) 
of second-generation dust \citep{Backman1993, Currie2008,KenyonBromley2008}.  
Many of the Coronet Cluster stars are M dwarfs with very low masses.  
Even if these stars have tenuous emission and lack evidence for accretion, 
they also have grain removal timescales comparable to the host star's age.  Thus, their 
dust need not be second generation \citep{CurrieKenyon2009}, they could be 
very late-stage transitional disks, and at best can only be considered 
debris disk \textit{candidates}.  Our basis for identifying sources 
with tenuous 24 $\mu m$ excess emission as debris disk candidates follows \citet{Carpenter2009} 
and is purely empirical.  They identify debris disk candidates around late-type stars as those 
with 0.25--0.5 24 $\mu m$ excesses.  In our analysis, we consider the evolutionary state 
of these tenuous disks around Coronet Cluster stars as indeterminable and simply label them 
as transitional disks/debris disk candidates\footnote{While some debris disks have much stronger 
emission, nearly all of them surround what are (or will be) main sequence A stars or 
early F stars \citep[e.g.][]{Rieke2005,Currie2008} that have masses $\gtrsim$ 1.5 M$_{\odot}$.  
Debris disks detected around lower-mass stars (the focus of this paper) are far more infrequent 
and typically have 24 $\mu m$ excesses of only a few tenths of magnitudes \citep[e.g.][]{Plavchan2009}.  
Our results are insensitive as to whether or not stars with tenuous excesses are considered to have debris 
disks or transitional disks.}.

For most sources, only photometric data is available.  For sources with IRS data from \citet{SiciliaAguilar2008}, 
we add the IRS flux densities, which when combined with our photometry yields flux densities
 at equally spaced wavelengths (Table \ref{irsdenspoint}).
The inclusion of the IRS data provides better 
sampling of the SED from 5 to 24 $\mu m$ and extends the wavelength coverage.
\subsection{Method for Disk Identification}
\subsubsection{Fiducial SEDs}
\begin{figure}
\plottwo{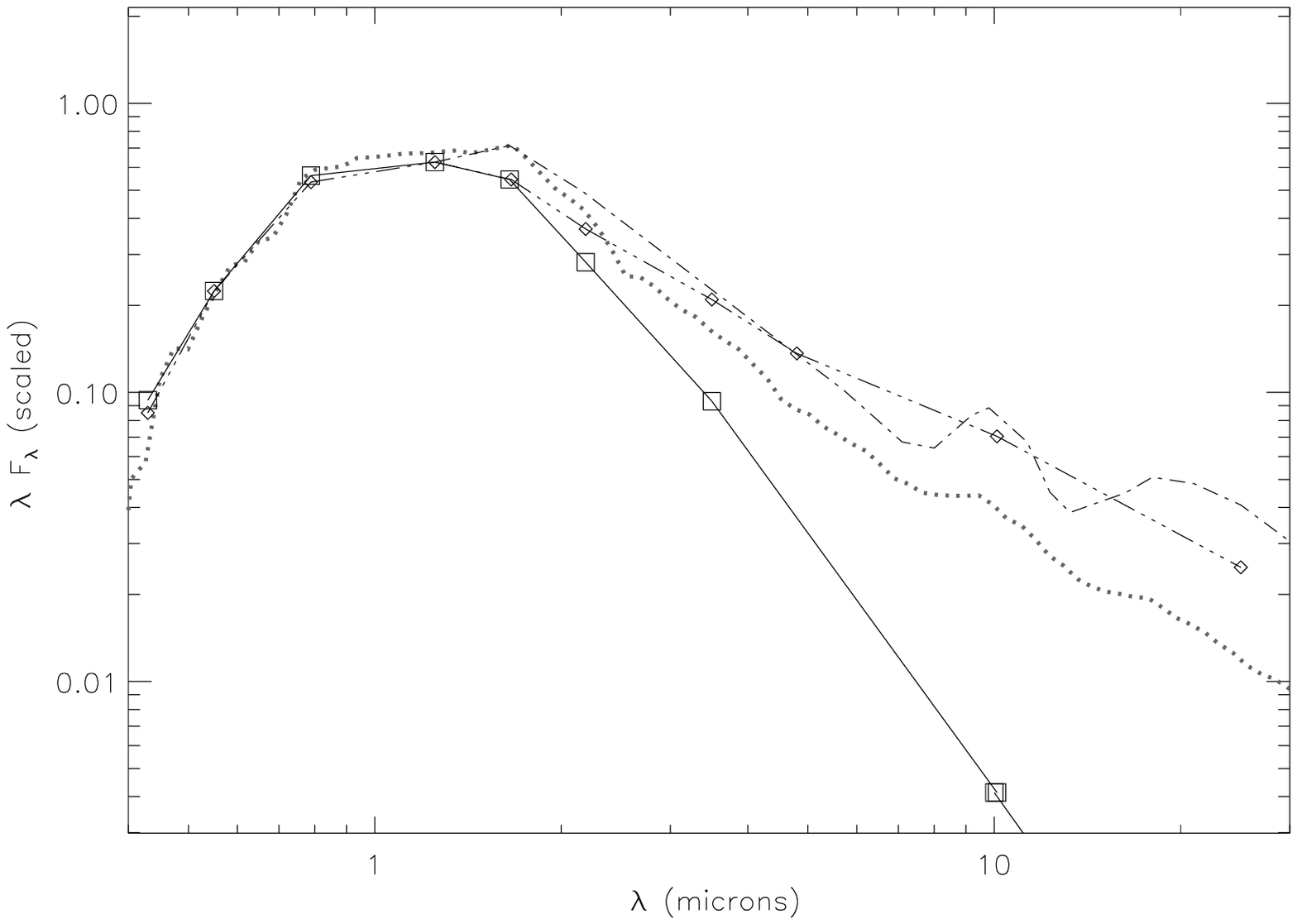}{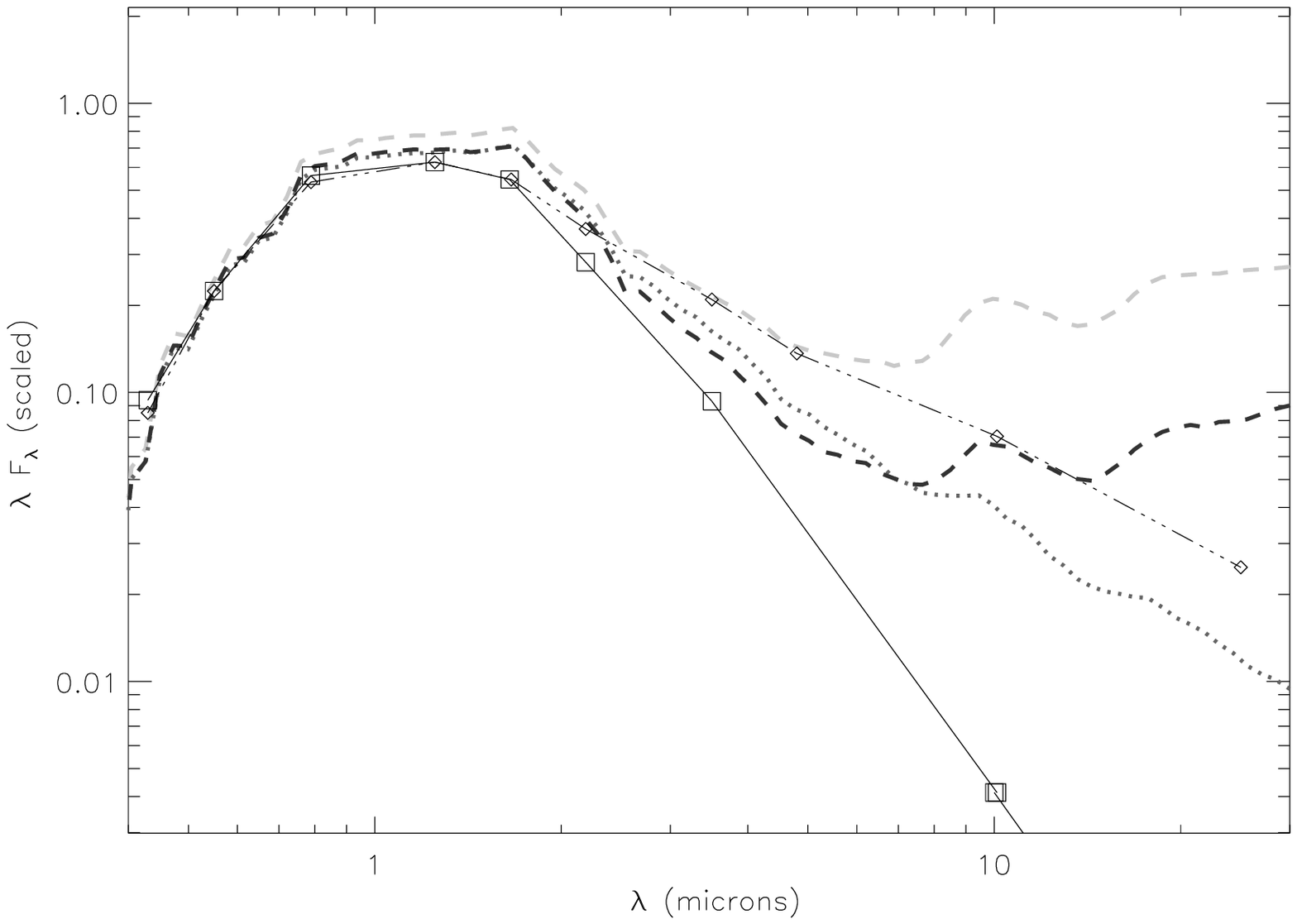}
\caption{Comparisons between various empirical and theoretical SEDs and 
our fiducial flat disk models used for analysis-- the \citet{KenyonHartmann1987} 
flat, reprocessing disk and our non-flaring, optically thick reprocessing disk from
the \citet{Whitney2003} Monte Carlo radiative transfer code.  The models shown 
are for M0, T$_{e}$ = 3850 K stars.  In both panels, the 
\citet{KenyonHartmann1987} disk is shown as a dashed-three dots--diamonds locus, 
the \citeauthor{Whitney2003} model is shown as a grey dotted line, and 
the stellar photosphere is shown as a solid line connected by squares.  (left) 
Comparisons with the lower-quartile Taurus SED from \citet[dash/dots][]{Furlan2006}.
(right) Comparisons with a standard flared disk model (grey dashed line) and a 
flared but "settled"/shadowed disk model (black dashed line).
The Whitney radiative transfer models deviate from the stellar photosphere and 
\citeauthor{KenyonHartmann1987} flat disk in the near-IR because it includes strong, 
hot, optically-thick emission from the puffed up inner disk regions. 
}
\label{modelcompare}
\end{figure}

We first compare source SEDs to geometrically flat, optically-thick models 
to identify disks with evidence for a reduced optical depth
of emitting material ($\tau_{IR}$ $\lesssim$ 1).  Figure \ref{modelcompare} shows 
two flattened reprocessing disk models appropriate for a T$_{e}$ = 3850 K 
primary (M0 spectral type), though SEDs for earlier and later stars (e.g. K5 to M6) 
yield similar mid-IR fluxes relative to the stellar photosphere.   
The first model (dash-three dots/diamonds) is 
the standard, razor-thin flat reprocessing disk model from \citet[][see also 
\citealt{Adams1987}]{KenyonHartmann1987}, truncating the 
inner disk at a dust sublimation temperature of T = 1500K.  
The second model (thick grey dashed line) is produced from the Whitney Monte Carlo radiative transfer 
code \citep{Whitney2003,Whitney2003b,Robitaille2006}.  The model assumes a disk mass of 0.05 M$_{\odot}$, 
no flaring (H/r = constant), no accretion, and no protostellar envelope emission.  The disk is 
optically-thick to its own radiation over spatial scales relevant for our study ($\lesssim$ 10 AU).

The Whitney disk model consistently has weaker disk emission than the simple \citeauthor{KenyonHartmann1987} 
flat disk model, because stellar photons in the Whitney model are partially 
attenuated before they reach the $\tau$ = 1 surface, due to the disk's finite thickness.
  Thus, the disk is colder at a given location and reradiates 
less energy.  To provide an empirical comparison for source SEDs, we overplot the lower-quartile 
Taurus SED (Figure \ref{modelcompare}, left panel) 
from \citet{Furlan2006}, which has been used in other work to help distinguish disks in 
different states \citep{Cieza2008,CurrieLada2009,CurrieKenyon2009}.
The lower-quartile Taurus SED almost perfectly tracks the 
\citeauthor{KenyonHartmann1987} flat disk model.

In the right panel of Figure \ref{modelcompare}, we compare our fiducial SEDs to a flared 
optically-thick disk model (thick grey long-dashed line) 
with H/r = r$^{2/7}$ \citep[e.g.][]{ChiangGoldreich1997} using the Whitney 
code and a disk that has undergone significant dust settling (thick black long-dashed line).  
Following \citet{Lada2006}, we quantify dust settling by reducing 
the disk scale height by a factor of 3 compared to the thermal equilibrium value 
adopted in the flared disk model\footnote{Technically, we assume less "settling" than 
in other "settled disk" models, like those presented in \citet{Dalessio2006}.  However, 
their treatment of settling assumes that some small, strongly emitting grains 
remain at larger heights above the disk midplane, whereas the Whitney code lacks 
this superheated grain population and thus may underpredict near-to-mid IR disk emission.  
Furthermore, some of the \citet{Dalessio2006} "settling" models likely produce weak 
emission because they assume low disk masses, in addition to being "settled".}.
The Whitney settled disk model predicts emission roughly equal to the Whitney flat disk model through 
$\sim$ 7 $\mu m$ and greater emission at longer wavelengths in spite of being flared: 
this probably happens because the puffed up inner wall effectively attenuates flux 
from disk regions slightly exterior (e.g. the disk only comes out of shadow at regions 
that effectively emit at $\lambda$ $>$ 7 $\mu m$).

\begin{figure}
\plottwo{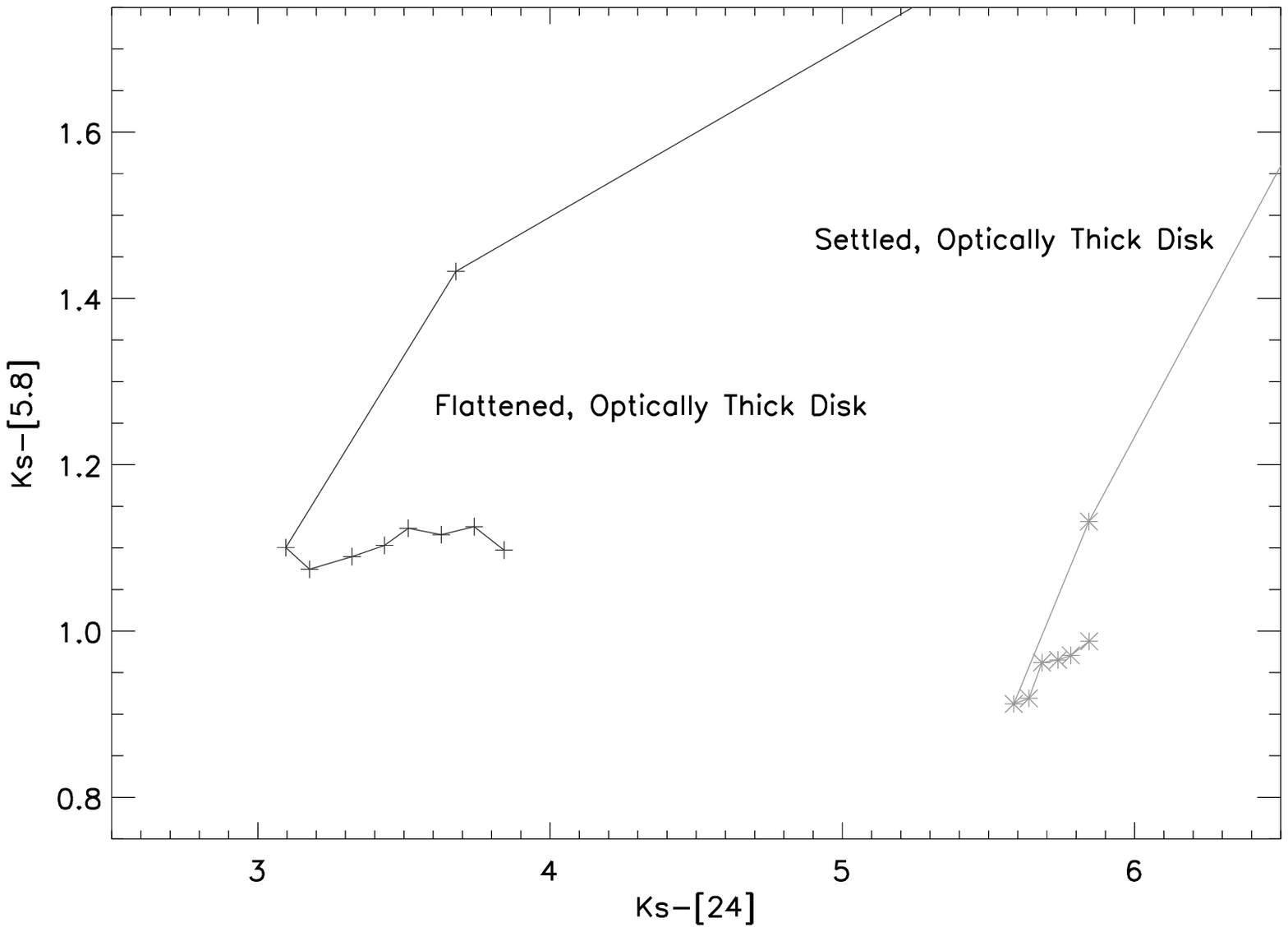}{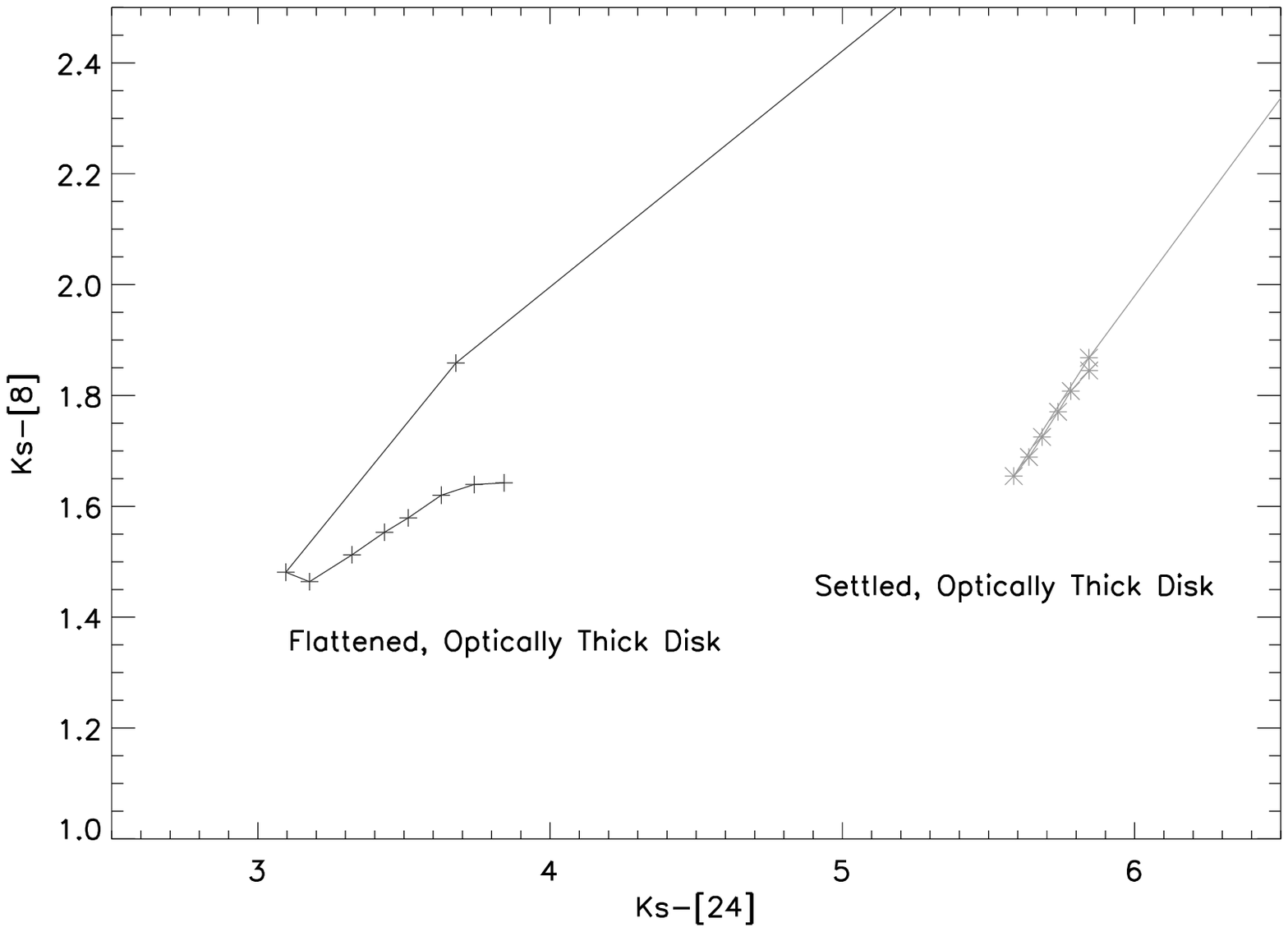}
\caption{Loci of colors for an optically-thick flattened disk around a M0 star (black solid line) and 
an optically-thick "settled" disk around a M0 star (grey solid line) as a function of disk inclination.  
Disks viewed edge on have extremely red colors and lie off the boundary of the plot.  The right-most 
point in each plot corresponds to the face-on case.  Note that since the model predicts some 
K-band excess, the colors \textit{compared to that of the stellar photosphere} are redder 
by an additional $\sim$ 0.1--0.2 mags.}
\label{modelcolors}
\end{figure}

In Figure \ref{modelcolors}, we plot the K$_{s}$-[5.8,8] vs. K$_{s}$-[24] colors for the 
flattened and settled disk models as a function of disk inclination, produced by
 convolving the model output with 2MASS and Spitzer filter functions provided 
as a part of the Whitney code.  There is little variation 
in color, especially for the flat model, except when the disk is viewed nearly edge on.  In this case, 
the disk heavily extincts the star, causing its observed colors to be far redder than in face-on 
cases.  The loci of colors indicate that flattened optically-thick disks should typically 
have K$_{s}$-[5.8,8, 24] $>$ 1.1, 1.5, 3.5.  Settled disks have slightly bluer K$_{s}$-[5.8] colors ($\sim$ 0.85--1) 
but redder K$_{s}$-[8,24] colors (1.6--1.9).  For the models where the disk is not viewed edge on, the 
disks have K-band excess emission of $\sim$ 0.1--0.2 magnitudes, consistent with a visual inspection of Figure \ref{modelcompare}.  
Thus, optically-thick, flattened disks lacking K band excess (e.g. because of submicron-sized grain growth),
should then typically have K$_{s}$-[5.8] $\gtrsim$ 1.2 and K$_{s}$-[8] $\gtrsim$ 1.6--1.7.  
The colors for an optically-thick flared disk lie off the plot range (e.g. K$_{s}$-[8] = 2.75, K$_{s}$-[24] = 6.75).  The disk emission 
at 24 $\mu m$ varies wildly with disk flaring, so MIPS-24 $\mu m$ does not effectively probe 
 the disk optical depth.  The colors from disk models for K5--M6 stars agree to within $\sim$ 0.05--0.1 mag .

Since the Whitney flat disk model produces weaker emission through 8 $\mu m$ than the other fiducial models,
 it defines a conservative limit for disks with optically-thick emission.
Because we identify primordial disks as those with optically-thick IR emission, 
disks with 2--8 $\mu m$ emission greater than this flat disk model are 
consistent with being primordial disks.   Conversely, disks with weaker emission 
correspond to later evolutionary stages: transitional disks and debris disk candidates.

We emphasize that our adopted optically-thick disk limit is conservative.  Emission in 
excess of this limit at $\lambda$ $\sim$ 5--8 $\mu m$ 
may be explained by disk flaring.  While the model 
used to produce this limit lacks flaring, real disks almost assuredly 
have some flaring.  Furthermore, the IRAC 8 $\mu m$ channel overlaps with the 10 $\mu m$ 
silicate feature, complicating comparisons with the flat disk limit \citep[see][]{Muzerolle2010}.  
  The model also lacks accretion luminosity, which contributes some IR flux, but many disks 
we model show unambiguous accretion signatures.

\subsubsection{SED Modeling}
After comparing source SEDs to our flat reprocessing disk SED, we model 
optical-to-mid IR fluxes with the grid of radiative transfer 
disk models from \citet{Robitaille2006}, which yields estimates for disk properties (e.g. 
mass, inner radius) consistent with other independent work \citep[see ][]{Robitaille2007}.  
By identifying the best-fitting models (min($\chi^{2}$)), we determine whether the disk plausibly has 
an inner region cleared of dust and is thus 
consistent with being a transitional disk with an inner hole.
The computed inner disk radius distinguishes disks that lack inner holes (primordial, 
homologously depleted transitional disk) and transitional disk with inner holes.
To be conservative, we set the division at ten times the dust sublimation 
radius: disks with R$_{in}$ $\ge$ 10 R$_{sub}$ are identified as transitional disks with inner holes.  
%The computed disk mass assumes a solar gas-to-dust ratio.
To account for flux calibration uncertainties, photometric errors, and variability, we 
assume a minimum 10\% flux uncertainty in each photometric filter and IRS monochromatic flux density.  

We restrict ourselves to disk models which provide a good fit, 
which we define as $\chi^{2}$-$\chi^{2}$$_{best}$ $<$ 3, where $\chi^{2}$$_{best}$ 
corresponds to the minimum $\chi^{2}$ \textit{per datapoint}.  This criteria is 
very similar to that used in previous studies \citep[e.g.][]{Robitaille2007,CurrieLada2009, Ercolano2009}. 
%For sources analyzed in this paper (in \S 3 and \S 4), this threshold typically corresponds to 
%the 4--6$\sigma$ confidence limit.  
Finally, we incorporate systematic uncertainties in 
the extinction A$_{V}$ and distance.  For the extinction, we nominally assume a 20\% systematic uncertainty  
with a "floor" of $\pm$ 0.5 mags.  For the distance, we assume 150 pc with an uncertainty of $\pm$ 20 pc.
%The sampling for the Robitaille grid is 
%stronger for disks without inner holes and disks with larger masses (M$_{disk}$ $\gtrsim$ 10$^{-2}$ 
%M$_{\odot}$) expected if their emission is optically thick.  

%We again stress that our disk classification presented in \S 3 is based on 1--24/70 $\mu m$ and, 
%especially for sources identified as having homologously depleted disks,  it is provisional.  
%Longer wavelength data provide much stronger constraints on disk mass \citep{Wood2002,Robitaille2007}, 
%which is important for determining whether sources identified as homologously depleted 
%transitional disks show evidence for a reduced mass of dust consistent with disk clearing 
%or whether extreme grain growth and dust settling/shadowing in primordial 
%disks may also explain their SEDs \citep[e.g.][]{Furlan2006}.    
%Modeling restricted to near-to-mid IR data is unable to conclusively 
%break this degeneracy, but we will address it in \S 4.

\subsection{Results}
Tables \ref{diskstatenew} and \ref{diskstateold} describe our provisional disk modeling results.   
The combined sample from both tables consists of 5 protostars, 16 stars with primordial disks, 6 
with homologously depleted transitional disks, 4 that have transitional disks with 
inner holes, and 6 bare stellar photospheres.  Sources not analyzed here lack 
sufficient photometric data to model their SEDs and/or determine their 
plausible range of spectral types (e.g. FP-8, FP-33, FP-37, and FP-38), making it impossible to identify 
their disk evolutionary states.  Below we discuss 
modeling of several sources illustrating how we determine disk states.

\subsubsection{Disks in Different Evolutionary States}
\begin{figure}
%\plottwo{S-CrASED.eps}{S-CrAmass.eps}
\plotone{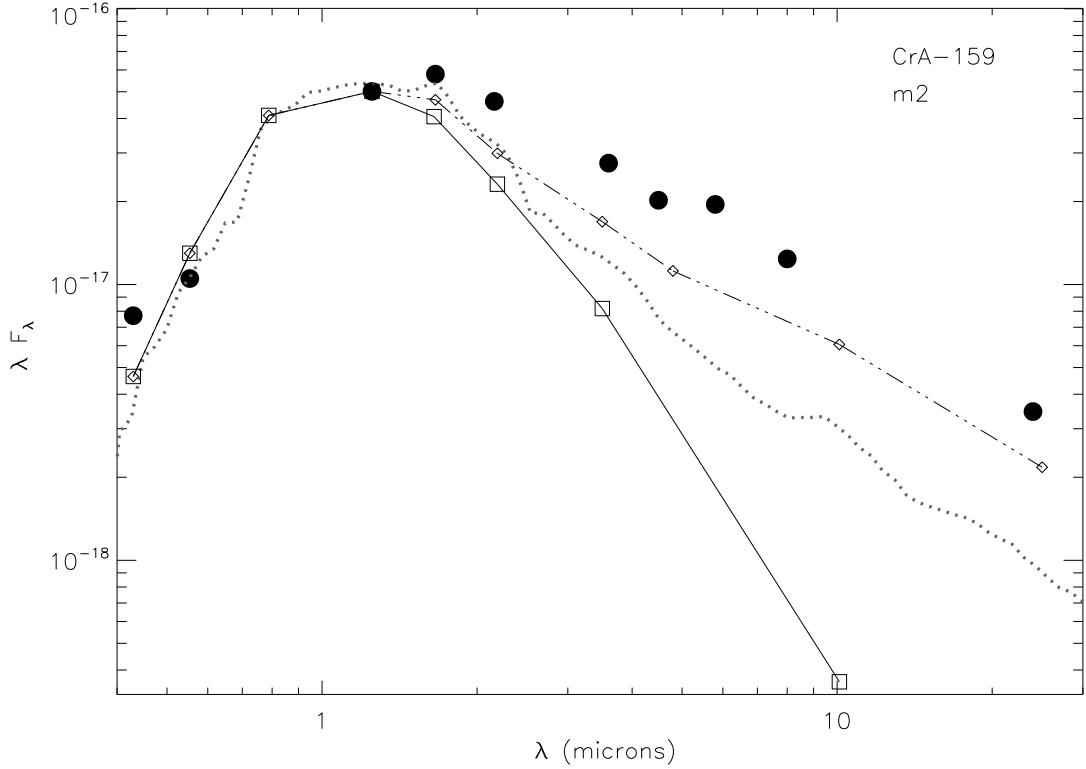}
\caption{
SED analysis for CrA-159, comparing its emission to 
the razor-thin flat optically thick disk model (e.g. 
\citealt{KenyonHartmann1987}, dash-three dots/diamonds) and a flat, optically-thick 
disk model constructed from the Whitney radiative transfer code (dotted line).
For reference we overplot the stellar photosphere as a solid line.  CrA-159 has 
disk emission consistently lying above the Whitney flat disk model and thus we 
classify it as having a primordial disk.
}
\label{cra159}
\end{figure}

Figure \ref{cra159} compares the SED for CrA-159 from the new sample to 
the \citeauthor{KenyonHartmann1987} razor-thin flat disk model and the 
flat disk model produced from the Whitney radiative transfer code.  
The observed SED is dereddened from A$_{V}$ = 3.  CrA-159 has an infrared excess 
detectable from 1.6--2 $\mu m$ to 24 $\mu m$, which consistently lies well 
above the Whitney flat disk model, indicating that its emission is optically 
thick.  According to our criteria, this source then has a primordial disk. 
With the exception of one star (V709), all disks in the new Coronet sample, 
predominately comprised of higher-mass stars, yield emission lying above the flat, optically-thick disk limit.

\begin{figure}
%\plottwo{CrA8.ps}{CrA4110mass.eps}
\epsscale{0.5}
\plotone{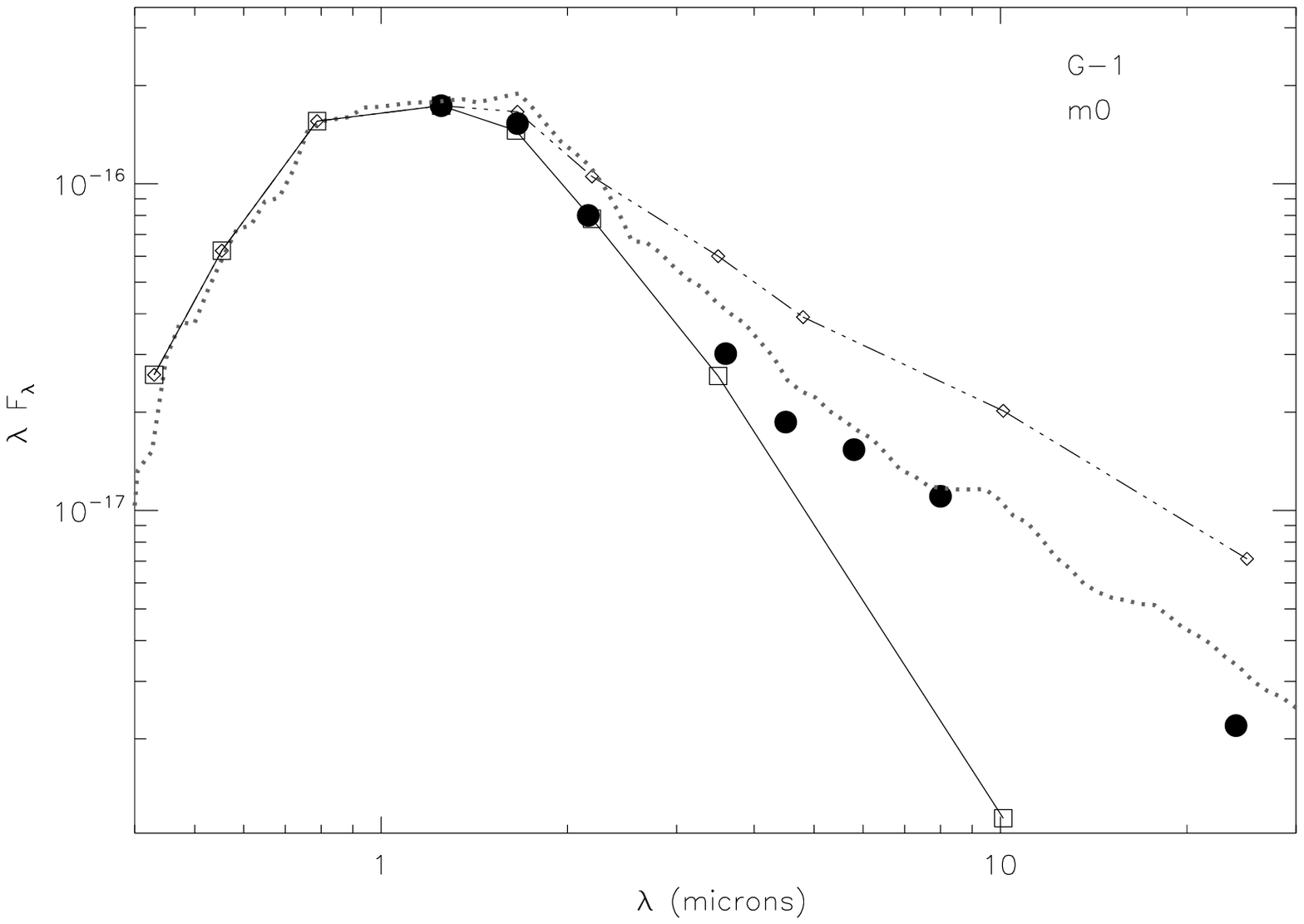}
\plotone{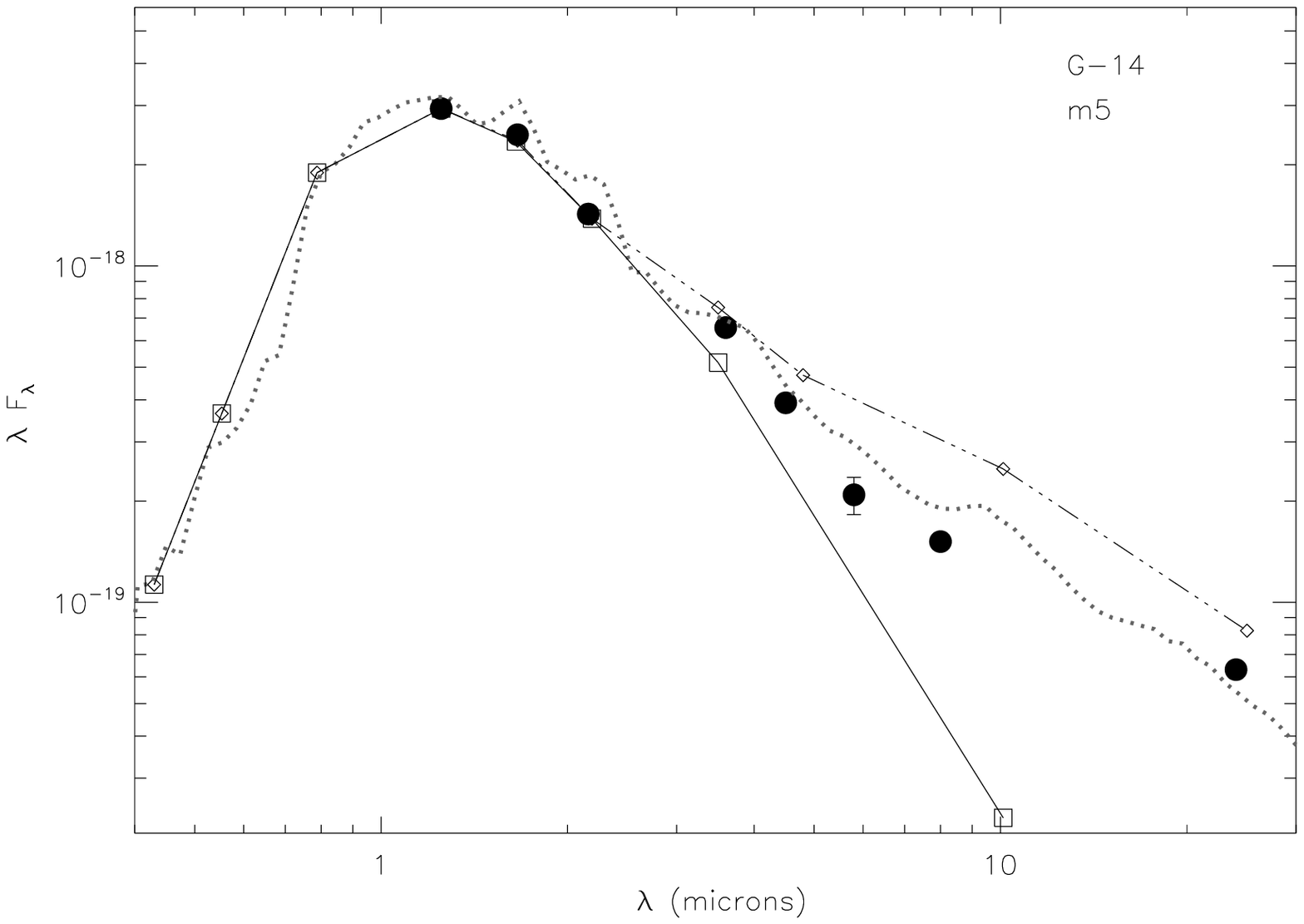}
\plotone{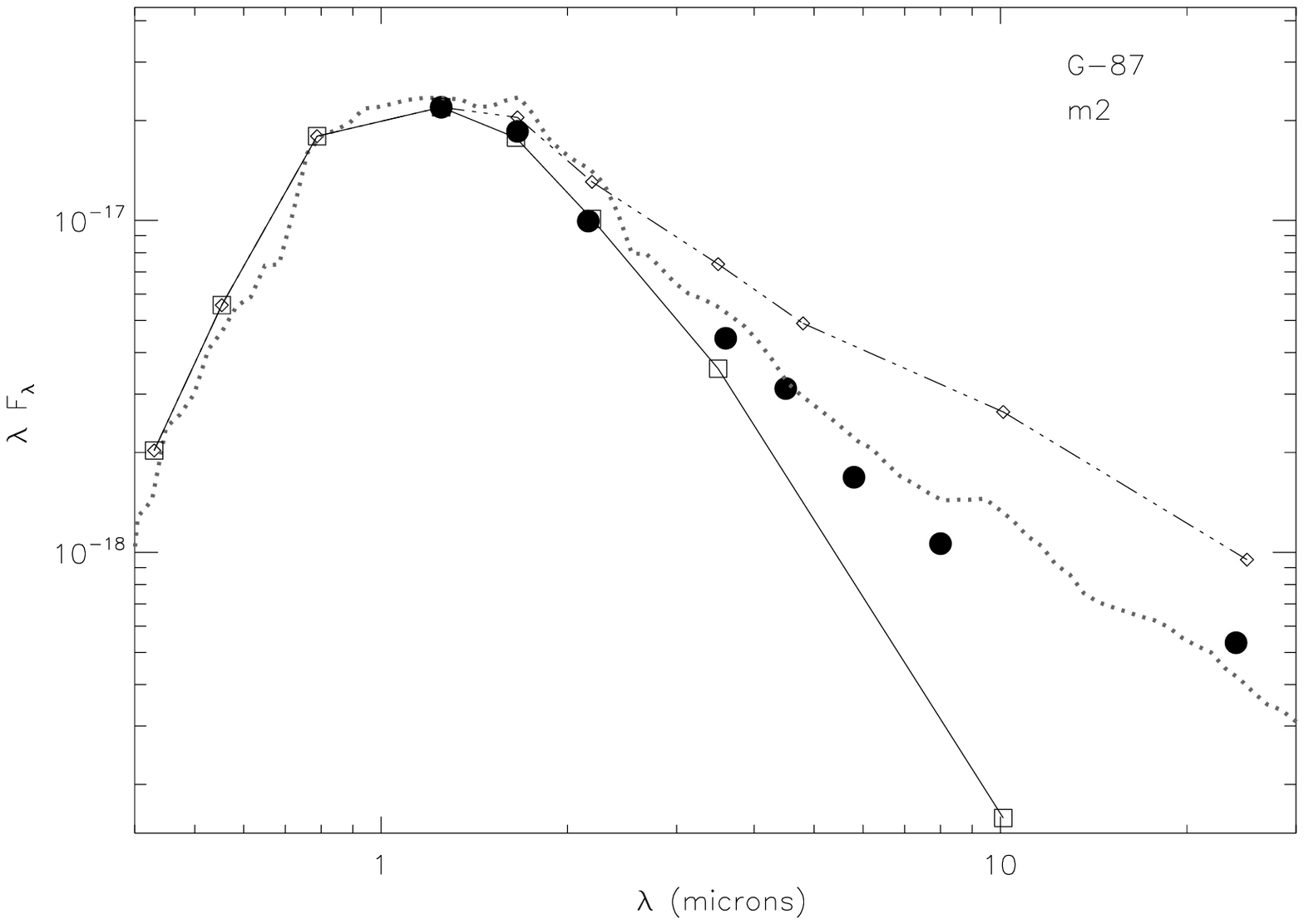}
\caption{SEDs G-1 (top), G-14 (middle), and G-87 (bottom).  We fit the optical/near-IR portion 
of the SED to a stellar SED constructed from the \citet{Currie2010} intrinsic colors (solid line), 
overplot the flat, optically-thick disk model constructed from the Whitney radiative transfer code 
(dotted line) and 
overplot the flat, reprocessing disk SED from \citet[][dash-three dots]{KenyonHartmann1987}.  
All three stars have homologously depleted transitional disks.}
\label{homog}
\end{figure}

The disk population for the \citet{SiciliaAguilar2008} sample is far more diverse, 
including many transitional disks. 
We show comparisons between three homologously depleted transitional disk SEDs 
and flat disk models in Figure \ref{homog}.  
Each source has a reduced level of disk emission compared 
flat disk models.    Likewise, all transitional disks with 
inner holes have weak/negligible 3.6--5.8 $\mu m$ emission but optically-thick 
24 $\mu m$ emission.  

\begin{figure}
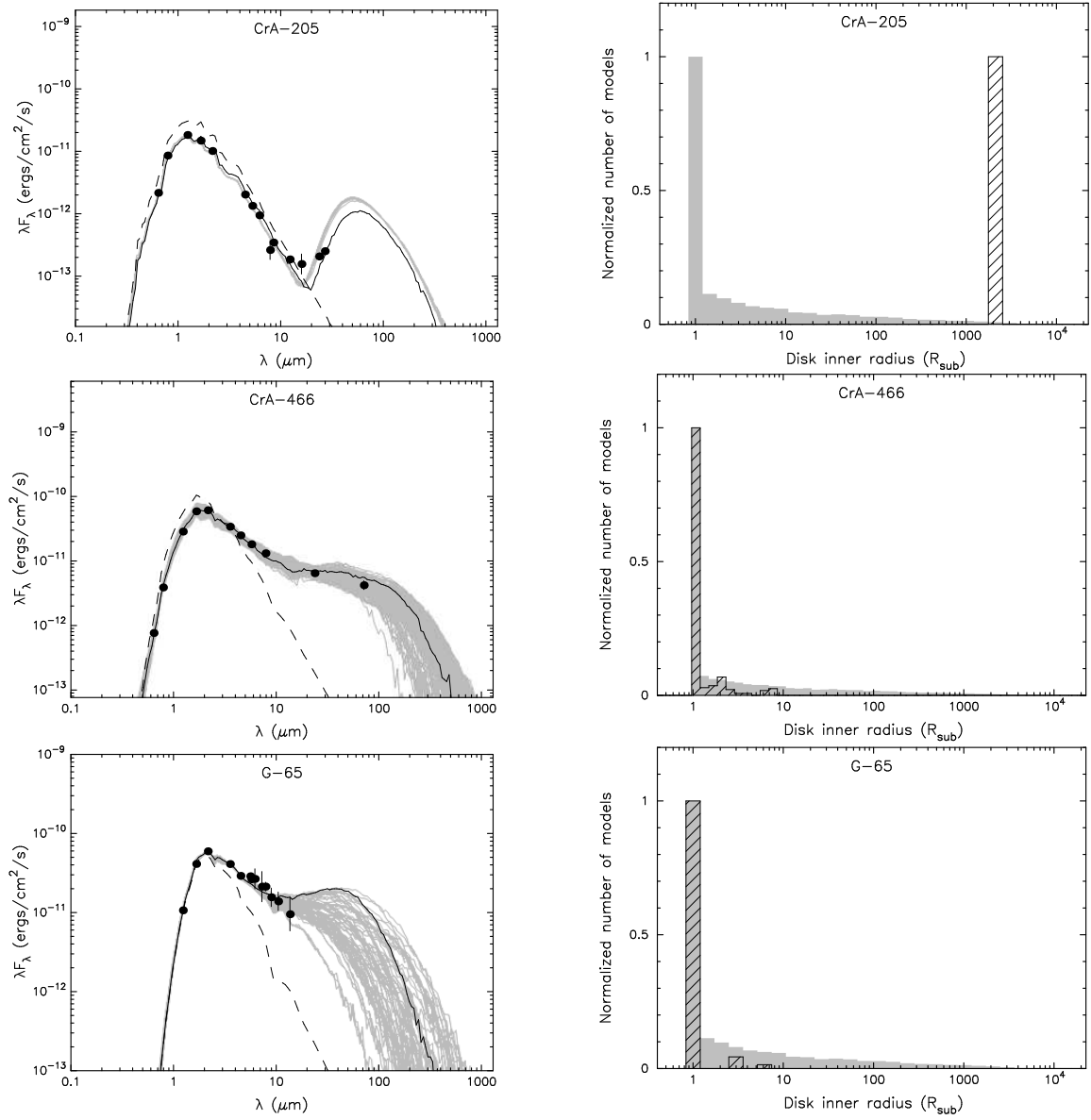

\epsscale{0.9}
\centering
\plottwo{CrA205sed.eps}{CrA205inner.eps}
\plottwo{CrA466sed.eps}{CrA466inner.eps}
\plottwo{G65sed.eps}{G65inner.eps}
\caption{Spectral energy distributions and histograms of inner disk radii for 
CrA-205, CrA-466, and G-65.  For the righthand panels, the hatched region shows the distribution of 
disk inner radii from best-fit models; the grey shaded region shows the normalized distribution of inner radii from 
all models.  CrA-466 and G-65 do not show evidence for 
an inner hole, in agreement with \citet{Ercolano2009}. CrA-205 
has an inner hole and thus is a transitional disk, in agreement with \citet{SiciliaAguilar2008}.}
\label{ercomp}
\end{figure}

The top panel of Figure \ref{ercomp} shows model fits for CrA-205, 
which has a transitional disk with an inner hole.  
CrA-205 exhibits no excess emission shortwards of $\sim$ 10 $\mu m$ but has a rising SED 
clearly departing from the stellar photosphere by $\sim$ 15 $\mu m$.  All 
of the best-fitting Robitaille models have inner disk radii (R$_{in}$ $>$ 1000 R$_{sub}$) 
much larger than our threshold identifying transitional disks with 
inner holes (10 R$_{sub}$).  Moreover, the grid sampling of inner disk 
radii (grey shaded region) is heavily peaked at 1 R$_{sub}$ (e.g. no inner hole).  
In spite of this intrinsic bias against selecting disk models with holes, 
all disk models lacking inner holes fail our $\chi^{2}$ threshold.
%There is good correspondence between the disk masses inferred from the Robitaille 
%radiative transfer models and the disk states inferred by flat disk models/empirical 
%comparisons: all disks with mid-IR emission less than the lower quartile Taurus 
%SED and flat disk models have predicted disk masses ($<$ 10$^{-4}$ M$_{\odot}$) 
%from the Robitaille grid.  

Our results for the \citeauthor{SiciliaAguilar2008} sample are intermediate 
between those of \citet{SiciliaAguilar2008} and \citet{Ercolano2009} 
who disagreed over the fraction of transitional 
disks with inner holes.  CrA-466 and G-65 have best-fit models that do not require an 
inner hole (Figure \ref{ercomp}, middle and bottom panels).
Like \citet{Ercolano2009}, we identify them as primordial disks.
  However, our modeling supports the claim by \citet{SiciliaAguilar2008,SiciliaAguilar2009} 
that CrA-205 has an inner hole, in fact the largest one in our sample.  
\citet{Ercolano2009} did not to identify CrA-205's inner hole probably because they do not 
include the IRAC 8 $\mu m$ data nor any IRS data in their fitting.  

While we also agree with \citet{Ercolano2009} that  G-14 and G-87 lack evidence 
for an inner hole, our modeling suggests that both have 
weaker emission than a perfectly flat, optically-thick reprocessing disk.
  Thus, like \citet{SiciliaAguilar2008} we identify these as transitional disks, albeit ones 
that are homologously depleted.  The \citet{Ercolano2009} study focused on 
identifying transitional disks with inner holes.  Therefore, we identify 
more transitional disks because we adopt a more expansive definition for
what constitutes a transitional disk.

%Additionally,
%they obtained the data by scanning the figures in \citet{SiciliaAguilar2008}: 
%all of these choices increase their analysis uncertainties.  
%While they claim that the IRS spectra has low signal-to-noise, due to many "spikes" in 
%the flux density at $\lambda$ $>$ 8 $\mu m$, what they presume to be "noise" are nebular gas 
%emission lines at well-constrained wavelengths \citep{SiciliaAguilar2008}.  Flux densities 
%measured at wavelengths uncontaminated by gas emission provide reasonable estimates 
%for the disk flux at wavelengths probed by IRS and, when combined with IRAC and MIPS data, 
%clearly indicate the presence of an inner hole.  

Two of the sources labeled as homologously depleted transitional disks may be debris disk candidates, while 
one of the transitional disks with inner holes may be a debris disk candidate.
The identity of V709 is particularly questionable: while we identify it as a disk-bearing star, 
it is possible that its weak excess could be due to nebular contamination or an offset 
between the near-IR data and the MIPS data/IRS spectra caused by variability such 
as that produced by an undetected eclipsing companion.  
More importantly, if it has a disk, it is unclear whether it is a debris disk or transitional disk.

Cluster members with new Spitzer photometry are mostly protostars or stars with primordial disks 
(5 and 8, respectively, Table \ref{diskstatenew}).  
The disk population for M stars sample studied in \citet{SiciliaAguilar2008} is more evenly 
divided between primordial disks (8) and transitional disks (3--4 with inner holes, 
4--6 homologously depleted).  Unlike the new sample, many stars in 
the \citet{SiciliaAguilar2008} sample have homologously depleted transitional disks.
Thus, based on modeling 1--24/70 $\mu m$ data, 
the percentage of protoplanetary (primordial + transitional) disks in our 
combined sample that are transitional disks is $\sim$ 30\% (7/23).  The 
percentage rises to 41\% if debris disk candidates are 
classified as transitional disks, including V709 (11/27).  

\subsubsection{Disk Properties As a Function of Stellar Mass}
To examine the spectral type/stellar mass dependence of 
disk properties, we bin our sample into three spectral 
type groups: earlier than K5, K5--M2, and later than M2.  Assuming the 
\citet{Baraffe1998} isochrones (mixing length = 1.9 H$_{p}$) and the 
T$_{e}$ vs. spectral type scale from \citet{Currie2010b}, the corresponding 
stellar mass division is M$_{\star}$ $>$ 1 M$_{\odot}$, M$_{\star}$ = 0.5--1 M$_{\odot}$, and 
M$_{\star}$ $<$ 0.5 M$_{\odot}$\footnote{In making these divisions, we are grouping together 
intermediate-mass (M$_{\star}$ $\sim$ 2--4 M$_{\odot}$) and solar-mass T Tauri stars.  
Grouping together these stars as the first spectral type bin may hide additional 
evolutionary trends since disk evolution for intermediate mass stars ($\gtrsim$ 2--3 
M$_{\odot}$) and solar-mass stars may proceed at different rates even by 1--3 Myr \citep[e.g.][]{CurrieKenyon2009}. 
Any differences between 1--4 M$_{\odot}$ stars and lower-mass stars then applies only to the 
\textit{ensemble average} of 1--4 M$_{\odot}$ stars.}.

%Table \ref{stypediskstate} compares spectral types and disk evolutionary states.
The transitional disk frequency for Coronet Cluster members may be spectral type dependent.
At most one and probably none
 of the disks around stars earlier than K5 appear to be transitional disks (0--1/5).  Stars between K5 and M2 
also mostly have primordial disks (7/10) with 2--3 stars harboring transitional disks and 0--1 with debris disks.  
Accounting for the uncertain status of debris disk candidates,
the relative fraction of transitional disks to all protoplanetary disks for stars M2 
or earlier with plausible masses equal to 0.5 M$_{\odot}$ or greater is then f(TD)/(f(TD)+f(PD)) 
$\approx$ 0.15--0.27 (2/13, 4/15).  For K5--M2 stars only, 22--30\% of the protoplanetary disks 
are transitional disks.  Transitional disks are more prevalent around stars later than M2, comprising 
50--58\% of the protoplanetary disk population (5/10 or 7/12).  This high frequency is consistent with previous 
results from \citet{SiciliaAguilar2008}.  

However, given the small population of Coronet Cluster stars studied here 
it is unclear whether our analysis truly identifies a
\textit{statistically significant} stellar-mass dependent frequency of transitional disks 
characteristic of most young clusters.
Furthermore, it is possible that some transitional disks provisionally identified 
from modeling optical-to-mid IR data may instead be primordial disks whose 
weak emission is due to the growth and settling of submicron-sized grains \citep[e.g.][]{Furlan2006}.  
Analyzing data from clusters where far-IR and submm data is available is required to 
more conclusively break these degeneracies.  To explore these issues further, 
and provide a context for our results, we now analyze disks in other clusters.

\section{Transitional Disk Frequencies and Lifetimes as a Function of Stellar Mass: A Comparison with 
Taurus, IC 348, NGC 2362, and $\eta$ Cha} 
\subsection{Sample and Wavelength Ranges Used For Analysis}
To investigate the evolution of the transitional disk frequency with time, we analyze data in other 
clusters spanning an age range of 1--8 Myr.  Comparing our results 
for the Coronet Cluster with 1--2 Myr-old Taurus and 2--3 Myr-old IC 348 addresses whether a stellar mass-dependent 
transitional disk frequency is a general feature of 1--3 Myr-old clusters.  
By comparing these results with those for 5 Myr-old NGC 2362 and 6--8 Myr-old $\eta$ Cha, we
 investigate how the frequency of transitional disks evolves with time and 
estimate the typical transitional disk lifetime.  
%Combining these analyses 
%then places constraints on the morphologies and lifetimes of the transitional disk phase.

\textbf{Taurus} -- Members of the Taurus-Auriga star-forming region are listed in catalogs 
from \citet{Kenyon2008}, \citet{Rebull2010}, and \citet{Luhman2009b}, which 
are updates of the classic \citet{KenyonHartmann1995} catalog.  We reanalyze Taurus 
members studied in \citet{Luhman2009} to provide direct comparisons with their 
disk analysis.  For Spitzer photometry, we use the IRAC and MIPS 24 $\mu m$ 
data from \citet{Luhman2009} and MIPS 70 $\mu m$ and 160 $\mu m$ data from 
\citet{Rebull2010}.  We adopt optical photometry from \citet{KenyonHartmann1995}, 
\citet{WhiteGhez2001}, and \citet{Audard2007} and submillimeter measurements from \citet{Andrews2005}, 
\citet{Jewitt1994}, \citet{Jensen1994}, and \citet{Beckwith1991}.  
To deredden stars, we determine optical extinctions for Taurus members, using 
estimates from \citet{Furlan2006}, \citet{Rebull2010}, \citet{KenyonHartmann1995},
and \citet{Luhman2009} as starting points and then fit the optical/near-IR 
portion of the SED to a synthetic SED produced using the \citet{Currie2010b} 
intrinsic colors and the \citet{Currie2010b} and \citet{Luhman2003} effective 
temperature scales.  

A large number of Taurus sources have complete SEDs from optical to far-IR/submillimeter wavelengths. 
 We model the SEDs of 25 K5--M6 stars whose IR colors plausibly identify both primordial 
disks and transitional disks based on various color-color selection criteria in the 
literature \citep[e.g.][]{Lada2006,CurrieKenyon2009,Luhman2009,Muzerolle2010}\footnote{We do not present 
modeling results for sources such as DM Tau, whose classification 
as a transitional disk has broad agreement}.  Quantitatively, our selection criteria 
includes K5--M6 sources with K$_{s}$ -[8] (dereddened) = 1.25--2.
We further restrict our sample to sources whose far-IR/submillimeter data 
can yield good constraints on the range of plausible disk masses (see Appendix). 
Specifically, we require that the source has published submillimeter data 
\textit{or} a \textit{detection} in the far-IR MIPS bandpasses (70 $\mu m$ or 160 $\mu m$).
While this criterion may bias our selection against including the most depleted disks, 
especially for the lowest mass stars, we include it to be conservative.

\textbf{IC 348} --   For IC 348, we compile optical, near-IR and Spitzer/IRAC photometry, spectral types and 
extinctions presented in \citet[][and references therein]{Lada2006} and MIPS photometry 
and upper limits presented in \citet{CurrieKenyon2009} for 307 members listed in both studies as well 
as data for 41 additional spectroscopically confirmed members from \citet{Muench2007}. 
We restrict our analysis to sources with MIPS 24 $\mu m$ detections.
\citet{CurrieKenyon2009} identified mid-IR colors for disks in different evolutionary 
states by modeling the SEDs of newly-detected disks in IC 348, but their 
sample is comprised mainly of stars later than M3--M4.  Here, we model the SEDs of 25 IC 348 
stars previously detected by \citet{Lada2006} and \citet{Muench2007} with full SEDs through 24 $\mu m$ 
focusing on those with the same range of mid-IR colors analyzed in Taurus. 
Most of these 25 stars are earlier than M3--M4. 

As with the Taurus sample, we use the published extinction estimates as a starting point and fit 
the optical/near-IR portion of the SED to arrive at a final value.
In most cases, the best-fit optical extinction (A$_{V}$) matches that listed 
by \citet{Lada2006}.  Exceptions include IDs 26, 58, 97, 110, and 9024: their 
new extinction estimates are generally larger.  

\begin{figure}
\epsscale{0.75}
\plotone{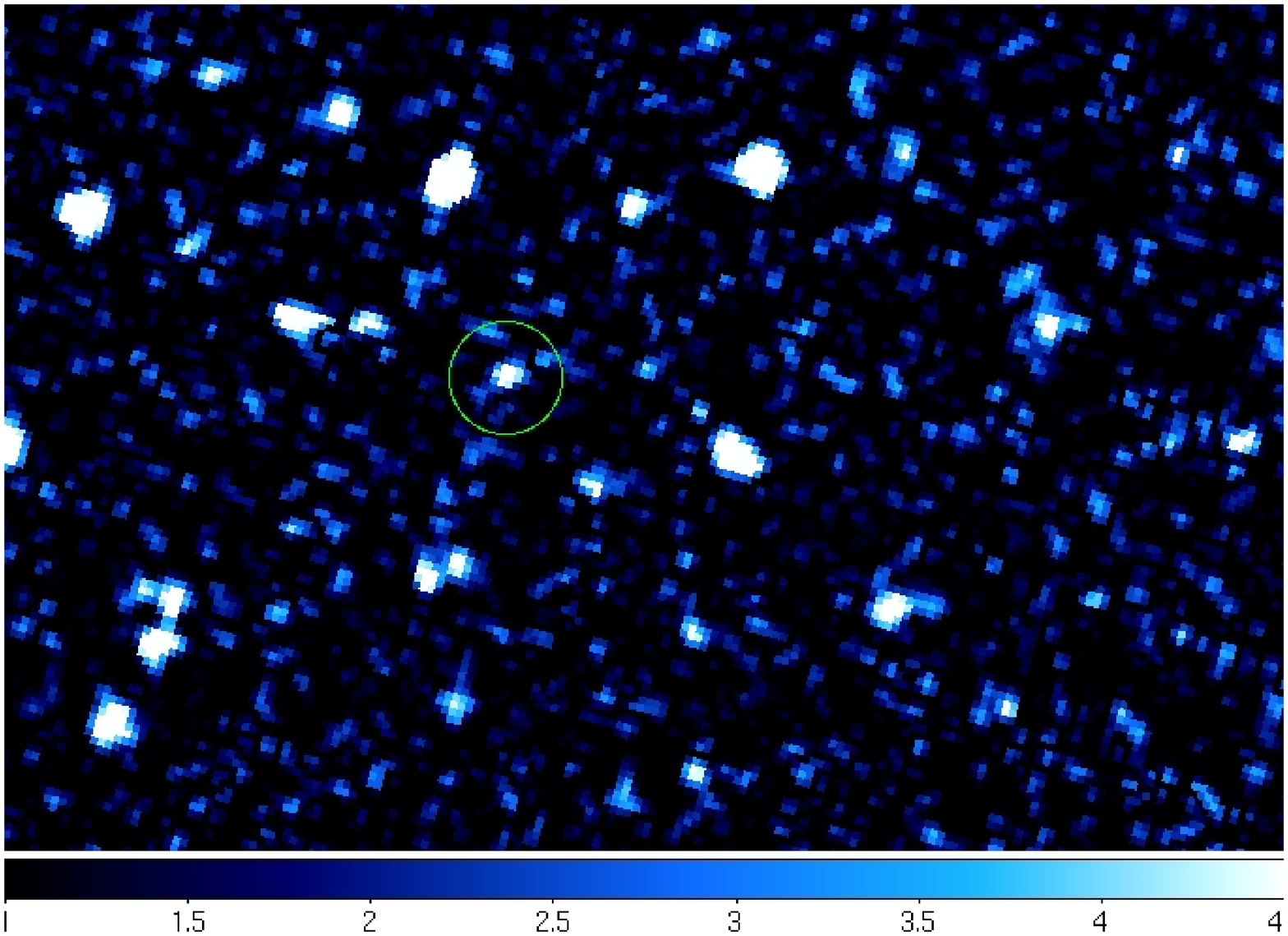}
\plotone{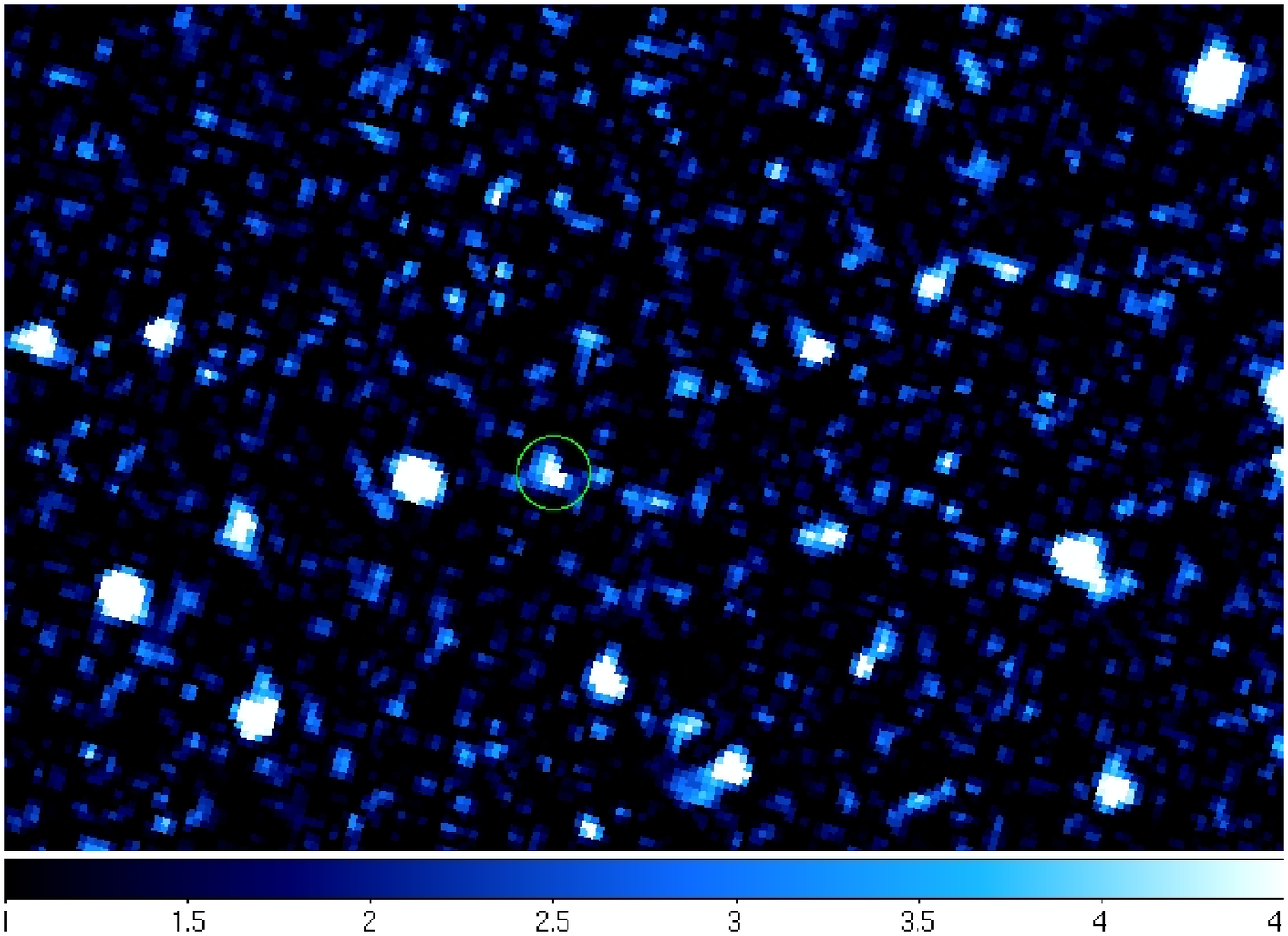}
\caption{Signal-to-noise maps constructed from our reduced NGC 2362 MIPS mosaic image with the positions of ID-41 (top) 
and ID-63 (bottom) circled.  The color stretch in each panel goes from 1 to 4.5 $\sigma$ (black to blue 
to white).  The signal-to-noise for the integrated flux is $\sim$ 5 for both ID-41 and 63.  \citet{Luhman2009} 
claim ``from inspection of the mosaic" that both of these sources have SNR $<$ 3.  If that 
were the case, none of the pixels centered on these targets would be light blue or white.  However, their removal of 
two stars (IDs 663 and 1091 but not IDs 809 or 931) from the \citeauthor{Irwin2008} sample 
not considered here is better justified:  one does has a low signal-to-noise and the other is
 blended with a different source.}
\label{ngc2362snrimage}
\end{figure}

\textbf{NGC 2362} -- For NGC 2362, we analyze the Spitzer IRAC and MIPS photometry presented by \citet{CurrieLada2009}.
\citet{CurrieLada2009} present data for two membership lists for 
NGC 2362: members/probable members identified by \citet{Dahm2005}
and candidate members identified by \citet{Irwin2008}.  
Here we analyze Spitzer data only for members/probable members identified 
by \citet{Dahm2005} to be conservative.  
We include the two sources -- IDs 41 and 63 -- not considered in analysis 
by \citet{Luhman2009}, because both have $\approx$ 5$\sigma$ detections and 
thus should be included in our analysis (Figure \ref{ngc2362snrimage}).

\citet{CurrieLada2009} already model the SEDs for NGC 2362 stars with the 
\citet{Robitaille2006} radiative transfer grid to identify disks with 
inner holes.  Therefore, we only compare NGC 2362  
SEDs through 24 $\mu m$ to the appropriate flat reprocessing disk models.  
As starting points for our flat disk comparisons, we use extinction estimates listed in \citet{CurrieLada2009} 
that were based off of comparisons with the \citet{KenyonHartmann1995} colors and 
off of output from fits with the \citeauthor{Robitaille2006} models treating the extinction 
as a free parameter.  

\textbf{$\eta$ Cha} -- For $\eta$ Cha, we use the IRAC and MIPS 24 $\mu m$ and 70 $\mu m$ photometry and 
spectral types from \citet{SiciliaAguilar2009}, MIPS 70 $\mu m$ upper limits 
and MIPS-160 $\mu m$ photometry/upper limits from \citet{Gautier2008}, and IRS data 
presented in \citet{Bouwman2006} and \citet{SiciliaAguilar2009}.
To model $\eta$ Cha sources, we compare source SEDs to appropriate flat disk 
models and fit their SEDs through 160 $\mu m$ with the Robitaille model grid 
to identify inner holes and estimate disk masses.  

\subsection{Classification Method}
Our disk classification method is an expanded, more thorough version 
of that used in \S 3.  As in \S3, we identify transitional disks with 
inner holes from fitting source SEDs with the Robitaille radiative transfer models.  
The difference between our method here and in \S 3 is in our 
differentiation between primordial disks and 
homologously depleted transitional disks.  
In \S 3, members of the latter group were identified as those having optically thin 
mid-IR emission (from comparing their emission to the flat disk 
limit) but lacking evidence for inner holes.  As stated in \S 3, we 
considered the classification of these disks as provisional 
since weak mid-IR emission could be due to a low overall 
 mass of submicron to submillimeter-sized dust \textit{or} the growth and settling of 
just the smallest, submicron-sized grains to micron sizes or larger.  
Modeling based on 1--24 $\mu m$ data alone is unable 
to break this degeneracy.  

However, SED modeling of sources 
with far-IR/submm data helps break this degeneracy 
\citep[e.g.][]{Wood2002}.  Taurus and $\eta$ Cha include many sources 
with far-IR and/or submm data. Thus, SED modeling of Taurus and $\eta$ 
Cha sources can identify which ones with optically-thin mid-IR emission 
but no inner hole have 1) a low inferred disk mass or 2) an inferred disk mass 
comparable to optically-thick primordial disks.  The former group 
are classified as homologously depleted transitional disks \citep{CurrieLada2009}; 
the latter group could be primordial disks undergoing extreme 
grain growth/dust settling/shadowing \citep{Furlan2006}. 

To identify the range of disk masses for optically-thick primordial disks
and thus most appropriate limit for dividing between primordial 
disks and homologously depleted transitional disks, we refer to 
SED modeling results for Taurus \citep{Andrews2005} which are discussed in detail in 
the Appendix.  Briefly, typical primordial disk masses as a fraction of 
stellar mass are M$_{disk}$ $\sim$ 0.01 M$_{\star}$ for Taurus: the interquartile range covers 
M$_{disk}$ $\sim$ 0.003--0.03 M$_{\star}$ \citep[Appendix, Figure 7][]{Andrews2007}\footnote{ 
\citet{Andrews2005,Andrews2007} use the term "Class II objects" to describe sources with 
optically-thick disk emission, though their sample includes some transitional disks with inner holes 
(e.g. UX Tau, CoKu Tau/4, GM Aur)}.  
%Also shown by Figure 7 of \citet[][]{Andrews2007}, 
About 92--96\% of the primordial disks in the \citeauthor{Andrews2005} 
Taurus sample have inferred masses greater than 0.001 M$_{\star}$ (cf. Appendix).  
As discussed in the Appendix, far-IR data ($\lambda$ $\gtrsim$ 70 $\mu m$) is sufficient 
to estimate the disk mass at a precision useful for our study.

Thus, if we define the range of 
primordial disk masses by the interquartile range, the appropriate limit 
differentiating optically-thick primordial disks from homologously depleted transitional disks 
is M$_{disk}$ = 0.003 M$_{\star}$.  If we use the full range of primordial disk 
masses, the appropriate limit is M$_{disk}$ = 0.001 M$_{\star}$.
As discussed in the Appendix, these 
estimates for typical masses of optically-thick primordial disks, and 
thus our criteria for separating primordial and transitional disks, are 
not compromised by incompleteness.
In our analysis, the disk mass is the median value of masses from the best-fit 
\citet{Robitaille2006} models; to determine the fractional disk mass, we 
use the \citet{Baraffe1998} isochrones and the effective temperature 
scales from \citet{Currie2010b} for K5--M2 stars and \citet{Luhman2003} 
for later stars.  

Combining our criteria from \S 3 with additional criteria presented in this section, 
we consider a disk in Taurus or $\eta$ Cha to be a primordial disk if it meets the following 
two conditions:
\begin{itemize}
\item It lacks evidence for an inner hole/gap as inferred from SED modeling.
\item Its mid-IR emission exceeds the optically-thick, flat disk limit \textit{or} 
it has an inferred disk mass of M$_{d}$ $>$ 0.003 M$_{\star}$.
\end{itemize}
Disks with inferred masses of greater than 0.001 M$_{\star}$ and 
up to 0.003 M$_{\star}$ but no inner hole are considered to be borderline cases, as 
their classification depends on whether the interquartile range or full range of 
optically-thick primordial disk masses is used to separate primordial disks from homologously depleted 
transitional disks.  Disks with masses $\le$ 0.001 M$_{\star}$, optically-thin mid-IR emission, 
and no inner hole are classified as homologously depleted transitional disks.  
We then determine the transitional disk frequency by dividing the number of transitional 
disks by the number of transitional + primordial disks.

To analyze disks in clusters lacking far-IR/submm data -- IC 348 and NGC 2362 -- 
we first assign provisional classifications based on \S 3 methods, which 
yield provisional frequencies for transitional disks.  Since mid-IR data 
alone is unable to differentiate between disks with a low dust mass and 
those with substantial dust settling, some the disks 
identified as transitional based on mid-IR data could instead be primordial 
disks.  Thus, we use the Taurus results to identify the mid-IR colors of 
transitional disks and determine the "contamination rate" of 
the transitional disk population selected by mid-IR data alone.
Transitional disk frequencies for IC 348 and NGC 2362 are then revised after taking into 
account this contamination rate.

\subsection{Transitional Disks in Taurus}
\begin{figure}
\centering
\plottwo{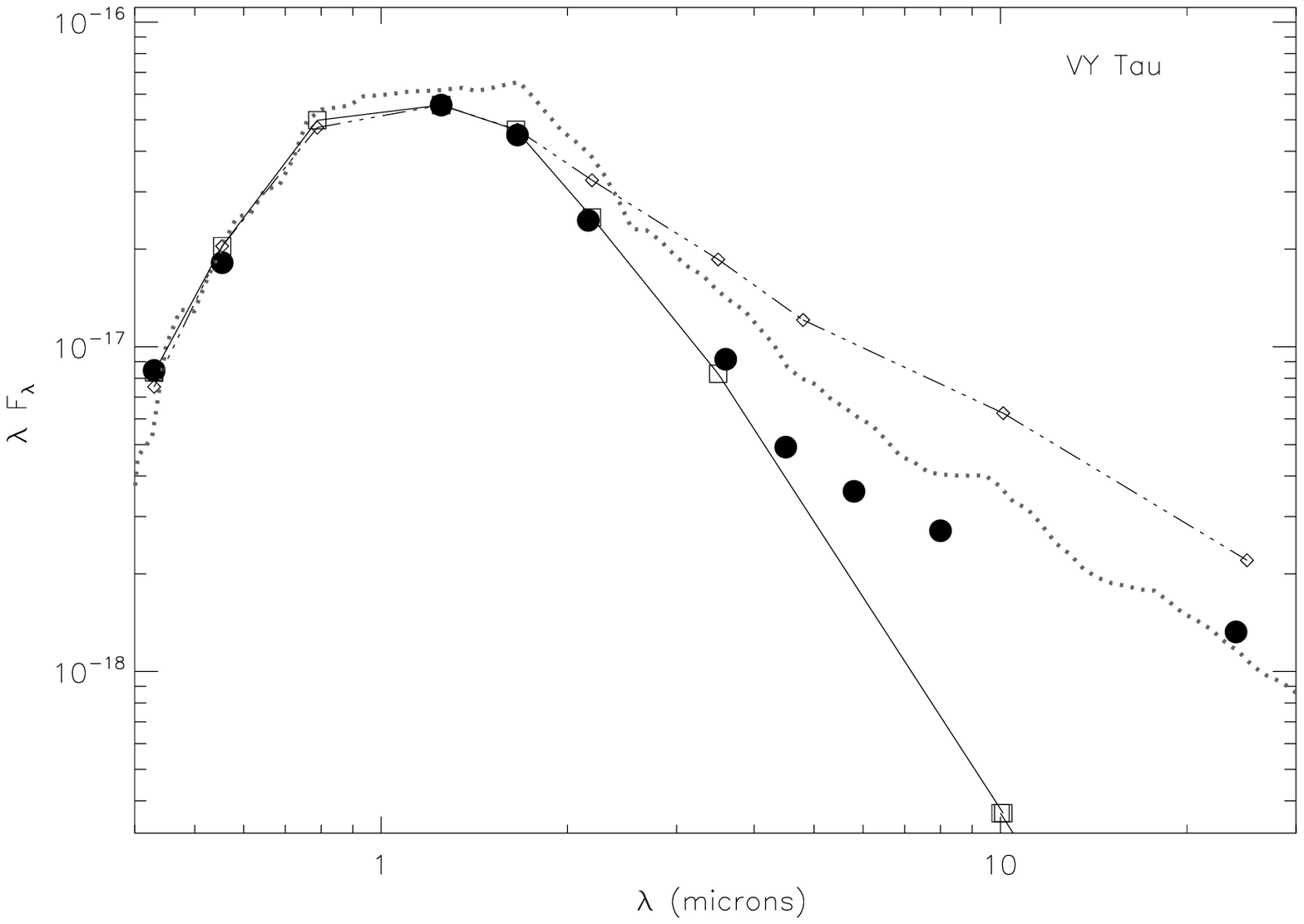}{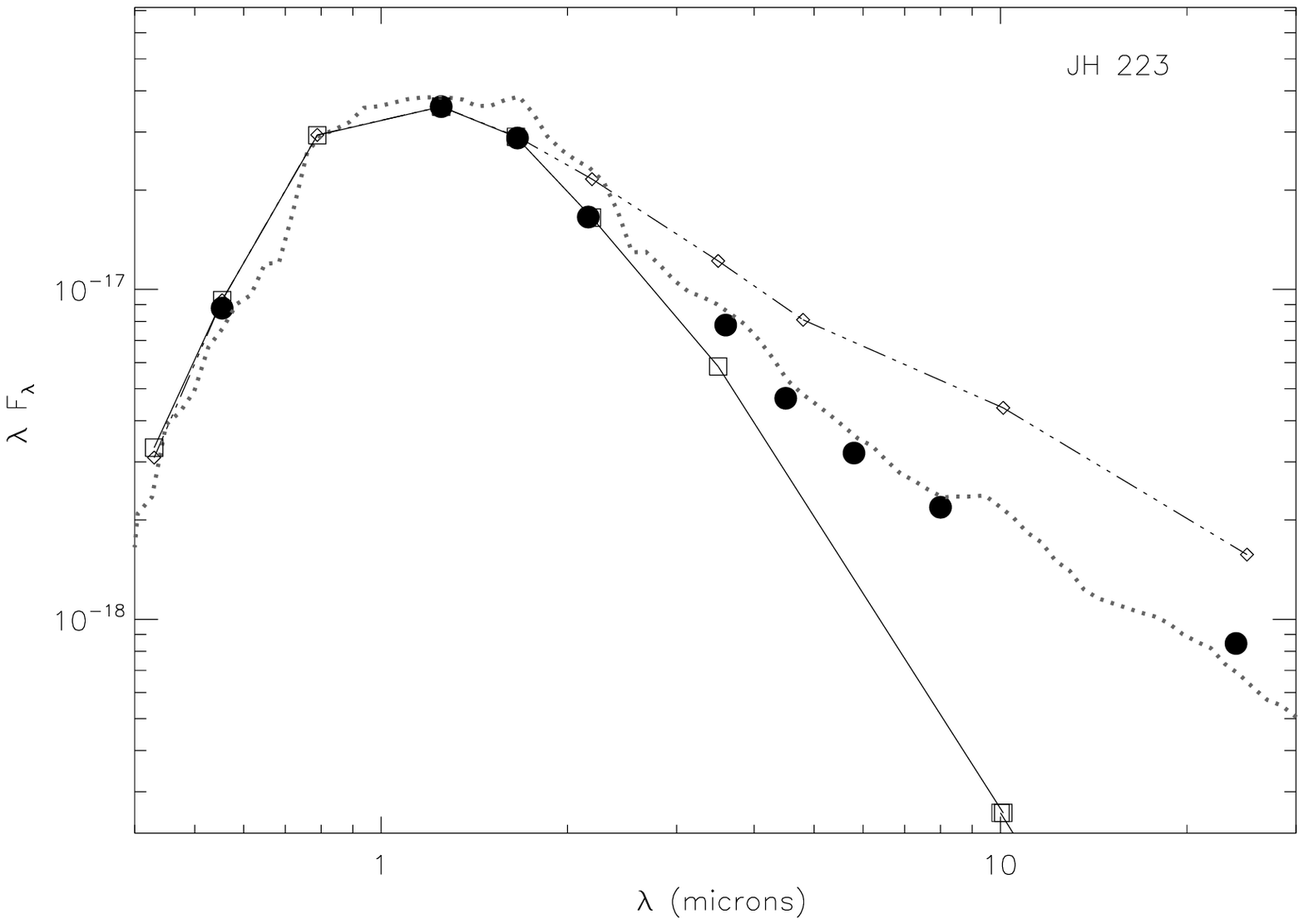}
\plottwo{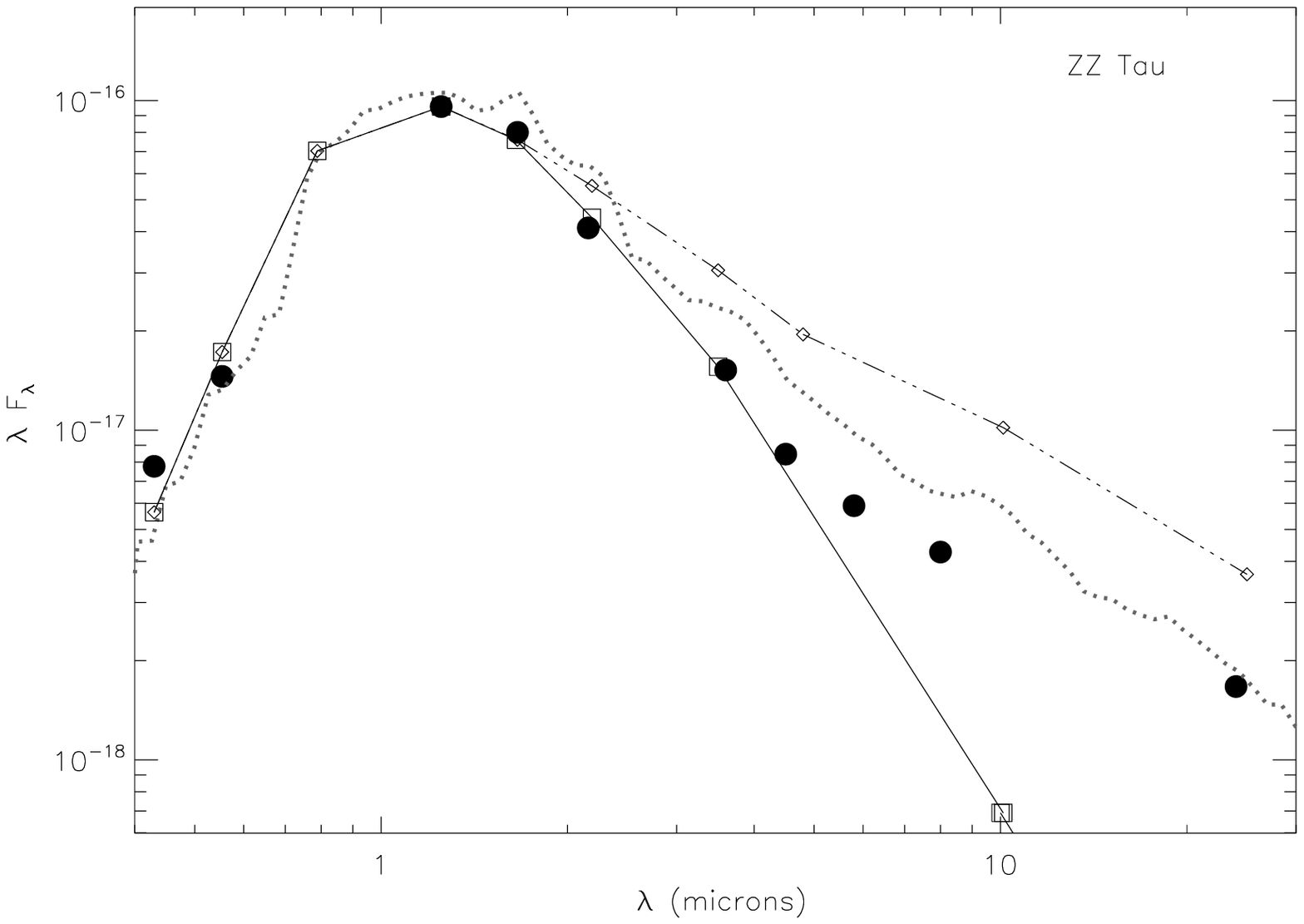}{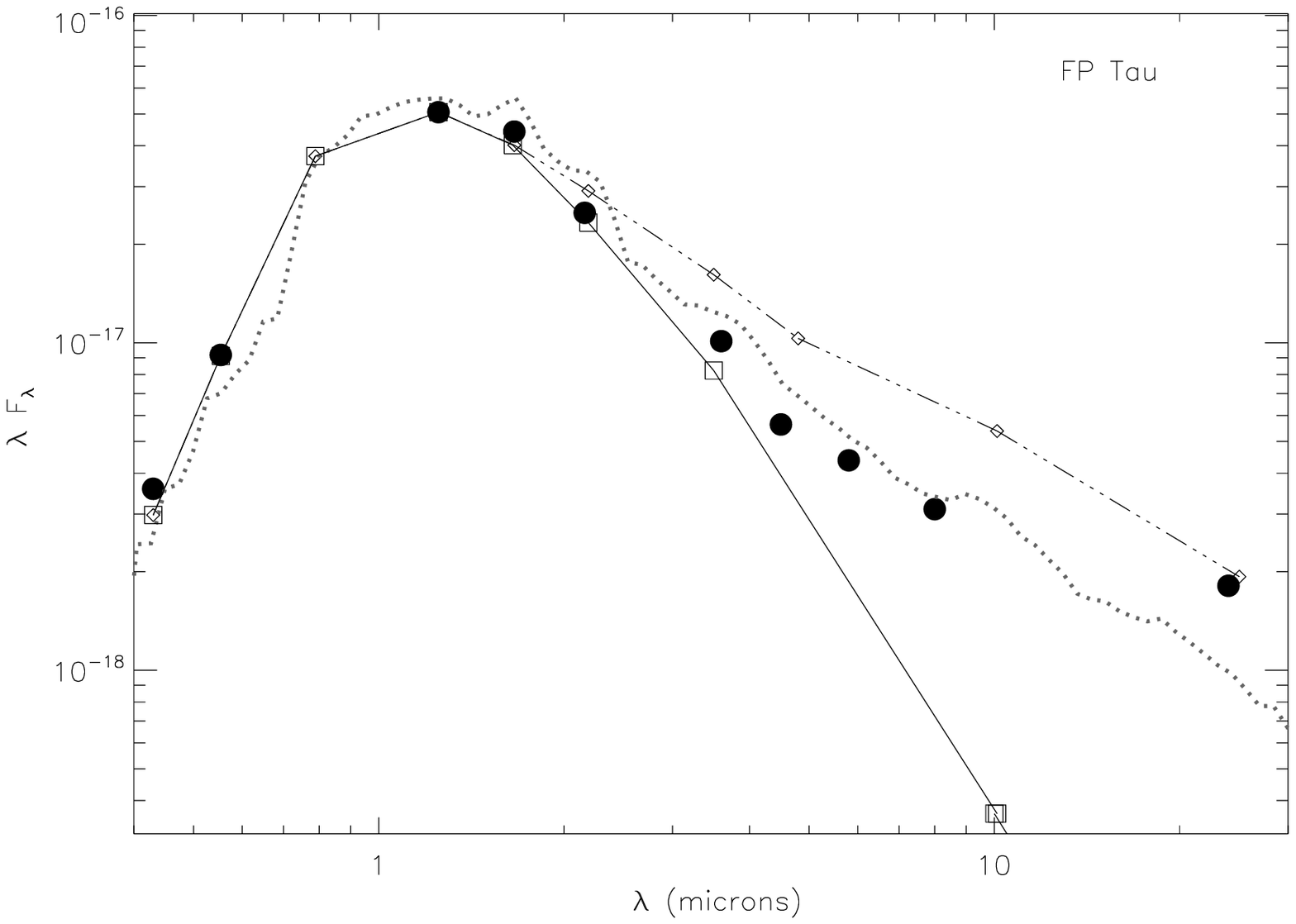}
\caption{SEDs for four homologously depleted transitional disks in Taurus, showing optical to 
mid-IR photometric data compared to the flat, optically-thick reprocessing disk limit and 
the lower-quartile Taurus SED.}
\label{taurussedcompare}
\end{figure}

Modeling results for 25 Taurus members are summarized in Tables \ref{taurusstate}  
and \ref{taurusstate2}.   
Several of these members have disks with inner holes - e.g. 
UX Tau, LkCa 15, and J04333905+2227207.  Many others -- e.g. VY Tau -- 
lack clear evidence for an inner hole but have emission falling below the 
flat, optically thick disk limit (Figure \ref{taurussedcompare}).  
They also have weak submm emission (Table \ref{taurusstate2}).
The weak near-IR through submm emission from these disks 
identifies a substantial depletion of the entire dust 
grain population from hot inner disk regions to cold outer disk regions.
Nearly all sources (94 \%) with K$_{s}$-[8] $\le$ 1.75 have emission 
 weaker than the optically-thick, flat disk limit.  

\subsubsection{Best-Estimated Masses of Disks in Taurus}
\begin{figure}
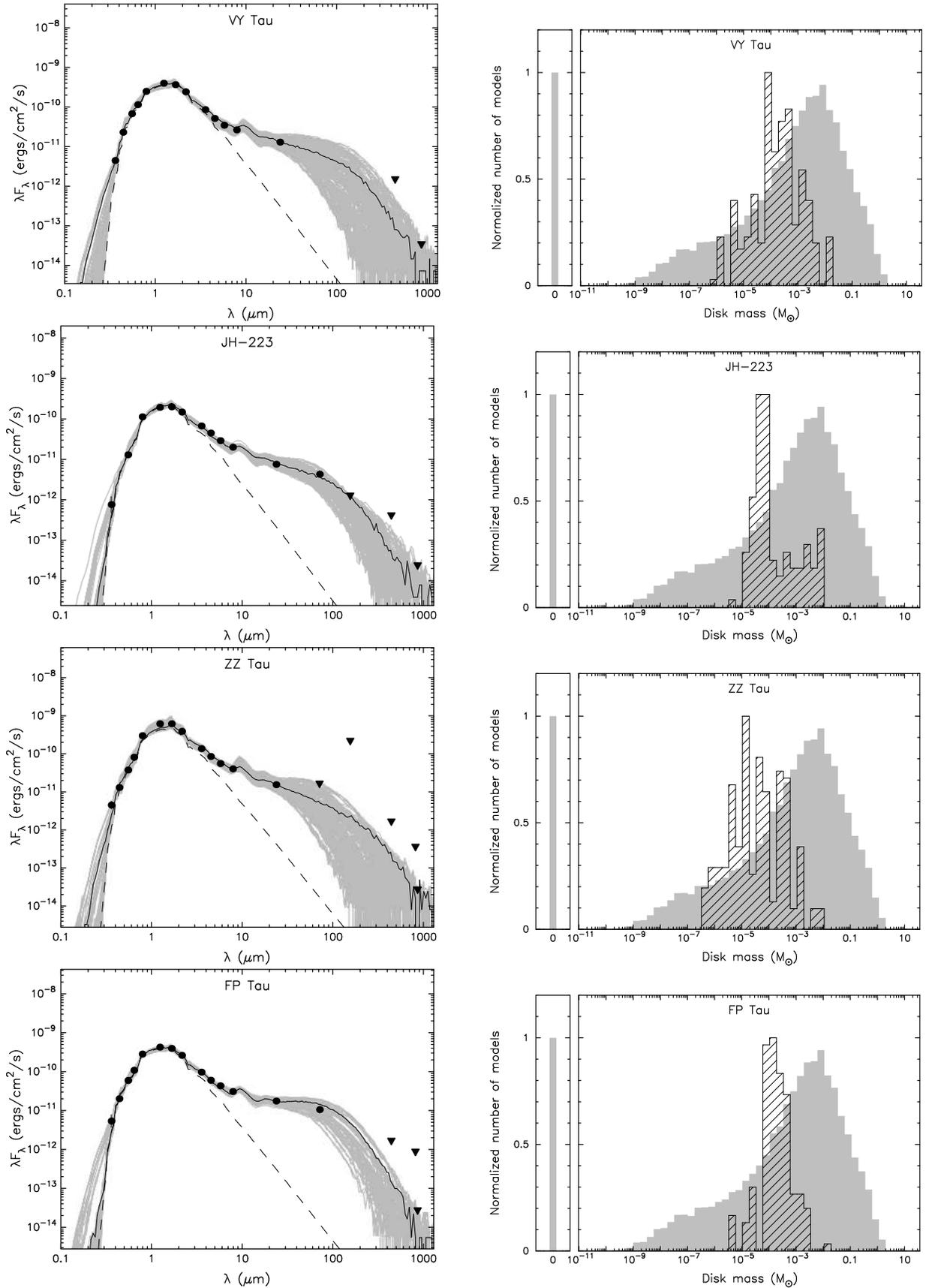

\centering
\plottwo{vytau_sed.eps}{vytau_mass.eps}
\plottwo{jh223_sed.eps}{jh223_mass.eps}
\plottwo{zztau_sed.eps}{zztau_mass.eps}
\plottwo{fptau_sed.eps}{fptau_mass.eps}
\caption{SED modeling of sources shown in previous figure, 
displaying the best-fit SEDs from the Robitaille grid along with the 
corresponding masses for the best-fit models.  For the righthand panels, 
the hatched region shows the normalized distribution of masses from best-fit models; the 
shaded grey region shows the normalized distribution of masses from all available models.  
The x-axis of the righthand side figures lists the disk mass, not the fractional 
disk mass.  
The division corresponding to M$_{disk}$/M$_{\star}$ = 10$^(-3)$ for VY Tau, 
JH-223, ZZ Tau, and FP Tau occurs at M$_{disk}$ = 7.5$\times$10$^{-4}$, 5.5 $\times$ 10$^{-4}$, 
4 $\times$ 10$^{-4}$, and 2.5 $\times$ 10$^{-4}$, respectively}
\label{taurusmodeling}
\end{figure}

Figure \ref{taurusmodeling} displays the SEDs and histogram 
distribution of masses from best-fit models for several 
 disks that lack inner holes but have weak near-to-mid IR 
emission -- VY Tau, JH-223, ZZ Tau, and FP Tau -- 
and indicates that they have low inferred masses (M$_{disk}$ $\le$ 0.001 M$_{\star}$).  
Such sources fit our criteria for being a homologously depleted transitional disk.
Several others (e.g. V807 Tau) lacking inner holes 
have optically-thin near-to-mid IR emission and 
M$_{disk}$ = 0.001--0.003 M$_{\star}$: these are borderline cases.
The intrinsic sampling of the Robitaille grid is heavily 
weighted towards disk models with masses $\sim$ 0.001--0.01 M$_{\odot}$ 
(shaded region).  In spite of this bias, the inferred disk masses for sources like 
those in Figure \ref{taurusmodeling} are at least an 
order of magnitude lower.

Mass estimates based on submillimeter data alone \citep{Andrews2005} 
provide a simple but independent check on the estimates  
 derived from radiative transfer modeling.
All of the homologously depleted transitional disks with submillimeter 
data that are listed in Table \ref{taurusstate2} have non detections at 350--850 $\mu m$, 
and thus their  masses have a wide range of possible values.  
However, the submm-derived upper limits constrain the mass of optically thin 
emitting dust, which is useful since all of these sources have 
optically thin near-to-mid IR dust emission and optical depth decreases with increasing 
wavelength.  For the homologously depleted disks, submm-derived 
upper limits range from 5.5$\times$10$^{-4}$ M$_{\star}$ to 
10$^{-3}$ M$_{\star}$.  In other words, submillimeter data 
alone provides evidence that these disks are less massive 
than primordial disks.

\subsubsection{Frequency of Transitional Disks in Taurus}

To compute the relative frequency of transitional disks 
around K5--M6 stars, we combine the transitional disk population identified 
by \citet{Luhman2009} and newly identified 
transitional disks from Table \ref{taurusstate2}.  
For the primordial disk population, we consider 
all Taurus members with Spitzer photometry that 
are not Class I protostars. Based on mid-IR 
colors, \citet{Luhman2009} identified eighteen transitional disks 
around K5--M6 stars:  nine around K5--M2 stars and nine 
around later stars.  To these, we add 6--7 transitional 
disks around K5--M2 stars and 3--7 transitional disks 
around later stars identified from SED modeling.  

Combining both transitional disk samples, the frequency of transitional disks 
is 0.19--0.22 (27--32/145) for the entire 
population, 0.20--0.21 (15--17/77) for K5--M2 stars, and 0.18--0.24 (12-16/68) 
for M3--M6 stars.  Two transitional disks 
identified by \citet{Luhman2009} have extremely tenuous 24 $\mu m$ excess 
emission (K$_{s}$-[24] $\sim$ 0.5--1) and could be 
young debris disks \citep{Carpenter2009}.
If we remove these sources, the transitional 
disk frequency for the entire population drops to 0.18 and 0.22, while 
the frequency for K5--M2 stars drops to $\sim$ 0.16--0.18. 
Thus, we find a slightly higher percentage of transitional disks 
around K5--M6 stars than \citet{Luhman2009} find for K5--M5 stars ($\sim$ 20\% vs. 
their 13\% (15/113)).  

\subsubsection{Mid-IR Colors of Primordial Disks and Transitional Disks}
\begin{figure}
\plottwo{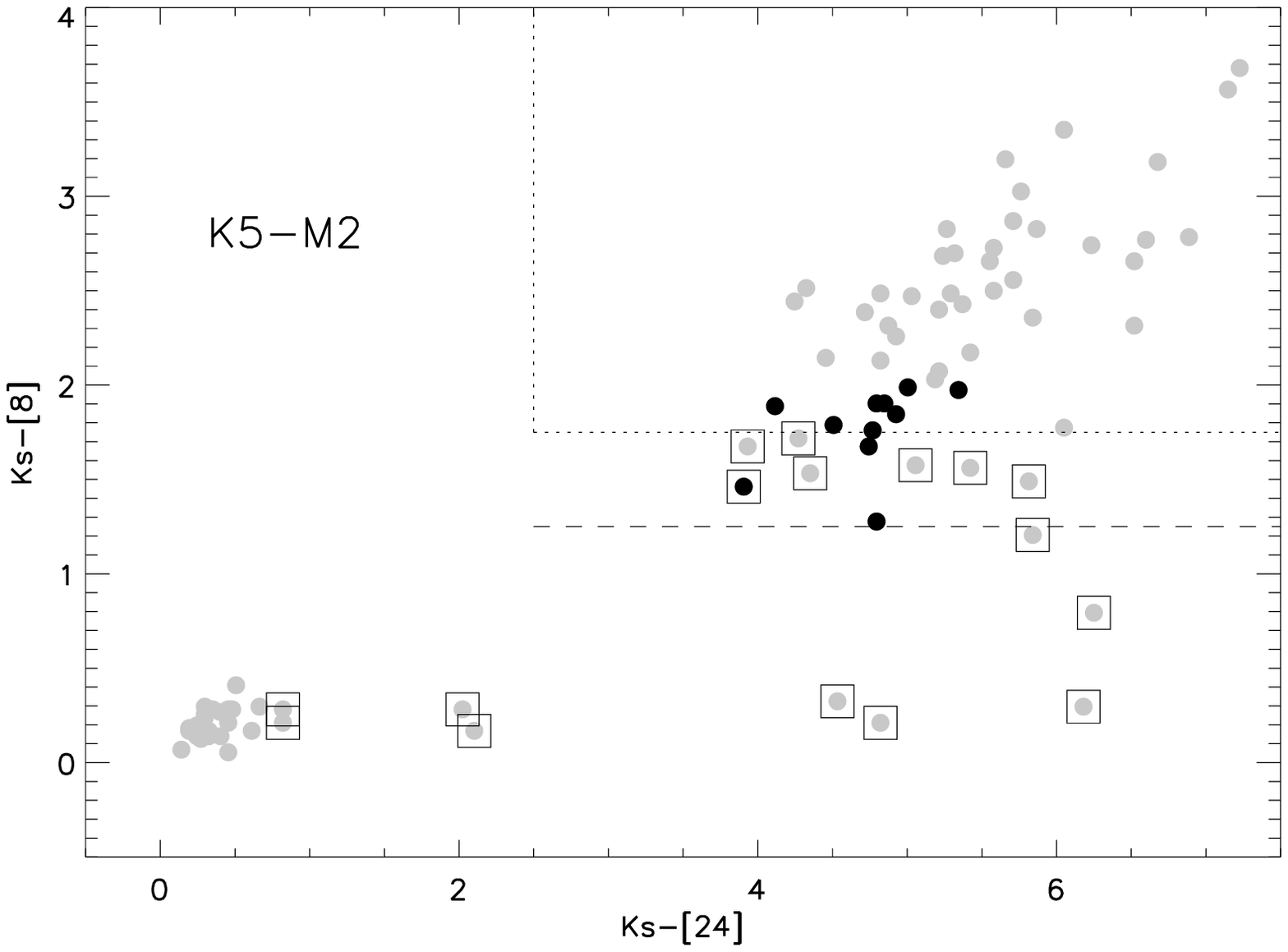}{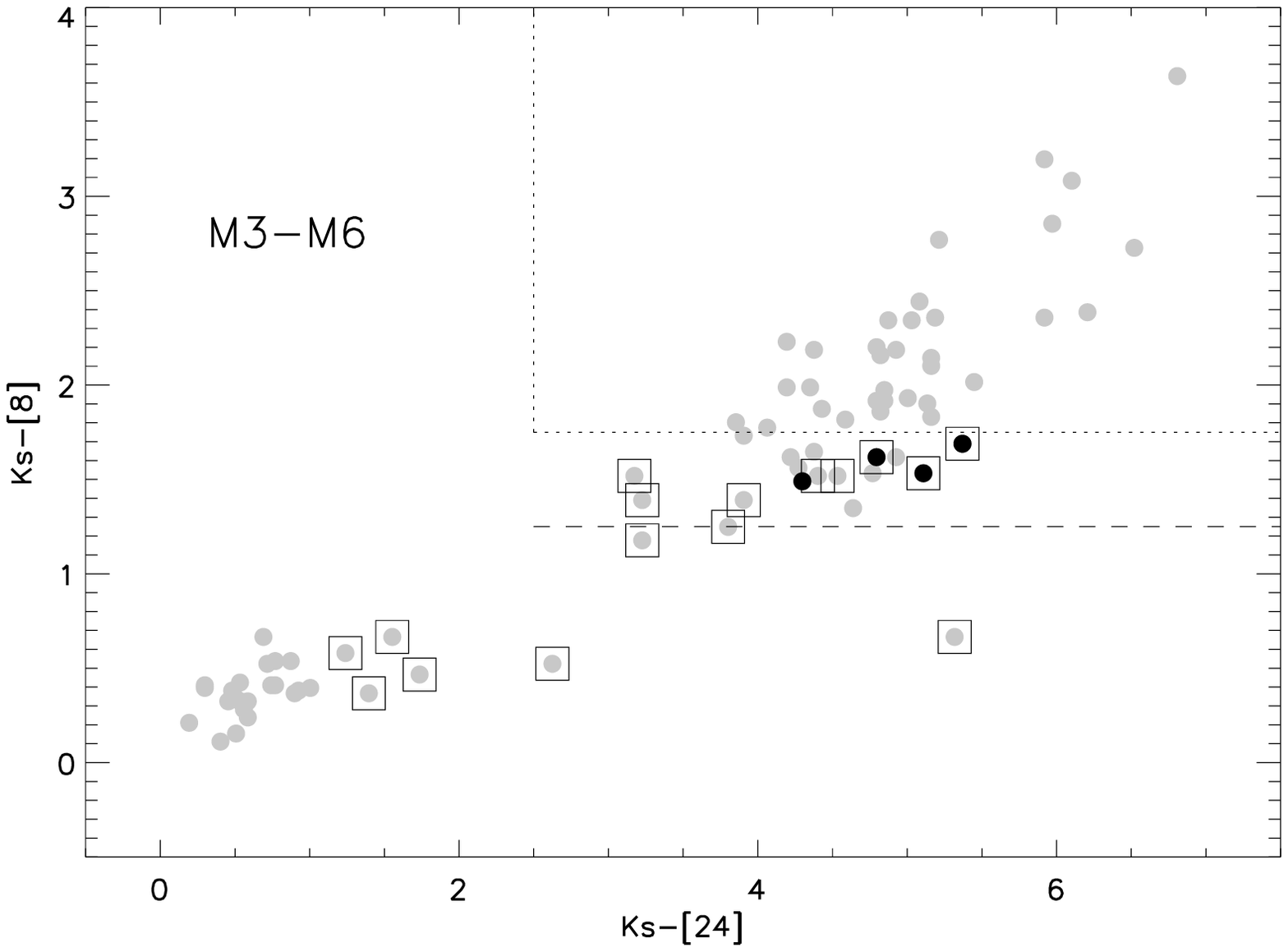}
\caption{K$_{s}$-[8] vs. K$_{s}$-[24] colors for K5--M2 (left panel) and 
M3--M6 (right panel) Taurus members with data 
from \citet{Luhman2009} and \citet{Rebull2010}.  Grey circles surrounded 
by squares represent transitional disks identified from SED modeling, 
black dots represent primordial disks identified from SED modeling, and 
black dots surrounded by squares identify disks that could either be 
primordial or transitional.  SED modeling was not performed for members 
represented as grey dots only.  Sources between the horizontal dashed line (K$_{s}$-[8]=1.25) and 
the horizontal dotted line (K$_{s}$-[8]=1.75) include both primordial disks and transitional disks.  
Those above the horizontal dotted line include only primordial disks while primordial disks are 
absent below the horizontal dashed line.}  
\label{tauruscolors}
\end{figure}

To guide our analysis of other clusters lacking far-IR/submm data (e.g. IC 348, NGC 2362), 
we determine the mid-IR colors (K$_{s}$-[8]) that identify 
primordial disks and transitional disks.
Figure \ref{tauruscolors} plots the mid-IR colors of Taurus members separated by 
disk evolutionary state.  Only members with primordial disks (black circles) have K$_{s}$-[8] $>$ 1.75. 
  Taurus members with K$_{s}$-[8] bluer than the horizontal dashed line (K$_{s}$-[8] = 1.25) do not have primordial disks.  
Members with intermediate colors include primordial disks, transitional disks (grey circles 
enclosed by squares), and borderline cases (black circles enclosed by squares).
Thus, there are two regions of mid-IR color-color space containing transitional disks: 
an "uncontaminated" region with blue colors consisting only of transitional disks and a 
"contaminated" region with redder colors which includes both transitional disks and primordial disks.  

Comparing the number of transitional disks and primordial disks in the "contaminated" 
 color region for Taurus members with far-IR/submm data 
provides an estimate for the level of contamination in other clusters 
lacking far-IR/submm data.
As shown by Figure \ref{tauruscolors} and Table \ref{taurusstate}, 
there are 9 K5--M2 members with K$_{s}$-[8] = 1.25--1.75: 
6 sources identified as transitional disks, two identified 
as primordial disks, and one borderline case (V807 Tau).  
Thus,  22--33\% (2--3/9) of the sources with K$_{s}$-[8]=1.25--1.75 
and provisionally identified as transitional disks from mid-IR data alone
could in fact be primordial disks with substantial 
dust settling.  For M3--M6 stars with K$_{s}$-[8]=1.25--1.75, 3 sources 
listed in Table \ref{taurusstate2} plus two identified by \citet[][]{Luhman2009} -- 
J04330945 and J04213459 -- are transitional disks, one is a primordial disk, and 
four are borderline cases.  Thus, the contamination rate for M3--M6 
stars with K$_{s}$-[8] = 1.25--1.75 is 10--50\% (1--5/10).

\subsubsection{Incorporating Uncertainties in Disk Mass Estimates}

Here we quantitatively assess how the intrinsic uncertainty in disk mass estimates 
affects our conclusions about the frequency of transitional disks in Taurus 
and the contamination level from selecting transitional disks by mid-IR colors. 
First, we compare the interquartile range of disk masses from best-fit models 
for each star.  
Figure \ref{quartilemass} shows that the range of disk masses for homologously depleted disks 
and primordial disks are distinct (see also Table \ref{taurusstate}).  The upper quartile of disk masses 
for homologously depleted disks ranges from $\sim$ 9$\times$10$^{-4}$ M$_{\star}$ to 
4$\times$10$^{-3}$ M$_{\star}$.  The lower quartiles of masses for primordial disks 
is systematically larger and does not overlap, ranging from 4$\times$10$^{-3}$ M$_{\star}$ to 
10$^{-2}$ M$_{\star}$.  The masses of disks in borderline cases are also systematically 
low, though the upper quartiles for three of the five stars -- J04231822+264116, J04153916+2818586, and ITG 15 -- 
overlap slightly with the lower quartile masses for several primordial disks (e.g. GH Tau).

Second, we consider the full distribution of disk masses from best-fit 
models for each star to identify the fraction of models yielding M$_{disk}$/M$_{\star}$ 
$\le$ 0.001 and 0.003 (f $<$ 1, 3$\times$10$^{-3}$), our thresholds identifying homologously depleted transitional disks 
and borderline cases.  Our goal here is to derive a purely probabilistic estimate 
of the transitional disk population from Table \ref{taurusstate} to combine with 
the transitional disks listed by \citet{Luhman2009} and then to compare with previous estimates of 
the total population in this section.  
Here, the number of transitional disks around \textit{n} stars listed in Table \ref{taurusstate} each 
with f(TD$_{i}$) of their disk models fulfilling our definition of a transitional disk is
\begin{equation}
n(TD) = \sum\limits_{i=0}^n f(TD_{i}).
\end{equation}
For disks with inner holes -- e.g. LkCa 15 -- determined by imaging and SED modeling 
f(TD$_{i}$) =1, otherwise it equals f($<$ 1, 3 $\times$ 10$^{-3}$)
as listed in Table \ref{taurusstate}.  This exercise yields n(TD) = 10.6--13.8 or $\sim$ 11-14 (n(TD) = 
5.5--7.1 for K5--M2 stars, 5.2--6.7 for M3--M6 stars).

Combining our probabalistic estimate for the transitional disks population drawn from Table \ref{taurusstate} and 
the 18 transitional disks identified by \citet{Luhman2009} yields total population of 29--32 (18+ [11,14]), 
in agreement with estimates derived in \S 4.3.2.  The estimates divided by spectral type range are likewise 
similar: 15--16 (9+[6,7]) for K5--M2 stars and 14--16 (9+[5,7]) for M3--M6 stars.
Finally, the resulting contamination estimates agree well with previous ones: 22--33\% (2--3/9) for K5--M2 stars 
and 10--30\% (1--3/10) for M3--M6 stars\footnote{For K5--M2 stars, there are nine sources within our contamination region, 
our exercise predicts that $\sim$ 6--7 are transitional disks, so the contamination rate is 2--3/9.  For M3--M6 stars, 
there are 10 sources within the contamination region.  Two of these are identified as transitional disks from \citet{Luhman2009}, 
whereas our exercise predicts that 5--7 out of the eight others are transitional disks.  Thus, the contamination rate is 1--3/10}.   
%Thus, our transitional disk frequencies are statistically robust against modeling uncertainties.

\subsection{Frequency of Transitional Disks in IC 348}
Table \ref{ic348state} summarizes our modeling results for selected IC 348 members; 
SEDs for members representing a range of mid-IR disk emission 
and provisional evolutionary states are shown in Figure \ref{ic348example}.  
As with Taurus, we find that all but one of the members with K$_{s}$-[8] $\le$ 1.75 
have optically-thin mid-IR disk emission.  Since our Taurus analysis finds 
that most members with these colors have low disk masses or inner holes (29--32/35), we infer 
that most IC 348 members with these colors also have low disk masses or inner holes.  
Of the 25 IC 348 stars modeled, 17 are consistent with being transitional disks 
based on 1--24 $\mu m$ data, while 8 are primordial disks.

To determine the frequency of transitional disks in IC 348, 
we use contamination estimates based on mid-IR colors from Taurus.  Following 
the Taurus results, we adopt K$_{s}$-[8] = 1.75 
as our fiducial division between primordial disks and transitional disks and 
derive a provisional frequency for transitional disks.  Then we derive a revised 
frequency taking into account contamination by primordial disks with 
K$_{s}$-[8] = 1.25--1.75.  

%Based on this method, 10/36 K5--M2 stars 
%and 42/77 M3--M6 stars fit our transitional disk criteria.  
%Corrected for contamination, 19--25\% of disks around K5--M2 
%stars have transitional disks (7/36 to 9/36), while $\sim$ 
%30--40\% of M3--M6 stars have transitional disks (21/77 to 31/77).
%Thus, the relative frequency of transitional disks 
%for K5--M2 stars is marginally lower than that for later stars, 
%consistent with results for the Coronet Cluster and Taurus.

Based on this method, 8/33 K5--M2 stars 
and 43/81 M2.5--M6 stars have evidence for a disk with K$_{s}$-[8] $<$ 1.75, and 
thus fit our nominal near-to-mid IR transitional disk 
criteria.  Of the 8 (44) disks around K5--M2.5 (M2.5--M6) stars provisionally 
classified as transitional, 4 (19) have mid-IR colors within the "contaminated" 
region.  Thus, corrected for contamination, about 32--40\% of K5--M6 
IC 348 disks are transitional disks.  The transitional disk population is 
heavily weighted towards later-type stars: 18--24\% of disks around K5--M2 
stars have transitional disks (6/33 to 8/33), while $\sim$ 
42--51\% of M2.5--M6 stars have transitional disks (34/81 to 41/81).

\subsection{Frequency of Transitional Disks in NGC 2362}
Table \ref{n2362state} summarizes our modeling results for NGC 2362.  Near-to-mid 
IR emission from most cluster members lies well below 
 the optically thick, flat reprocessing 
disk limit, indicating that the mid-IR emission is at least marginally optically thin 
 ($\tau$ $\lesssim$ 1).  Comparing this modeling to radiative 
transfer modeling performed for NGC 2362 in \citet{CurrieLada2009} and 
for Taurus in this work shows that many sources with 
weak/negligible 1--8 $\mu m$ emission and optically-thick 24 $\mu m$ emission 
require inner holes while many with marginally optically-thin near-to-mid IR emission 
 have low inferred disk masses.  Figure \ref{ngc2362example} displays 4 representative 
SEDs comparing the observed emission to the optically thick, flat disk limit.

Our results are consistent with those from \citet{CurrieLada2009} who argue that transitional disks 
are equal to or greater in number than primordial disks.  \citet{CurrieLada2009} find that 
81\% of disks around K0--M3 members are transitional disks (17/21).  Here 
 four disks are primordial disks (IDs 111, 139, 187, and 202), and one (ID 177) 
appears to be a borderline case (since it may have an inner hole) but 
is counted as a primordial disk in the analysis below to be conservative
(since it has K$_{s}$-[8] $>$ 1.75).
The rest (16; 14 of which are K5--M3 stars) have near-to-mid IR 
emission consistent with being transitional disks.  Of these, five K5--M3 stars 
either have K$_{s}$-[8] $\le$ 1.25 (thus lying outside the range of colors 
contaminated by primordial disks) or show evidence for inner holes.  
The other 9 K5--M3 stars have mid-IR colors between K$_{s}$-[8] = 1.25 and 1.75, 
within the contaminated region.  Based on our Taurus analysis,
  2 K5--M2 stars and 0--1 M2.5--M3 stars provisionally identified as having 
transitional disks instead have primordial disks with dust settling.  Therefore,
we arrive at a final transitional disk frequency of 58\%--63\% (11/19--12/19) for 
K5--M3 stars.  
The frequency for K5--M2 stars (63\%) and M2.5--M3 stars (33-66\%) are statistically 
indistinguishable given the small number in the latter group (3).

\subsection{Frequency of Transitional Disks in $\eta$ Cha}
Table \ref{etachastate} summmarizes our modeling results for $\eta$ Cha.  
Of the 15 late-type members with IRAC and MIPS photometry, 8 have evidence for a disk.  
Only three sources (RECX-11, J0843, and J0844) fulfill our definition of a primordial disk.
  Three members have transitional disks with inner holes 
(RECX-3, 4, and 5); two have homologously 
depleted transitional disks (RECX-9 and J0841).  
  Thus, $\eta$ Cha has a high transitional disk frequency (5/8 or 63\%).  
Our results agree with those of \citet{SiciliaAguilar2009} who 
separately analyze Spitzer IRS data to find that transitional disks 
comprise 50--75\% of the disk population.
A ``probablistic" estimate of the transitional disk fraction (e.g. as in Section 4.3) 
yields the same answer: n(TD) = 4.95 or $\sim$ 5.

%We strongly disagree with \citet{Luhman2009} who 
%find a much lower transitional disk frequency  
%based on comparing broadband K$_{s}$-[8] vs. K$_{s}$-[24] colors 
%to those for a fiducial model.  Aside from our differences 
%in the adopted definition of transitional disks,
%our modeling results are more reliable because 
%we fit the observed SEDs based on data with a 
%much wider and better sampled wavelength range, which 
%provides far superior constraints on disk properties 
%than can be determined from a single color-color diagram (see Appendix). 

Gas diagnostics are consistent with our argument that most disks in $\eta$ Cha are transitional disks.  
 RECX-5 and 9 are accreting at extremely low rates (4--5 $\times$ 10$^{-11}$ M$_{\odot}$ yr$^{-1}$) \citep{Lawson2004}.  
RECX-3, RECX-4, and J0841 lack evidence for accretion circumstellar gas.
RECX-3, 5, and 9 lack evidence for rovibrational H$_{2}$ emission from warm 
circumstellar gas \citep{RamsayHowat2007}.  
Only J0843, identified as a primordial disk from SED modeling, shows evidence for 
H$_{2}$ emission and for substantial accretion (10$^{-9}$ M$_{\odot}$ yr$^{-1}$).

\subsection{Revised Transitional Disk Frequency for the Coronet Cluster}
Based on our analysis of Taurus data in \S 4, some fraction of Coronet Cluster stars 
determined to have transitional disks based on mid-IR data probably are primordial disks 
with dust settling.  In particular, G-1 (M0), G-14 (M4.5), and G-87 (M1.5) have K$_{s}$-[8] 
colors placing them within the contaminated region: other transitional disks 
have inner holes or lie within the uncontaminated region.  
Accounting for contamination, the size of the transitional disk population drops 
to 11--22\% for K5--M2 stars (1/9, 2/9) and 30--50\% for M2.5--M6 stars (3/10, 6/12),
where the ranges account both for contamination by primordial disks with substantial dust settling 
and for possible misclassification of debris disk candidates.
Thus, our transitional disk frequencies for all five clusters now account for 
uncertainties in disk classification based on mid-IR data alone.

\subsection{The Evolution of the Transitional Disk Population}
Combining our results for the Coronet Cluster, Taurus, IC 348, NGC 2362, and $\eta$ Cha, 
we now investigate the frequency of transitional disks as a function of 
age and stellar properties.  
Table \ref{freqtranall} lists the relative frequency of transitional 
disks as a function of time in three spectral type bins:  mid-K to mid-M stars (K5--M6), K5--M2 stars, 
and M3--M6 stars.  
%To compare with previous estimates for transitional disk frequencies, we overplot the 
%results from \citet{Muzerolle2010} in the top left panel, which list the frequencies for 
%transitional disks with inner holes and those that may also have a more homologous depletion of 
%disk material (grey squares and diamonds).
To compare with theoretical predictions, we overplot the approximate loci (Figure \ref{freqtranvstime} top panel, solid 
and dotted lines) 
expected from the \citet{AlexanderArmitage2009} disk evolution/planet formation models, which 
yield frequencies for transitional disks as a function of time and assume a disk clearing timescale 
of $\sim$ 0.5 Myr.   
%Because of small sample sizes, in Figure \ref{freqtranstype} we do not plot 
%any data for $\eta$ Cha and do not plot data for M3--M6 stars in NGC 2362.

The transitional disk frequency for K5--M6 stars increases as a function of time from $\sim$ 20\% at 
1--2 Myr to $\sim$ 50--60\% by 5 Myr.  The frequencies are consistently 1.5--3 times higher than those 
predicted by \citet{AlexanderArmitage2009} for a 0.5 Myr transition timescale.  Within 
the context of the \citet{AlexanderArmitage2009} study, our results indicate that 
the transitional disk phase on average lasts longer than 0.5 Myr.  

The bottom panel of Figure \ref{freqtranvstime} compares 
our frequencies with those from \citet{Muzerolle2010} who use a more 
empirically-based approach to identify transitional disks based primarily on IRAC and MIPS flux 
slopes (bottom panel).  
Our transitional disks frequencies are substantially larger than those \citeauthor{Muzerolle2010} derive 
based on their self-described \textit{classical} definition for what constitutes a transitional 
disk (diamonds): a disk with an inner hole and optically-thick outer disk.  
However, \citeauthor{Muzerolle2010} argues for an expanded definition for 
a transitional disk, including disks whose mid-IR emission plausibly 
identifies a more homologous depletion of emitting dust and 
disks with inner holes/cavities and more optically thin outer disks.  Adopting this definition,
their transitional disk frequencies (triangles) are in good agreement with those we derive 
for clusters older than 2 Myr.  For the two clusters 
we both analyze, IC 348 and $\eta$ Cha, our frequencies are nearly identical.

To derive a timescale for the transitional disk phase, we follow a parametric, 
Monte Carlo approach, similar to that from \citet{Muzerolle2010}, 
evolving a population of 10$^{5}$ stars to 
simulate the frequency of disks in different states for clusters with 
ages less than 10 Myr.  
We adopt a 3 Myr e-folding timescale for the protoplanetary disk stage, 
consistent with Spitzer observations (Figure \ref{timetranall}, top panel).  Some fraction of 
the protoplanetary disk lifetime is spent in the transitional disk phase.  To 
assess the duration of this phase,  we vary the transitional disk e-folding timescale (t$_{TD}$); correspondingly, 
the e-folding timescale for the primordial disk phase is 3 Myr - t$_{TD}$.  Through the first 0.54 Myr, 
we fix the disk fraction at 1 to account for the lifetime of Class I protostars \citep{Evans2009}.  
Finally, we assume a 1 Myr age dispersion at each cluster age.

As shown by Figure \ref{timetranall} (bottom panel), transitional disk frequencies in all clusters 
are most consistent with a 1 Myr transition timescale, lying well above predictions 
for a 0.1--0.5 Myr timescale and below predictions for a 2 Myr timescale.
Thus, our analysis confirms earlier results based on NGC 2362 and $\eta$ Cha 
from \citet{CurrieLada2009} and \citet{SiciliaAguilar2009} that the mean transitional 
disk lifetime must be an appreciable fraction of the total protoplanetary disk lifetime, 
estimated to be $\sim$ 3--5 Myr \citep[][]{Hernandez2007,CurrieLada2009}.  Adopting \citet{Muzerolle2010}'s 
model for the frequency of transitional disks with time yields the same 
qualitative result.  They predict smaller transitional disk frequencies at 3--10 Myr 
than we do for a 1 Myr transition timescale.  The identification of a long 
transition timescale based on our predictions is then conservative.   On the other hand, our 
results are in conflict with other recent claims, particularly those 
of \citet{Luhman2009} who argue for a transitional disk lifetime 
less than 0.5 Myr.  In \S 5, we compare our methodology with their's and 
those from other recent studies.

We emphasize that our disk frequencies account for disks with
 \textit{both} transitional disk morphologies: disks with inner holes and 
homologously depleted disks.  Homologously 
depleted transitional disks generally appear to be more numerous than those 
with inner holes, up to a factor of 1.5--2 larger in number for the Coronet 
Cluster and NGC 2362.  However, these differing frequencies do not constrain 
the relative lifetimes of disks following these evolutionary 
paths: we cannot tell between a morphology with a shorter lifetime and 
one that operates less frequently.  
We also note that these results are statistical.  In other words, 
while there is some uncertainty in the state for individual objects, 
these uncertainties are averaged out at the end.
%Moreover, the typical lifetime may 
%be a strong function of the disk clearing mechanism \citep[see Discussion;][]{AlexanderArmitage2009,
%SiciliaAguilar2010}. 

At least for the Coronet Cluster and IC 348, 
the transitional disk frequency is spectral type dependent (Figure \ref{freqtranstype}).
The transitional disk frequencies for M3--M6 stars in these clusters is 
marginally but consistently higher than those for K5--M2 stars by about a factor of 
$\sim$ 1.5.  Formally, M3--M6 stars in Taurus also more frequently have transitional disks 
than do K5--M2 stars, though the difference is not statistically significant.  
\citet{Muzerolle2010} also find that the transitional disk population is dominated by 
later-type stars, though their correlation is stronger than what we find.

A spectral-type dependent transitional disk frequency may help explain the 
high transitional disk frequency found by \citet{SiciliaAguilar2008}.
Since the \citet{SiciliaAguilar2008} sample is drawn heavily from stars later than M2, 
their sample is predisposed towards having a high transitional disk fraction compared 
to cluster-averaged values and compared to those for brighter, more easily detectable 
K to M2 stars.  The transitional disk frequency for the bulk 
cluster population (e.g. including members in Table \ref{diskstatenew}) is 
similar to that for other 1--3 Myr-old clusters. At least one other cluster, IC 348, 
also has a very high transitional disk fraction for very low-mass stars.  

\section{Discussion}
\subsection{Summary of Results}
This paper expands upon previous Spitzer analysis of the Coronet Cluster disk population 
around very low-mass stars performed by \citet{SiciliaAguilar2008}, presenting new photometry for other 
cluster members, including many with higher stellar masses.
We use SED modeling and comparisons with simple disk models/empirical metrics to 
assess the Coronet Cluster disk population, focusing on identifying and characterizing 
candidate transitional disks.   By analyzing the disk population for other 1--8 Myr-old clusters, 
we investigate the utility of using mid-IR data to probe 
disk evolutionary states, determine how the frequencies of transitional disks change with 
time, and investigate how this lifetime may depend on stellar properties.  
Our study yields the following major results:
\begin{itemize}
\item The Coronet Cluster 
contains a high frequency of transitional disks: $\sim$ 30\% (formally, 21--36\%) of the disk 
population around K5--M6 stars.  Although the samples are small, our analysis 
hints at a spectral type-dependent frequency for transitional disks, 
confirming earlier suggestions by \citet{SiciliaAguilar2008,SiciliaAguilar2009}: they 
appear more frequently around the very low-mass stars that were the focus 
of \citet{SiciliaAguilar2008}.  IC 348 exhibits the same 
spectral type/stellar-mass dependent transitional disk frequency. 

\item Based on optical to submm SED analysis of Taurus sources 
with high-quality photometry, we confirm that many disks lacking 
clear evidence for inner holes also have a low dust mass.  Their mid-IR emission 
is due to a depletion of dust, not dust settling: these represent 
a more homologous depletion of disk material with time instead of an 
inside-out dispersal.   Many transitional disks can only be identified 
from SED modeling as their mid-IR colors overlap with primordial disks.  
About 22--33\% (10--50\%) of disks 
around K5--M2 (M3--M6) stars identified as homologously depleted transitional disks 
based on optical to mid-IR data alone may in fact be primordial disks with a 
heavy depletion of submicron-sized dust but a large total dust mass.
Disks with very different morphologies can occupy the same region of color-color 
space: SED modeling is required to avoid misclassifying disks.
\item Combining analysis of the Coronet Cluster disk population with that for 
Taurus, IC 348, NGC 2362 and $\eta$ Cha shows that the relative frequency 
of transitional disks around K5--M6 stars increases from $\sim$ 20\% at 
1--2 Myr to $\sim$ 50--60\% by 5--8 Myr.   Parametric modeling shows that 
this trend is implies a mean transitional disk lifetime near 1 Myr, not $\sim$ 0.1--0.5 Myr. 

%Thus, disks around 
%later type, lower-mass stars Assuming that 
%the total disk lifetimes for stars in these spectral type ranges are comparable, 
%our results suggest that the transitional phase lasts longer for very 
%low-mass stars than for solar-mass stars.  
\end{itemize}

\subsection{Transitional Disk Identification in Other Recent Work: The Importance of SED Modeling}

%\subsubsection{Cieza et al. (2008,2010), Merin et al. (2010)}
Other recent studies of transitional disks 
provide evidence that the weak-IR-to-submm dust emission from homologously 
depleted transitional disks is likely due to disk clearing, consistent with 
our results.  \citet{Cieza2008} selected transitional disks as those with 
mid-IR emission weaker than the lower-quartile Taurus SED and 
low levels of accretion, including many that lack any 
evidence for inner holes.  They found that disks with 
weak mid-IR emission have low submmm-inferred disk masses. 
\citet{Cieza2008} interpret this trend as evidence in favor 
of UV photoevaporation models for disk clearing: photoevaporation 
should begin to clear the inner disk once the total disk mass (probed 
by submm data) drops significantly.  More generally, though, 
this trend implies that for some disks the mass of emitting 
dust drops simultaneously over a wide range of stellocentric 
distance, consistent with a homologous depletion of disk material. 

\citet{Cieza2010} found that the disk mass and accretion rate for 
transitional disks, including those we would identify as homologously depleted,
 appear to be correlated \citep{Cieza2010}: disks with lower accretion rates 
or small upper limits have lower inferred disk masses.  
%\citet{Muzerolle2010} also 
%find that disks we would call homologously depleted ("warm excess" disks in their terminology) 
%have a lower frequency of accretion.  
There is no clear reason why the accretion rate should drop because 
submicron-sized grains had grown; however, disks undergoing a homologous 
depletion of their total mass in gas and dust should have 
low inferred masses \textit{and} a lower frequency and 
rate of accretion.   

Other studies, particularly the C2D survey of transitional disks 
presented in \citet{Merin2010}, conclusively show that SED modeling is 
needed to properly characterize disks.    By modeling optical through 
submm photometry and spectroscopy of a mid-IR selected sample of candidate transitional disks 
("cold disks" in their terminology), they determine which disks have inner holes and 
which lack evidence for an inner hole and investigate typical mid-IR colors for 
both kinds of disks.   Their analysis shows that the mid-IR colors of 
transitional disks overlap with primordial disks (e.g. their Figure 15).  
Computing the frequency of transitional disks 
using mid-IR colors is then error prone unless SED modeling of well 
characterized sources is used to identify uncontaminated regions of color-color 
space and quantiatively assess contamination where transitional disks and 
primordial disks have the same colors.  This finding is qualitatively consistent with our 
Taurus modeling results.  

%\subsubsection{Luhman et al. (2010)}
\citet{Luhman2009} compile data for numerous 1--10 Myr-old clusters to 
uniformly identify transitional disks and quantify their population 
self consistently.
Their modeling follows a different approach than that followed here 
and in some previous work \citep[e.g.][]{CurrieLada2009,Merin2010}, as they select synthetic 
K$_{s}$-[8] and K$_{s}$-[24] colors produced from the \citet{Dalessio2006} 1+1D disk code as fiducial colors 
separating primordial disks from transitional disks.  The model they use has two grain populations:
where the "big" grain population extends from submicron to millimeter sizes and 
is confined to the disk midplane, while the "small" grains lie above the midplane 
and are $\lesssim$ 1 $\mu m$ in size.  It assumes a negligible
 accretion rate ($\dot{M}$ = 10$^{-10}$ M$_{\odot}$ yr$^{-1}$) and a "depletion 
factor", $\epsilon$=0.001, which removes 99.9\% of the small dust grains.  
They find lower transitional disk frequencies 
and thus derive a shorter lifetime for the transitional disk phase.  
They claim that previous work \citep[e.g.][]{CurrieLada2009,SiciliaAguilar2009}
arguing in favor of a longer transitional disk lifetime improperly identifies 
sources with emission weaker than the median Taurus SED as transitional disks 
or conflates a lack of near-IR emission with evidence for disk clearing\footnote{
As can be seen from a simple inspection of the \citet{CurrieLada2009} analysis,
 their disk classification was instead based on comparing SEDs to the 
\textit{\textbf{lower-quartile}} Taurus SED.  This was clarified in later 
work \citep{CurrieKenyon2009}.}.

We agree with \citet{Luhman2009} that some primordial disks can have 
very blue colors due to dust settling.  However, they did not quantitatively 
consider the possibility that some transitional disks may (in some color-color diagrams) 
have redder colors than the bluest primordial disks.  Based on a more detailed, multiwavelength analysis, 
we derive higher transitional disk frequencies than they do for nearly all clusters, 
especially for those older than Taurus.  Our disagreement primarily arises 
because many disks that \citet{Luhman2009} identifies as being optically-thick primordial 
disks from their color criteria 
are consistent with having optically thin/marginally optically-thin emitting dust 
in the mid-IR \textit{and} low inferred disk masses.  If their weak mid-IR emission 
 were due only to the growth and settling of the smallest grains, their far-IR to submm emission 
(and thus inferred disk masses) should be comparable to known primordial disks, 
which is not observed.  Therefore, 
these disks are more consistent with being transitional disks.

There are other aspects of \citeauthor{Luhman2009}'s analysis that 
weaken or undermine their conclusions.  First, while \citet{Luhman2009}'s adopted model results from a sophisticated code, 
their conclusions are prone to large uncertainties, mostly because they use color comparisons 
from a single model to identify transitional disks, not full SED modeling.
Within the context of their model, the \citeauthor{Luhman2009} conclusions are \textit{extremely} 
sensitive to assumed values for their free parameters, $\epsilon$ and $\dot{M}$, and 
thus highly selective.  Assuming either that disks remove 99\% of their small dust 
instead of 99.9\% or have a small accretion rate of $\dot{M}$ = 10$^{-9}$ M$_{\odot}$ yr$^{-1}$ 
revises their fiducial mid-IR colors redwards by $\sim$ 0.5 mags (e.g. K$_{s}$-[8] $\sim$ 1.75, 
K$_{s}$-[24] $\sim$ 4.5 for K5--M2 stars).  Using these colors to classify disks
yields results similar to those presented here and in \citet{CurrieLada2009} for 
15 of the 16 model combinations of $\epsilon$ (0.001-1) and $\dot{M}$ (10$^{-10}$--10$^{-7}$ 
M$_{\odot}$) studied in \citet[][e.g. their Figure 13]{Dalessio2006}.

Second, while $\epsilon$ may be a true free parameter,
measured accretion rates for many sources are much higher than that 
from the disk model used by \citet{Luhman2009}, rendering their 
adopted model inapplicable.  For example, ID-85 in NGC 2362, identified as a transitional 
disk by \citet{CurrieLada2009}, is a likely accretor based on its H$_{\alpha}$ emission.  
Using relations from \citet{Dahm2008} and \citet{HerczegHillenbrand2009}\footnote{Here we assume E(B-V) = 0.01 (1/10th the cluster
average listed by \citealt{Moitinho2001}), an age of 5 Myr, a stellar radius for an M2 star of 
this age from \citet{Baraffe1998}, and R$_{in}$ = 5 R$_{\star}$.  
The intrinsic uncertainty in the accretion rate is $<$ 0.5 dex \citep[see][]{Dahm2008}.
According to SED modeling presented here, the reddening 
is probably higher, meaning that the intrinsic H$_{\alpha}$ luminosity is higher and the 
resulting accretion rate is also higher.}, its accretion rate is 
$\sim$ 4--7 $\times$ 10$^{-9}$ M$_{\odot}$ yr$^{-1}$, or \textit{40 to 70 times greater} 
than what \citeauthor{Luhman2009} assume.   From comparing the SED of ID-85 to output from 
\citeauthor{Dalessio2006} disk models with realistic accretion rates, 
this source should be classified as a transitional disk, contrary to \citeauthor{Luhman2009}'s results.
Taurus-Auriga, Trumpler 37,and other clusters analyzed by \citeauthor{Luhman2009} include
 many sources with weak mid-IR emission (transitional disks/borderline cases) whose 
measured accretion rates render \citeauthor{Luhman2009}'s analysis inapplicable, such as
 ZZ Tau \citep[1.3 $\times$ 10$^{-9}$ M$_{\odot}$ yr$^{-1}$,][see also \citealt{SiciliaAguilar2010}]{WhiteGhez2001}.

Third, the model that \citet{Luhman2009} adopt likely 1) has an extremely low disk mass which 
2) is almost never optically thick and thus cannot distinguish \textit{optically-thick} 
primordial disks from other disks.   As shown in Tables 2.2 and 2.4 of 
\citet{Espaillat2009}, where the \citeauthor{Dalessio2006} input 
parameters are detailed, the disk mass for models with parameters 
adopted by Luhman is $\sim$ 1.38--1.45 $\times$10$^{-4}$ M$_{\odot}$: an order-of-magnitude 
(or more) smaller than our primordial/transitional disk division for all but the latest-type stars and 
the median primordial disk mass \citep[][and Appendix]{Andrews2005}.
Furthermore, Figure 2.6 of \citet{Espaillat2009} shows that almost none 
of this disk is optically thick to its own radiation ($\tau_{R}$ $>>$ 1), unlike 
our adopted model and other \citeauthor{Dalessio2006} models.
%\textit{Independent of depletion factor}, mid-IR emission predominately originates from 
%optically-thick regions of the disk \citep[cross ref. Figures 2.6, 2.15--2.18 from][]{Espaillat2009}.  
Thus, a major reason why the Luhman's adopted model yields weak mid-IR emission is because 
of its low dust mass, not just because of dust settling.  To better parse disks 
into different states, \citeauthor{Luhman2009} should have 
instead adopted a model yielding an optically-thick disk with a dust mass comparable to 
typical primordial disk masses (e.g. the fiducial model in \citealt{Dalessio2006} except with 
$\epsilon$ = 0.001 and the accretion rate set to a negligible value). 

Finally, SED modeling and imaging of individual sources shows that the \citeauthor{Luhman2009} color criteria 
fails to uniquely identify primordial/transitional disks.  For example, UX Tau and LkCa 15 are identified 
as primordial disks according to \citeauthor{Luhman2009}'s color criteria as they are 
redder than the authors' primordial disk limit.  But SED modeling 
and/or high-contrast imaging show that both have large gaps with optically-thin dust/no dust 
extending from $\approx$ 0.1 AU to $\approx$ 50 AU from the star \citep[][]{Thalmann2010,Mulders2010,Espaillat2007}.
SED modeling also shows that sources in NGC 2362 (e.g. ID-3), IC 348 (e.g. ID-97), and 
Trumpler 37 (e.g. ID 14-11) classified by \citeauthor{Luhman2009} as primordial disks probably 
have inner holes and thus are transitional disks \citep[e.g.][this work]{CurrieLada2009,SiciliaAguilar2007}.

Even though we disagree with \citeauthor{Luhman2009}'s conclusions for reasons described above, 
 their study reinforces a major point made here -- mid-IR colors of disks can be highly degenerate.  
The degeneracy highlighted from both of our studies shows that full SED modeling is needed to 
accurately assess the frequency of transitional disks and the duration of the transitional disk phase.
  While time intensive, this approach provides \textit{de facto} more 
information about disk properties than can be provided by comparing 
source colors with synthetic colors resulting from a single disk model in a single color-color diagram.

\subsection{Transitional Disks In Young Clusters: Why the Paucity of Transitional Disks 
in Taurus Does Not Imply a Short Timescale for Disk Clearing}
Many pre-Spitzer studies have argued that the small number of transitional disks compared to optically-thick 
primordial disks and disk-less stars in young clusters like Taurus provides 
evidence in favor of a short ($\lesssim$ 0.1 Myr) transition disk phase due to 
rapid disk clearing \citep[e.g.][and later work]{Skrutskie1990,Simon1995,Wolk1996}. 
Though transitional disks are infrequent compared to primordial disks and diskless stars 
in the youngest clusters (e.g. 20\% for Taurus), 
it does \textit{not} follow that the transitional phase must be very shortlived (e.g. $\lesssim$ 0.1 Myr).
Here, we review why this is the case, describe how our work and other recent studies 
support this position, and consider recent work that revives older claims.

As argued by \citet{CurrieLada2009}, stars in Taurus are typically 
too young compared to the mean protoplanetary disk lifetime (3--5 Myr) 
for the majority of them to begin clearing.  Thus, it is unsurprising that relatively few 
show evidence of active clearing.  Our parametric modeling quantitatively 
supports this argument, showing that transitional disks do not comprise the 
majority of disks in the youngest clusters even for lengthy transitional disk phases 
(e.g. t$_{TD}$ = 1 Myr).  Thus, the size of the transitional disk population at $\sim$ 1--2 Myr does 
not, by itself, clarify how long protoplanetary disks show signs of active clearing over a 
typical lifetime of 3--5 Myr.
To explain the large population of diskless stars in Taurus ($\sim$ 30\% of the total population), 
 \citet{CurrieLada2009} argue that stars with close binary companions lose their protoplanetary 
disks far more rapidly than stars without close binary companions \citep[e.g.][]{IrelandKraus2008}.

\citeauthor{Luhman2009} disagree with both of these claims.
They argue that age cannot explain Taurus's paucity of transitional disks,
since since diskless stars in Taurus are more dispersed and are thus older than stars with disks.
We agree with \citet{Luhman2009} that an age spread may also explain the large number 
of diskless sources in Taurus.  However,  this fact would simply confirm that 
that older stars are less likely to have disks 
\citep[e.g.][]{SiciliaAguilar2006,Hernandez2007,Currie2007,CurrieLada2009} and thus does 
not provide evidence in favor of rapid disk clearing.  

  \citeauthor{Luhman2009} also argue that binarity cannot explain the large diskless population in 
Taurus.  They note that the Taurus binary frequency declines with mass.  They claim that if 
many Taurus disks are cleared by binaries then the disk fraction should be highest at low masses, which 
is not observed.  However, the drop in binary frequency for low-mass stars is 
heavily weighted towards a drop in \textit{wide binary companions} with separations 
of $\sim$ 200 AU or greater, which are irrelevant here \citep[][]{Kraus2006,Kraus2011}.  Moreover, many diskless 
stars in Taurus are tight binaries (Kraus et al. 2011, ApJ submitted).  With tight binaries removed, 
the population of diskless stars shrinks significantly.  \citet{Cieza2009} also find that 
disks in short-period binary systems are less frequent than those around wide-separation binaries 
or single stars \citep[see also][]{Bouwman2006}.  

In summary, either binarity or an age spread explain the smaller number of Taurus stars with transitional 
disks relative to primordial disks \textit{and} diskless stars.  Moreover, parametric modeling shows that 
the frequency of protoplanetary disks in the transitional disk phase 
is perfectly consistent with an extended transitional disk phase.  
Thus, Taurus data \textit{provides no evidence} that the transitional disk phase must be rapid.

\subsection{Limitations and Uncertainties of Our Study and Future Work}
\subsubsection{Qualifications on Our Identification of Transitional Disks}
Our identification of some transitional disks (those labeled 
as homologously depleted) is predicated on 1) our assumptions about 
the range of primordial disk masses, 2) the reliability of 
the Robitaille models in estimating disk masses, and 3) the reliability 
of our assumed stellar masses.  Current data shows that the lower quartile of 
 masses for optically-thick primordial disks 
is 0.003 M$_{\star}$ and nearly all primordial disks have 
masses greater than 0.001 M$_{\star}$.  Since not 
all Taurus members have been targed by submillimeter observations, it 
is possible that future surveys will uncover more
disks with optically-thick near-to-mid IR emission but inferred 
masses that are systematically lower than the interquartile range/full 
range from current observations.   

Additionally, as shown in Figure \ref{taurusmodeling}, the best-fit disk models 
for homologously depleted transitional disks have median values 
for absolute disk masses that yield fractional disk masses 
lying below our adopted primordial disk limit.  However, 
some of the models for each disk (typically, 10--30\%) yield masses 
lying above 0.001 M$_{\star}$.  Thus, it is 
possible, albeit unlikely, that some of these disks are not low mass.

However, submillimeter data constrain the mass of emitting, optically 
thin dust to be less than these limits.  Moreover, the 
best-fit models for some primordial disks and all borderline cases 
include those with masses below 0.001 M$_{\star}$.  Thus, 
uncertainties in determining disk masses do not necessarily 
bias our results in favor of finding a high frequency of transitional disks.
Similarly, the classification of some sources 
with optically-thin near-to-mid IR emission but 
large inferred disk masses (e.g. V836 Tau) as primordial disks is conservative.
While these sources lack clear evidence for \textit{cleaned} inner holes/gaps,
their inner regions may be optically thin, and they may be undergoing 
the early stages of inner disk clearing.

More generally, the intrinsic sampling of the \citeauthor{Robitaille2007} grid 
is non-uniform across all fitting parameters, which in principle could 
 undermine our conclusions.  As mentioned previously, disks 
without inner holes are more frequently represented than those without holes 
and massive disks (M$_{disk}$ $\sim$ 0.01 M$_{\odot}$) are also better 
represented than lower-mass disks.  Figure 3 of \citeauthor{Robitaille2007} 
indicates the presence of ``coupled" non-uniform sampling between parameters: 
there are many more models with an envelope accretion rate greater than 10$^{-6}$ M$_{\odot}$ yr$^{-1}$ 
\textit{and} that these models predominantly have disk masses greater than 10$^{-3}$ M$_{\odot}$.  

\begin{figure}
\plotone{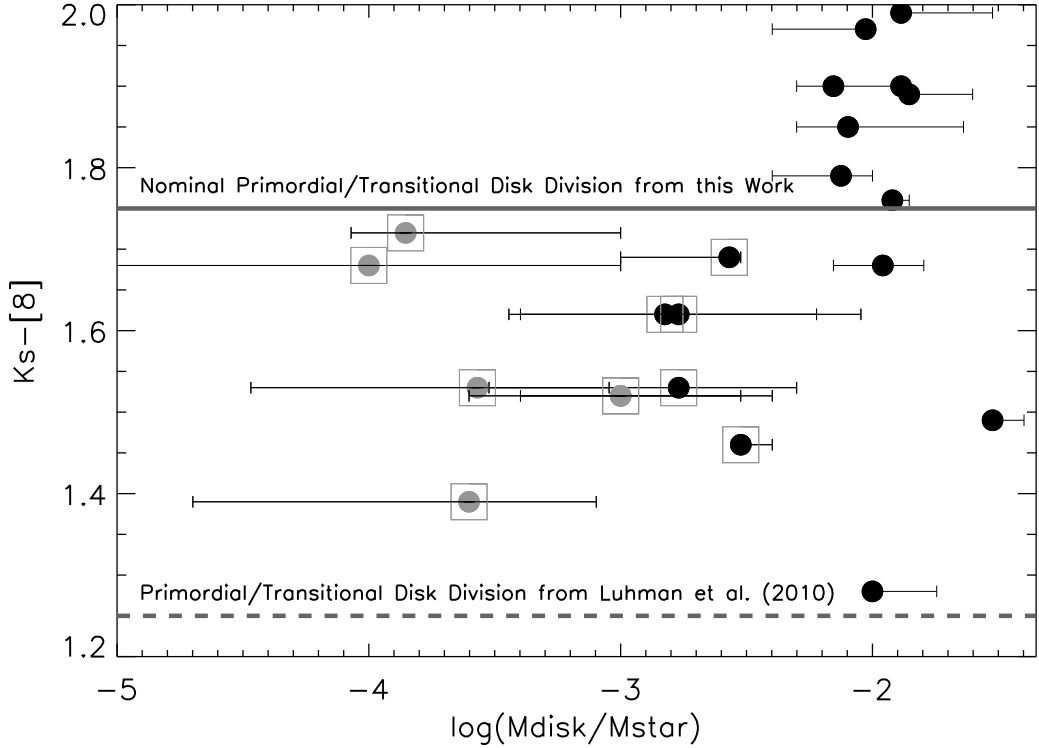}
\caption{K$_{s}$-[8] vs. fractional disk mass for homologously depleted 
transitional disks (grey circles surrounded by squares), "borderline" 
cases (black dots surrounded by squares),
and primordial disks (black circles).  The error bars identify the 
the interquartile range of disk masses for each star.  For reference, we show 
the threshold in K$_{s}$-[8] color below which the Taurus transitional 
disk population begins (horizonatal grey line, K$_{s}$-[8] = 1.75) and 
that assumed in \citet{Luhman2009} for K5--M2 stars 
(horizontal black dotted line, K$_{s}$-[8]=1.25).  This plot shows 
that the ranges of disk masses for the homologously depleted 
transitional disk and primordial disk populations are distinct.  
  The interquartile range of disk masses for the borderline 
cases slightly overlaps with the range for primordial disks.}
\label{quartilemass}
\end{figure}

\begin{figure}
\centering
\plottwo{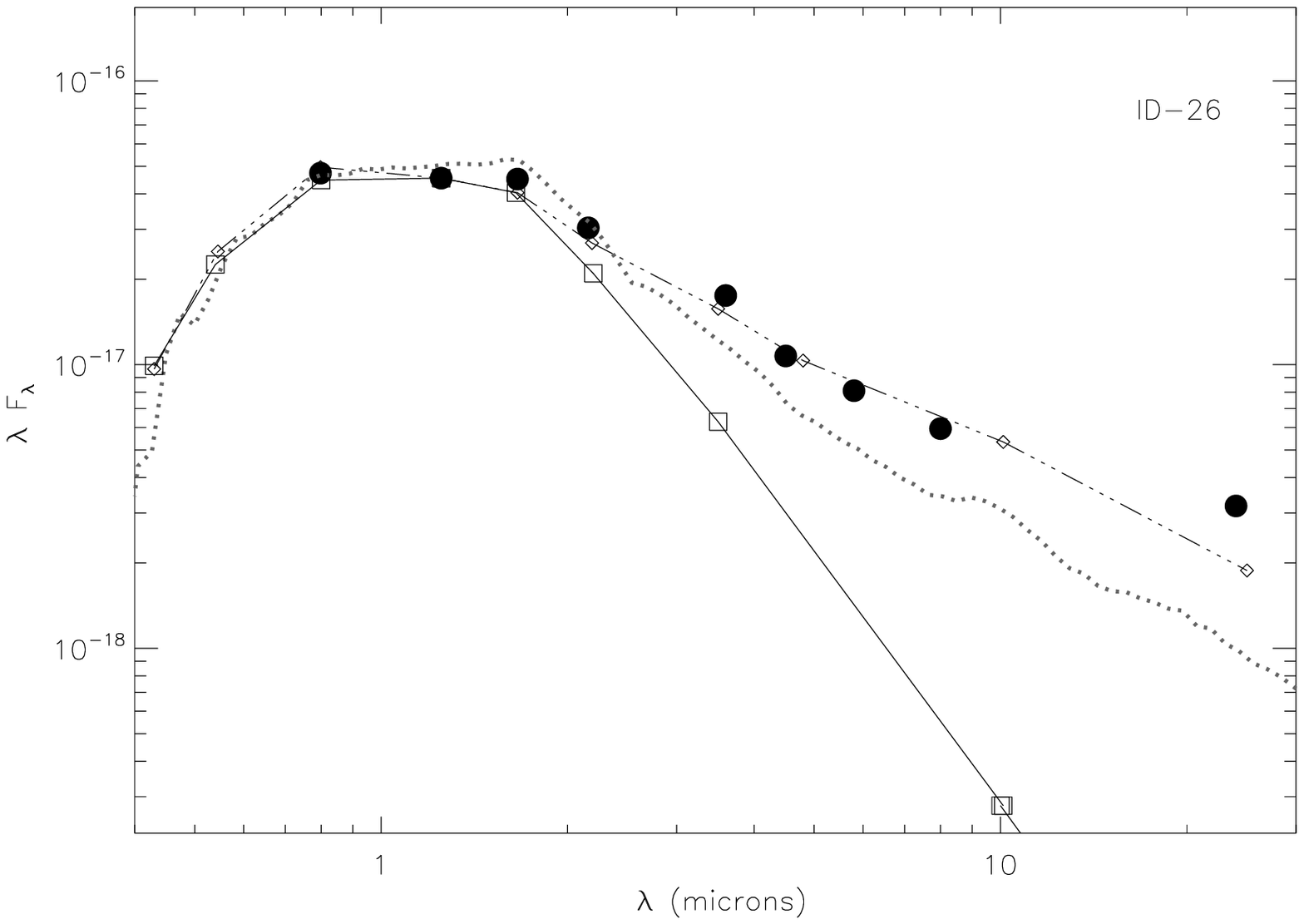}{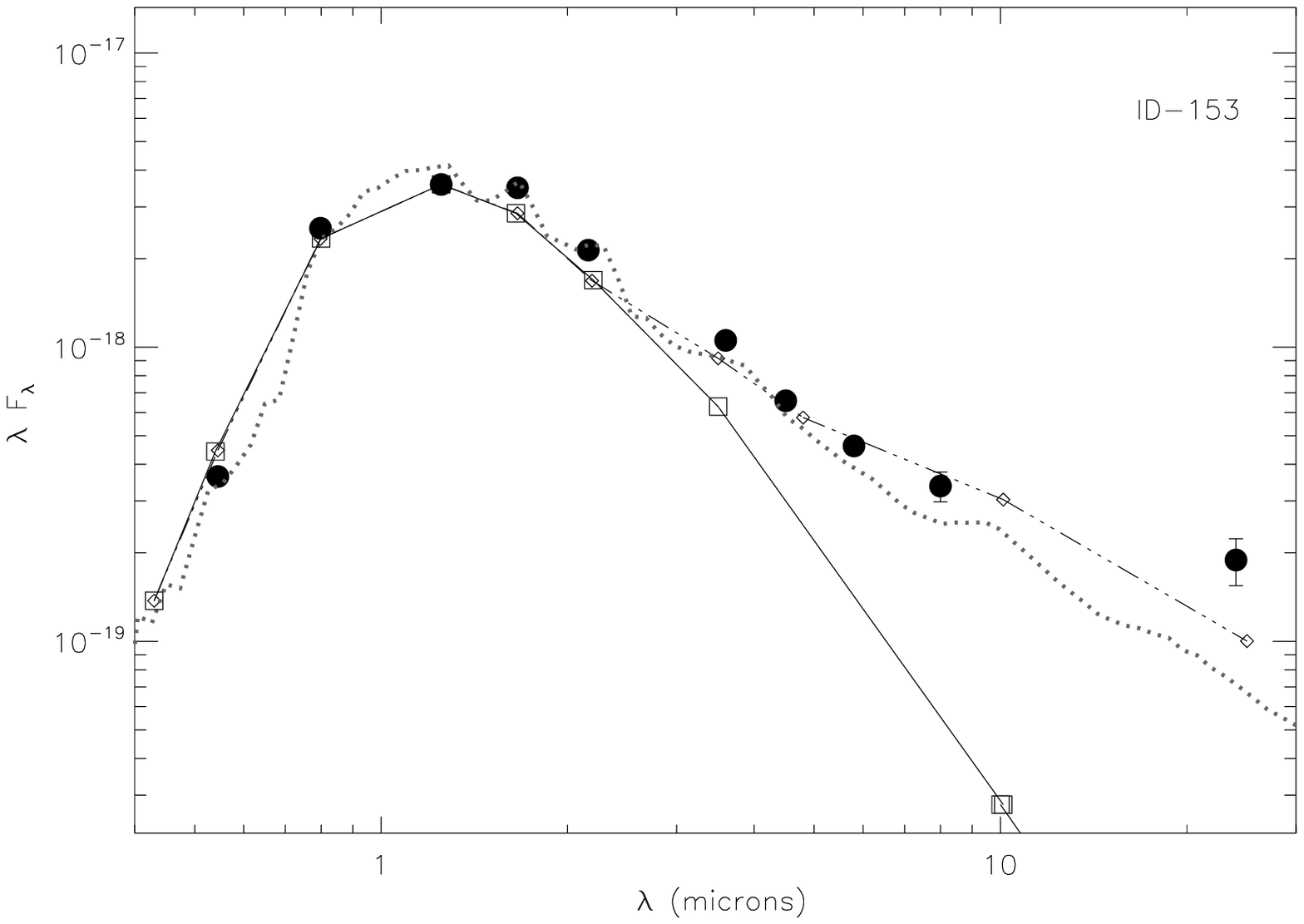}
\plottwo{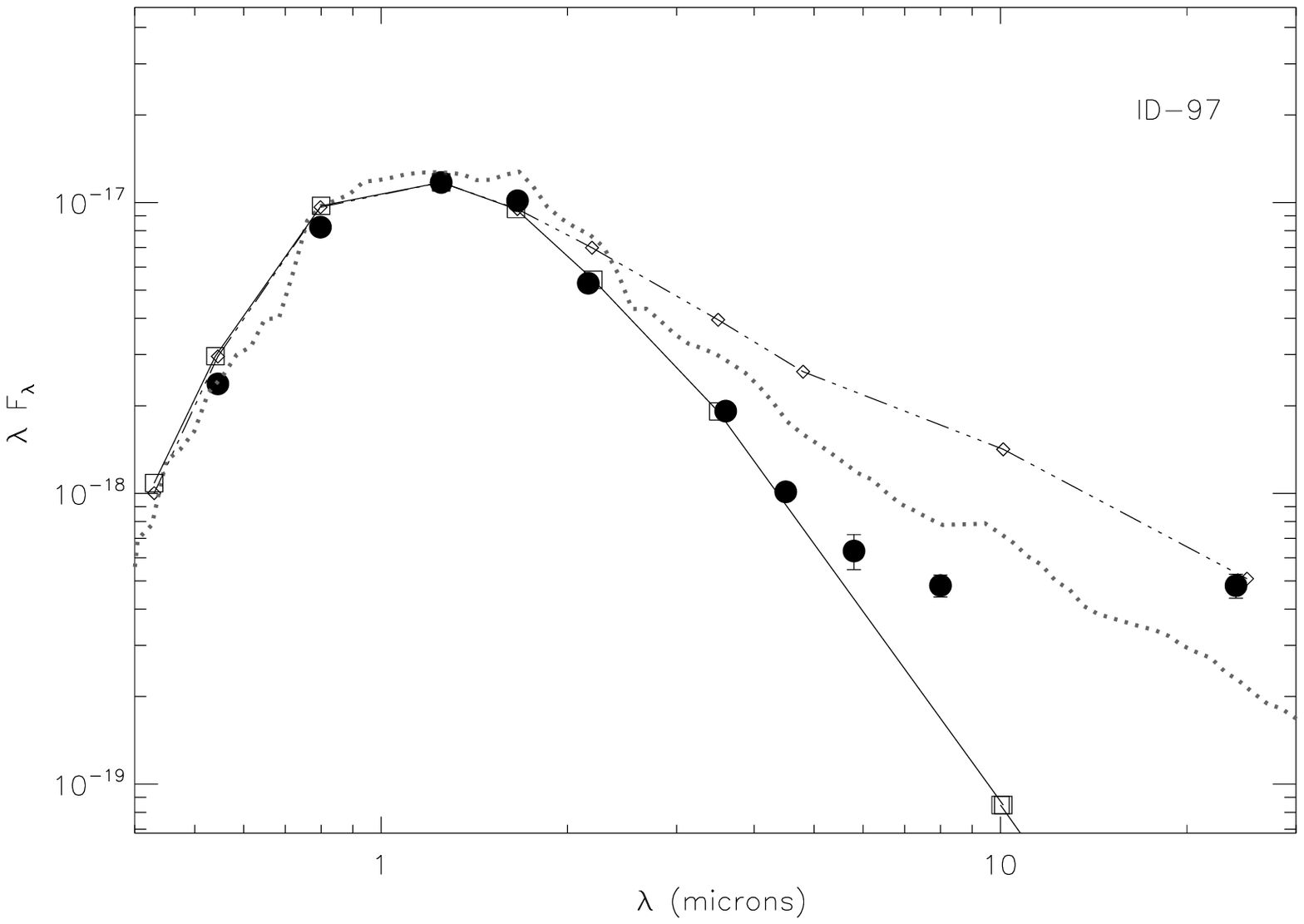}{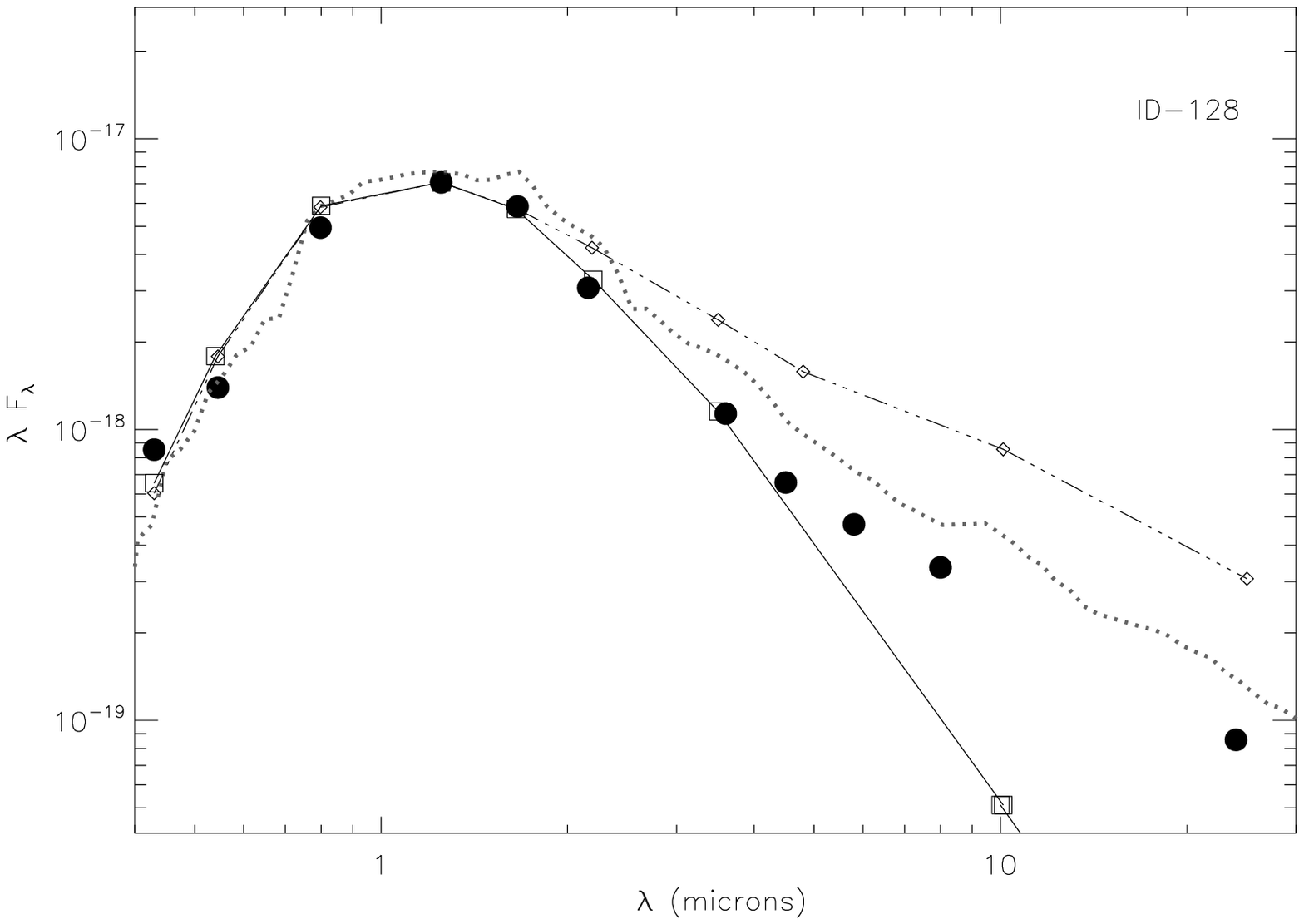}
\caption{Representative SEDs for IC 348 sources.
The provisional disk states for 
these members are (clockwise from top-left): primordial disk, primordial disk, 
homologously depleted transitional disk, and transitional disk with an inner hole.}
\label{ic348example}
\end{figure}

\begin{figure}
\centering
\plottwo{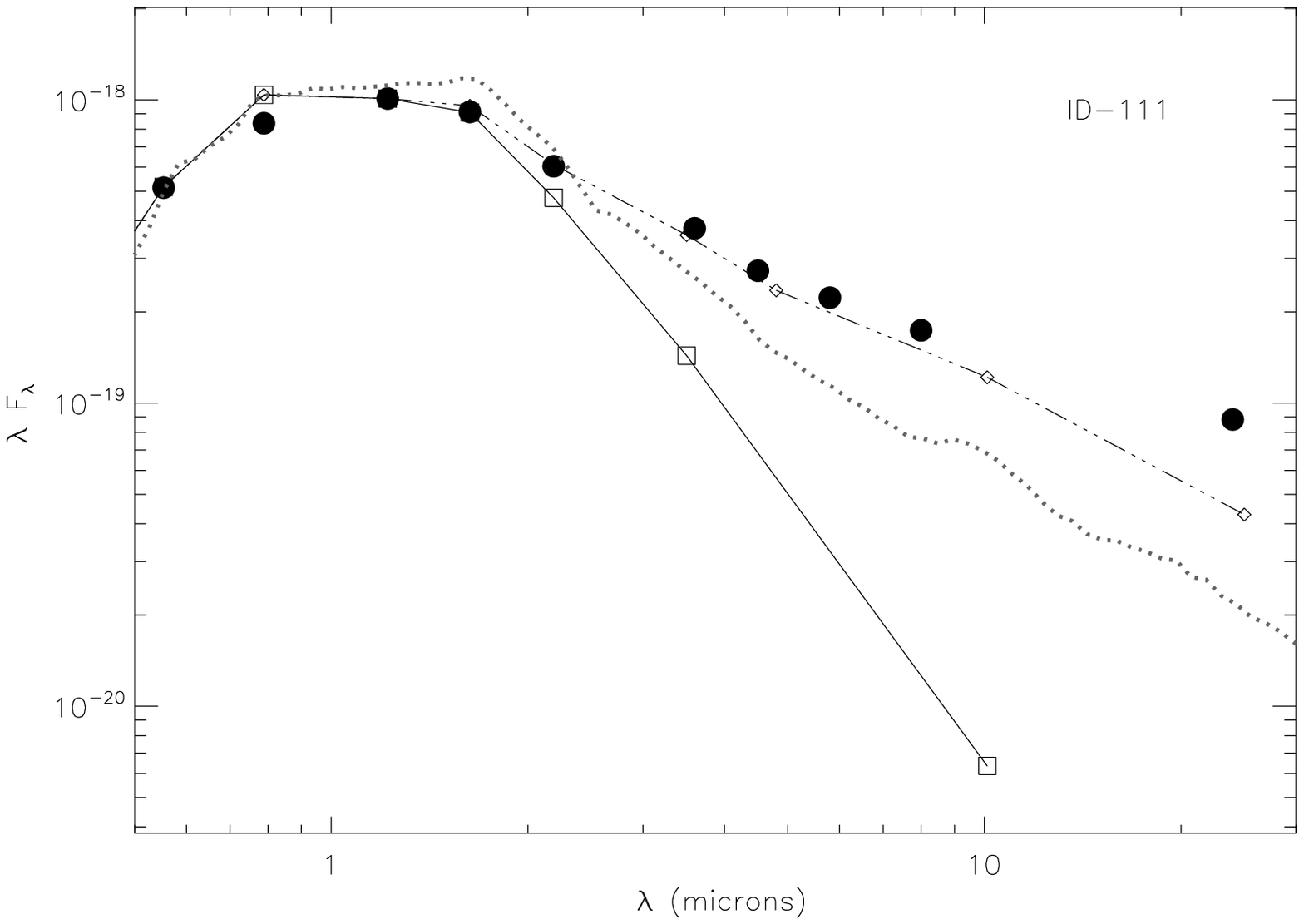}{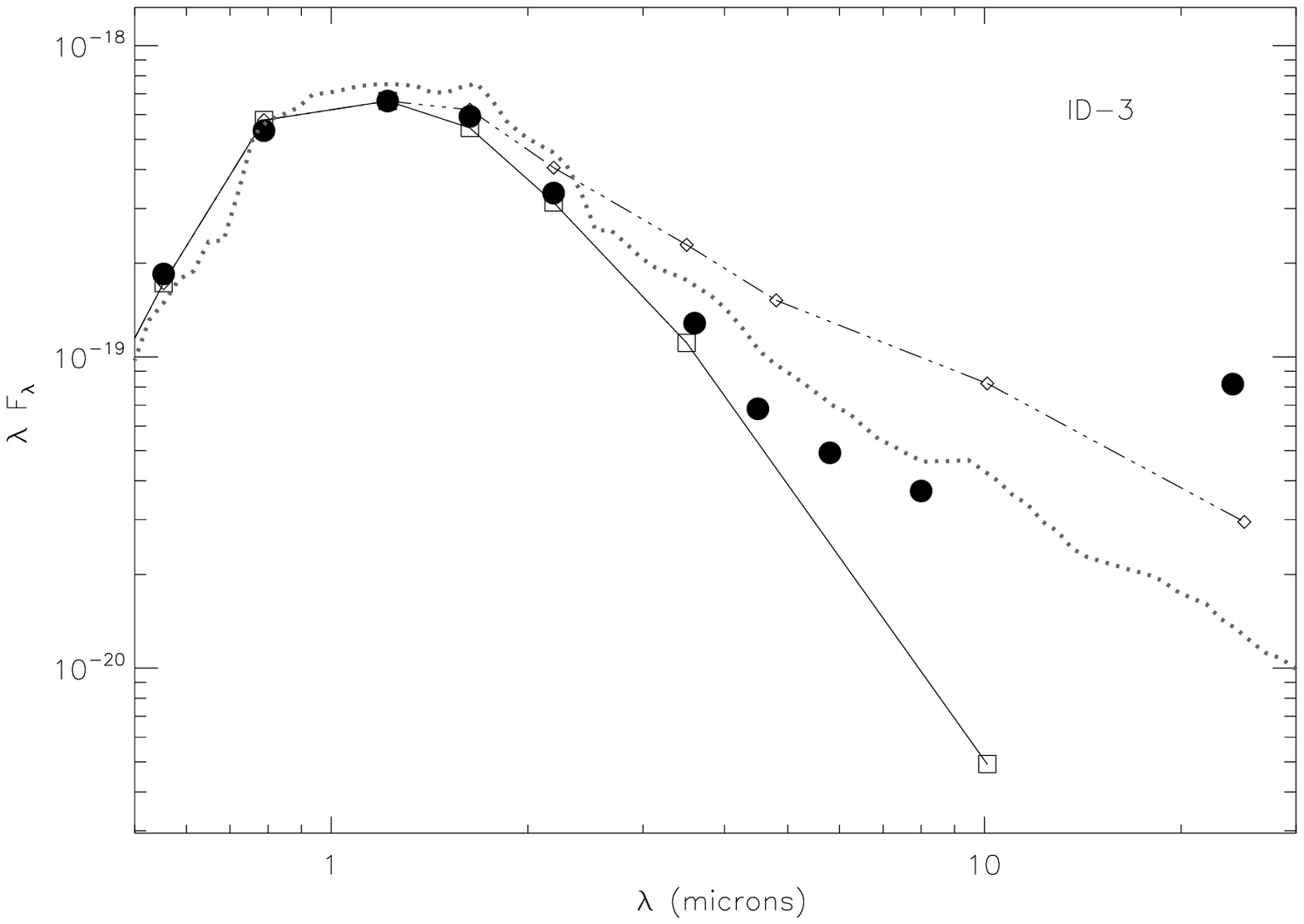}
\plottwo{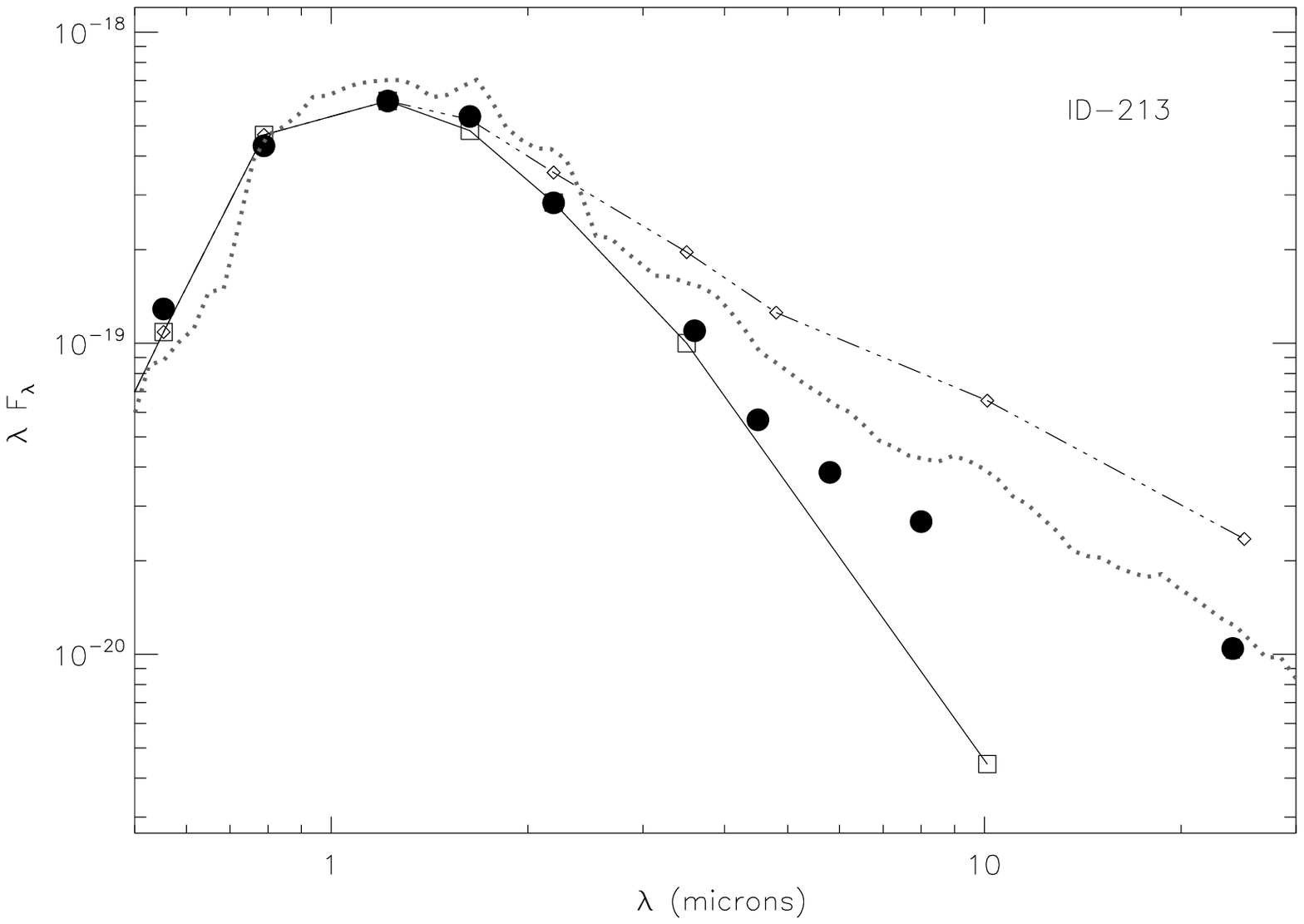}{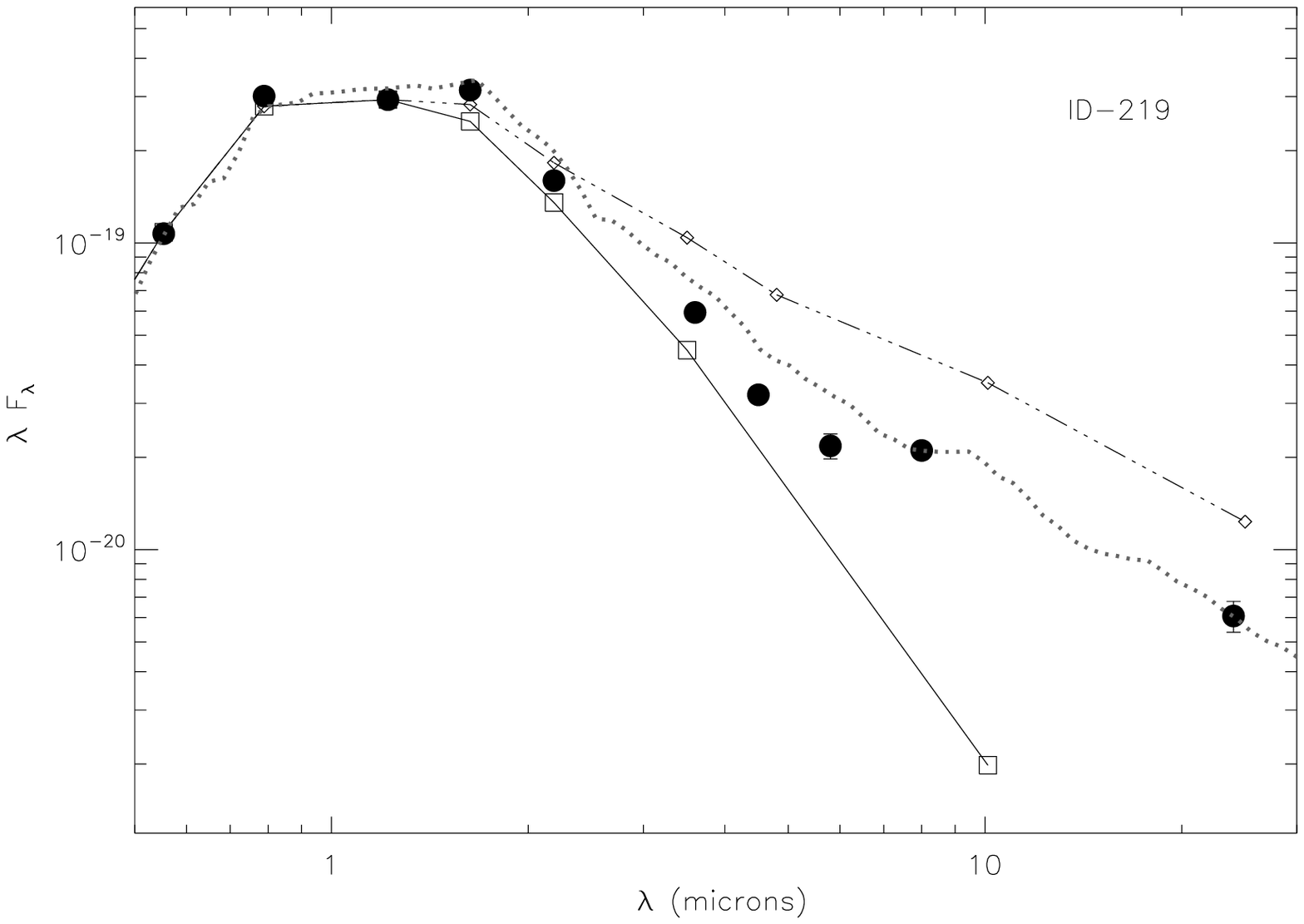}
\caption{Representative SEDs for NGC 2362 sources.  
The provisional disk states for 
these members are (clockwise from top-left): primordial disk, transitional disk with inner hole, 
homologously depleted transitional disk, and homologously depleted transitional disk.}
\label{ngc2362example}
\end{figure}

However, our disk mass estimates are not undermined by 
non-uniform sampling.  The lower two panels of \citeauthor{Robitaille2007}'s Figure 3 
and our Figure \ref{aatau} indicate that simply including far-IR and submm data 
eliminates vast swaths of parameter space and overcomes these sampling biases.  
The intrinsic sampling of the Robitaille grid is  
weighted towards massive disks, yet we find numerous homologously depleted disks 
inspite of this bias because far-IR/submm non-detections rule out nearly all massive disks.  
Conversely, the submm detections for primordial disks provide strict limits on 
the inferred disk mass \citep[see also][]{Beckwith1990,Andrews2005} and  
make the interquartile range of M$_{disk}$/M$_{\star}$ for these objects nearly 
single valued in many cases (see Figure \ref{quartilemass}).  On the other hand, 
uncertainties in deriving other disk parameters (e.g. flaring power) have not 
been fully explored here and may be seriously affected by non-uniform sampling.

Our analysis leverages not just on the precision with 
which we can estimate the mass of disks but also our accuracy.
  Estimating a mass of emitting dust from submm data alone 
requires assuming a value for the dust opacity, $\kappa$ \citep[e.g.][and later 
references]{Beckwith1990,Beckwith1991,Henning1996}.  Since the dust opacity is not 
strictly known (and is affected by grain growth), what can be constrained from such data is
the product of the dust mass and the opacity, not simply the dust mass.  
On the other hand, grain growth/dust settling (which affect $\kappa$) have 
far stronger effects on the near-to-mid IR portion of disk SEDs than in the submm 
\citep[e.g.][]{Wood2002,Dalessio2006}.  For instance, ``settled disk" models 
from \citet{Dalessio2006} have 250-1000 $\mu m$ emission equal to or \textit{larger} 
than that for disks without grain growth/settling \citep[see Figure 13, top two sets 
of panels in][]{Dalessio2006}.

The total disk mass is also uncertain.  Since continuum 
IR to submm emission comes predominantly from dust, a gas-to-dust 
ratio must be assumed in converting a submm-inferred dust mass to 
a total disk mass.  The models used here, in \citet{CurrieLada2009}, 
and other work (e.g. \citealt{Cieza2010,Andrews2005}) assume 
standard values from $\kappa$ and a solar gas-to-dust ratio 
\citep[see][]{Beckwith1991}.  However, comparisons with 
disk masses derived from accretion rates (assuming steady-state accretion) 
indicate that our methods may systematically underestimate true disk 
masses \citep[e.g.][]{Andrews2007}.  This may be due either to 
an inaccurate assumed value for $\kappa$ or a substantially non-solar 
gas-to-dust ratio, which may occur from grain growth and 
planetesimal formation.  Our analysis provides 
evidence that masses for homologously depleted transitional disks    
are systematically lower by a factor of 10-1000 compared to primordial disks; 
other work provides evidence that gas and small dust deplete on similar timescales 
for disks in general and for homologously depleted transitional disks in particular 
\citep{Fedele2010,KennedyKenyon2009,Cieza2010}. 
However, given the uncertainties in determining $\kappa$ and the gas-to-dust ratio,
our analysis does not provide an absolute calibration for disk masses that 
can inform models for planet formation \citep[e.g.][]{KenyonBromley2009,BromleyKenyon2011}.

Accurately quantifying uncertainties in stellar mass for each star is 
another challenge.  
Between 1 Myr and 5 Myr, the \citet{Baraffe1998} isochrones with a mixing length 
parameter of L$_{p}$ = 1.9 H$_{p}$ (``required to fit the Sun") predict 
that 1.4 M$_{\odot}$ stars increase in T$_{eff}$ 
by $\sim$ 125 K from 4663 K to 4786 K, thus changing in spectral type by 
$\sim$ 1 subclass \citep[cf. ][]{Currie2010b}.  The variation is smaller ($\sim$ 100 K) 
for solar-mass stars and inconsequential for very low-mass stars (M$_{\star}$ $\lesssim$ 
0.7 M$_{\odot}$).  Adopting the conversion from spectral type to mass from 
\citeauthor{Baraffe1998}, our stellar masses are then intrinsically uncertain by 
$\approx$ 0.1 M$_{\odot}$ or less due to the different mapping between spectral type 
and mass as a function of age.  The uncertainties in 
measured spectral types are up to 1--2 subclasses, considering the range of 
spectral types reported in the literature for various Taurus members \citep[e.g.][]{WhiteHillenbrand2004}.  
The age and spectral type uncertainties listed above 
indicate that our fractional disk masses could be uncertain by an additional 
10--20\%, and that considering these uncertainties could slightly broaden 
the interquartile range of fractional disk masses for some objects.

Finally, our analysis may miss some sources 
whose gaps cannot be inferred from photometric data alone.  Among our sample, 
the SED for DH Tau shows evidence for a $\lambda$F$_{\lambda}$ 
$\propto$ $\lambda$$^{-3}$ decline from 3.6 $\mu m$ to 8 $\mu m$, 
%which is confirmed by inspection of the 5--35 $\mu m$ IRS 
%spectrum \citep{Furlan2006}, 
similar to that for transitional disks UX Tau and LkCa 15, 
which have large (tens of AU) gaps separating a small optically-thick, $\sim$ 1500 K inner 
disk from an optically-thick outer disk.
The SED fit to this source from the Robitaille et al 
grid is poor.  Since the Robitaille et al grid provides good sampling for 
full disks with a range of masses (and thus mid-IR emission) 
\textit{and} disks with cleaned inner holes, it is plausible that 
DH Tau's disk has a different morphology such as a disk with 
a developing gap.  Since not all Taurus members have been targeted by mid-IR spectrographs, 
more disks with small holes/gaps may be identified by future observations.
Indeed, \citet{Espaillat2011} argue that IP Tau, which 
we identify as having a primordial disk in spite of its weak mid-IR emission, 
has a $\sim$ 2 AU gap.  Thus, the true size of the transitional disk population in Taurus, 
especially the subset with small gaps/inner holes, is plausibly larger than what we find.

\subsubsection{Limits on Our Study As a Probe of 
Transitional Disk Lifetimes/Disk Clearing}
While our analysis provides strong evidence in favor of a 
long transitional disk phase, especially for very low-mass stars, 
several important caveats qualify this conclusion.  First, 
 many different mechanisms may be responsible for 
explaining the morphologies of transitional disks 
\citep{AlexanderArmitage2009,Cieza2010,SiciliaAguilar2010}.  
Therefore, a "transition timescale/transitional disk lifetime" derived 
from the relative frequency of all transitional disks 
is only an averaged value of the timescales from a number 
of different mechanisms like gas giant planet formation and 
UV photoevaporation.  Some of these mechanisms may 
operate on fast, $\lesssim$ 0.1--0.5 Myr timescales 
but less frequently whereas others take $\gtrsim$ 1 Myr to 
clear dust in disks and operate more frequently.

The disk clearing time for many candidate mechanisms 
(e.g. gas giant planet formation, UV photoevaporation) 
depends on properties such as disk viscosity, initial disk mass and 
initial angular momentum that are poorly constrained and likely have a large 
intrinsic dispersion.  This dispersion in disk properties may 
produce a large dispersion in the duration of the transition phase similar to the 
observed dispersion in total protoplanetary disk lifetimes.  
While the mean transition timescale may be $\sim$ 1 Myr as 
found here, in \citet{CurrieLada2009}, and \citet{SiciliaAguilar2009}, 
the transitional disk phase for individual disks may last much shorter 
or longer.

%\subsection{Future Work}
Quantitatively assessing how transitional disks probe the clearing of 
both gas and dust requires diagnostics of circumstellar gas sensitive 
to the bulk gas content.  The \textit{Herschel Space Observatory} offers 
a sensitive probe of far-IR line emission to identify cool gas in 
planet-forming regions of the disk.  \textit{Herschel} programs such 
as GASPS will survey many 1--10 Myr-old stars for evidence of circumstellar 
gas and thus provide strong constraints on gas dissipation as a function of time.  
Comparing the gas properties of transitional disks with those for primordial 
disks will more definitively determine how transitional disks are clearing 
their gas, complementing studies of dust clearing investigated in our work.

\acknowledgements The anonymous referee provided comments and suggestions that 
greatly improved the organization and content of this paper. 
We thank Cornelis Dullemond, Thomas Henning, Scott Kenyon, 
James Muzerolle, Carol Grady, Adam Kraus, and Meredith Hughes
 for useful conversations and suggestions.  We also thank Richard Alexander for providing 
us with output from his disk evolution models presented in \citet{AlexanderArmitage2009}.  
Finally, we thank Jeroen Bouwman for assistance with our IRS data reduction.
T.C. is supported by a NASA Postdoctoral Fellowship; A.S-A. acknowledges support from the 
Deutsche Forschungsgemeinschaft (DFG) grant SI-1486/1-1. 
This work made extensive use of the \textit{Simbad Astronomical Database}.
{}
\clearpage
\input{ tab_list.tex}
\input{ tab_new.tex}
\input{ tab_diskstate.tex}
\input{ tab_diskstateold.tex}
\input{tab_irs.tex}
\input{ tab_taurus.tex}
\input{tab_taurus2.tex}
\input{tab_taurusextra.tex}
\input{ tab_ic348.tex}
\input{ tab_ngc2362.tex}
\input{ tab_etacha.tex}
\input{ tab_all.tex}
%\begin{figure}
%\plotone{laboca.ps}
%\caption{Submillimeter map of the Coronet Cluster, showing 

%\begin{figure}
%\plotone{FP_seds_2.eps}
%\caption{Same as previous figure.}
%\label{sedsnew2}
%\end{figure}
\clearpage

\clearpage

\clearpage

\begin{figure}
\plottwo{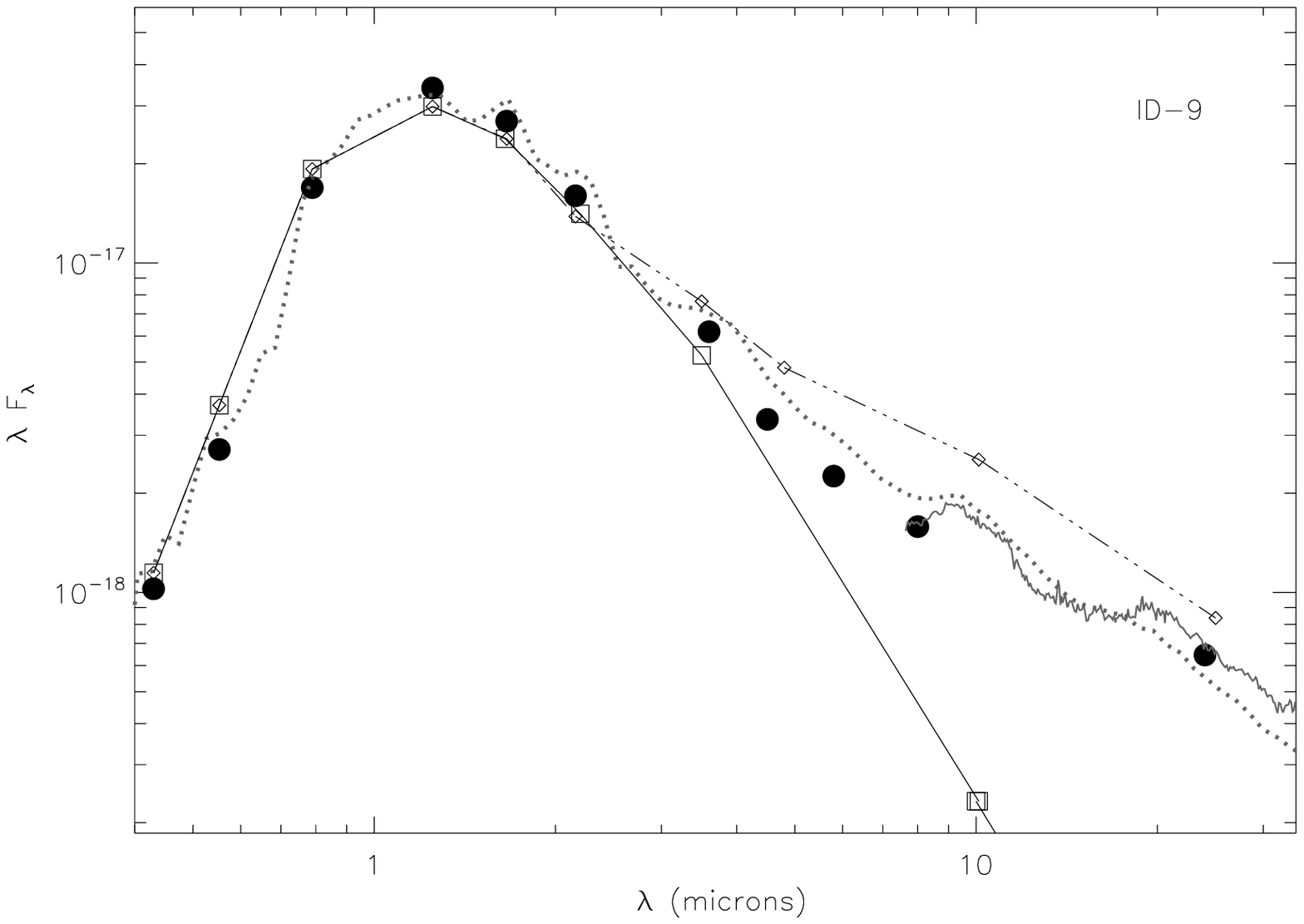}{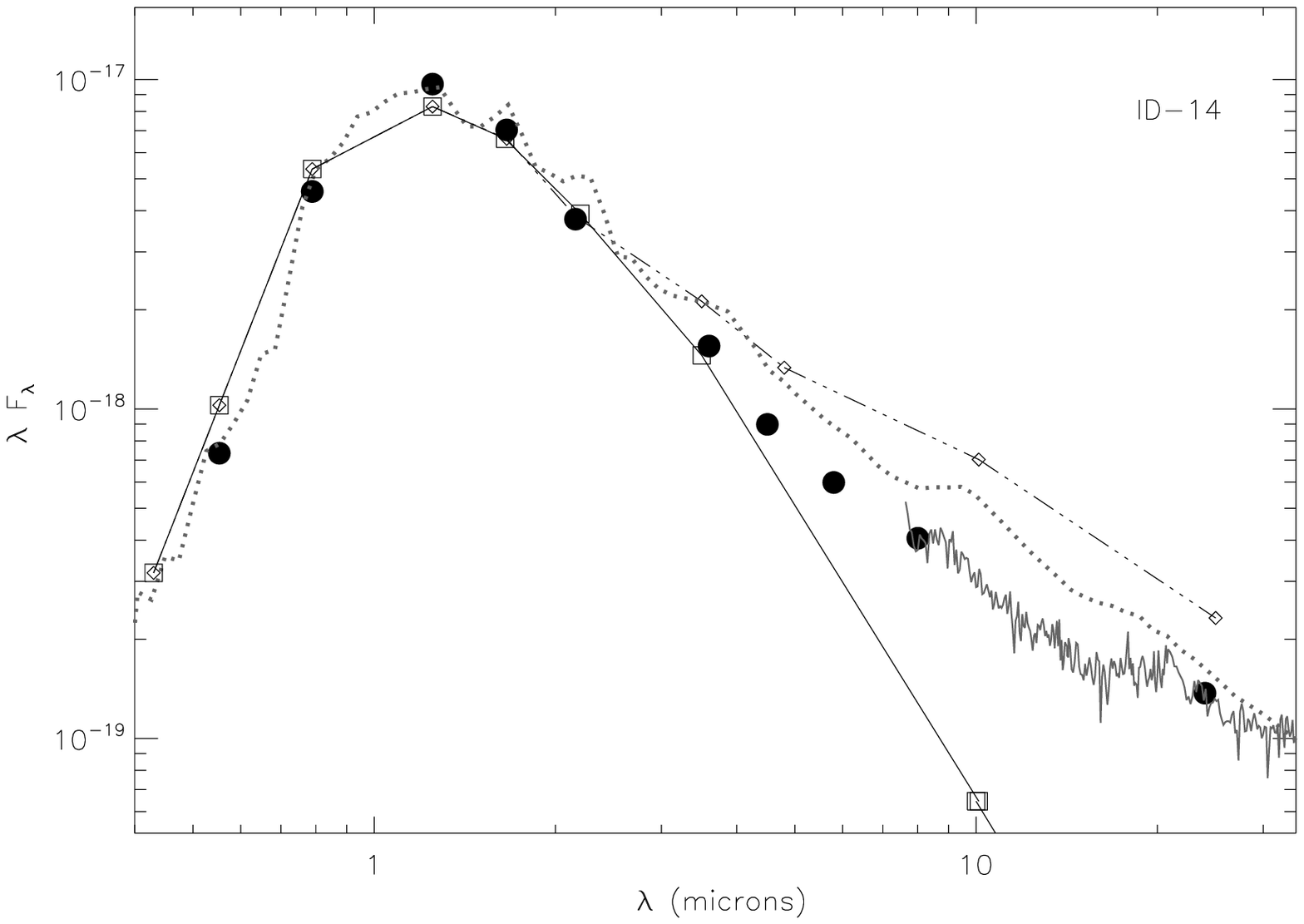}
\caption{SEDs for two $\eta$ Cha members identified as having homologously depleted 
transitional disks based in their photometric and spectroscopic flux densities (dots and 
thick grey line, respectively) compared to the appropriate flat, reprocessing disk models.}
\label{etachaexample}
\end{figure}

\begin{figure}
\epsscale{0.7}
\plotone{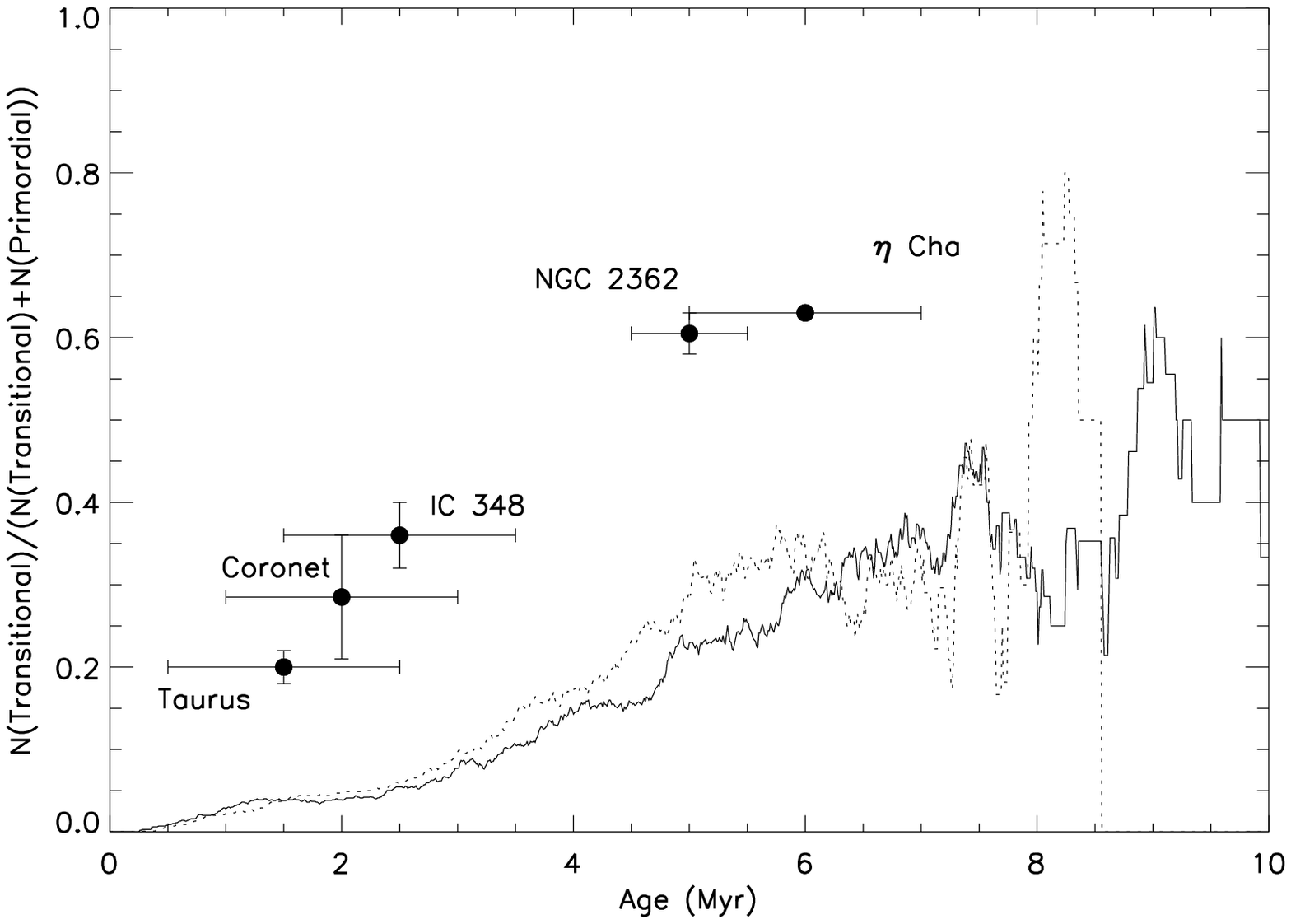}
\plotone{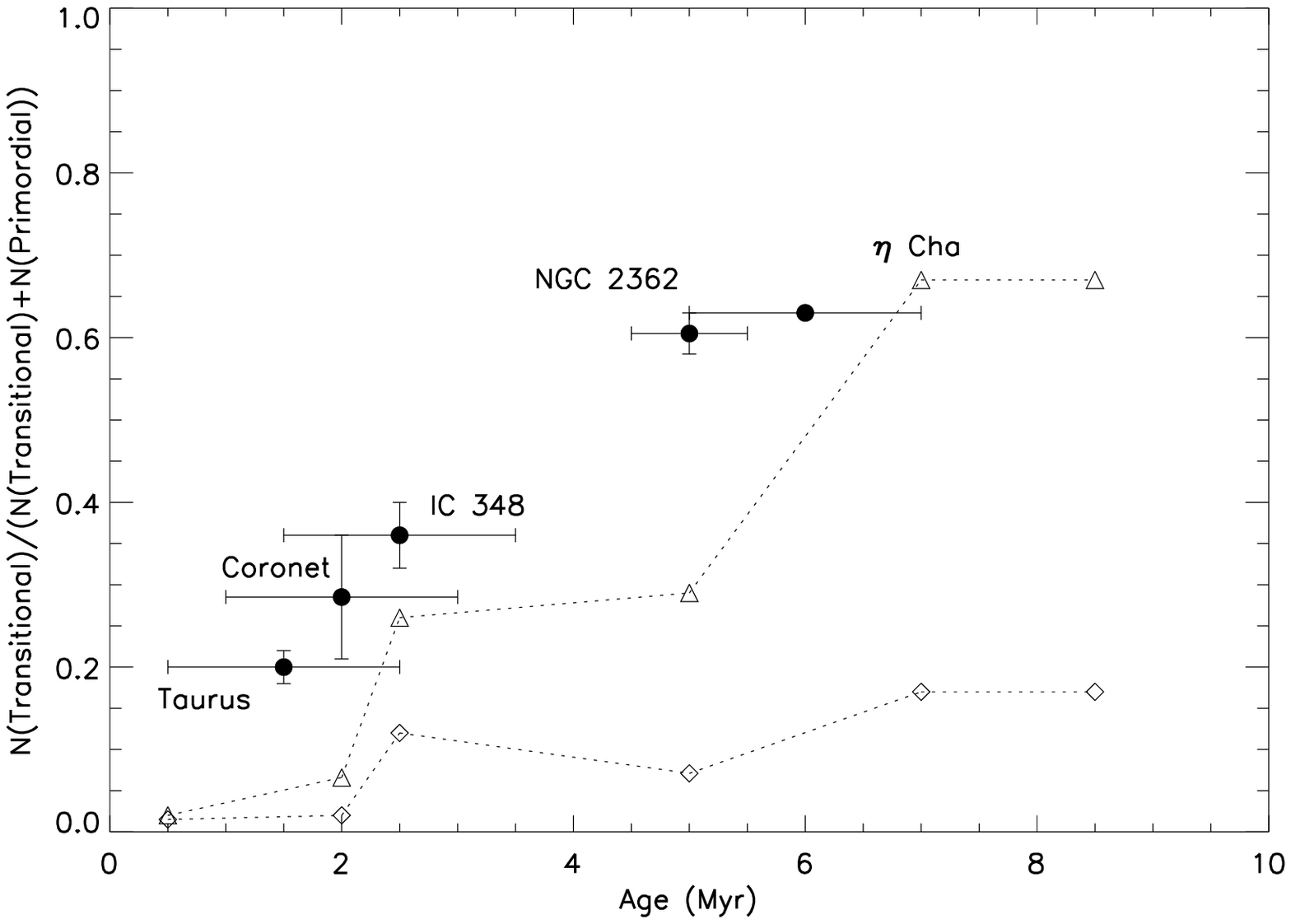}
\caption{Frequency of transitional disks vs. time.  (Top panel) We compare our 
frequencies to the theoretical models of \citet{AlexanderArmitage2009} who 
determine the frequency of transitional disks vs. time from a disk evolution/planet 
formation model with a disk clearing timescale of 0.5 Myr.  The dashed line corresponds 
to the reference model, while the solid line includes a dispersion in disk properties 
(initial angular momentum, disk lifetime, disk mass).  See \citet{AlexanderArmitage2009} 
for details.  (Bottom panel) Our frequencies compared to those derived by 
\citet{Muzerolle2010}.  Their loci correspond to frequencies assuming that 
all transitional disks have inner holes and optically thick outer regions (lower dashed line/diamonds) 
and allowing for a broader definition of transitional disk that would include 
many identified from this work (upper dashed line/triangles).  In this figure and the next one, 
the dots identify the average of the lower and upper limits for the transitional disk 
frequency as listed in Table \ref{freqtranall}.}
\label{freqtranvstime}
\end{figure}

\begin{figure}
\epsscale{0.75}
\centering
\plotone{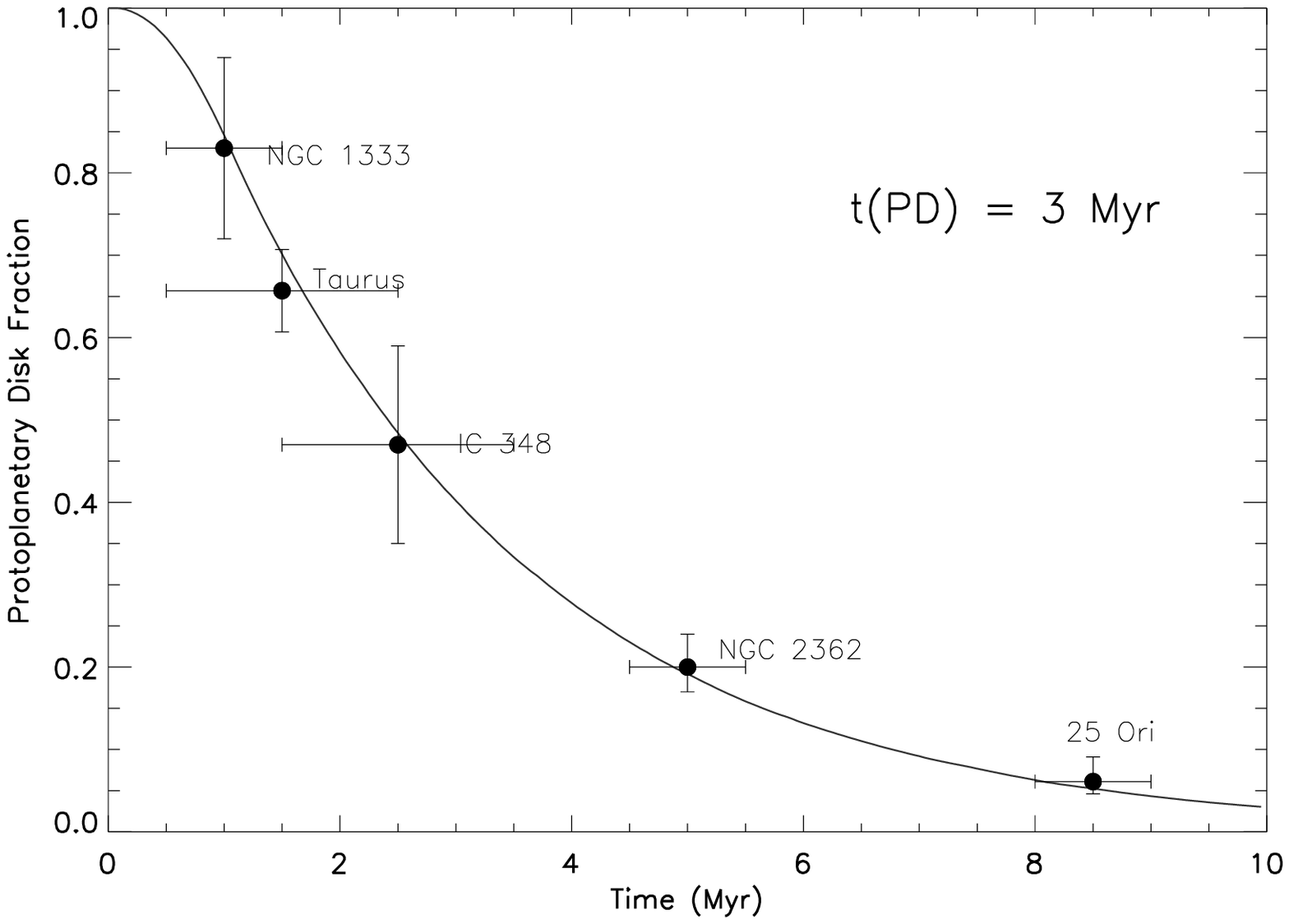}
\plotone{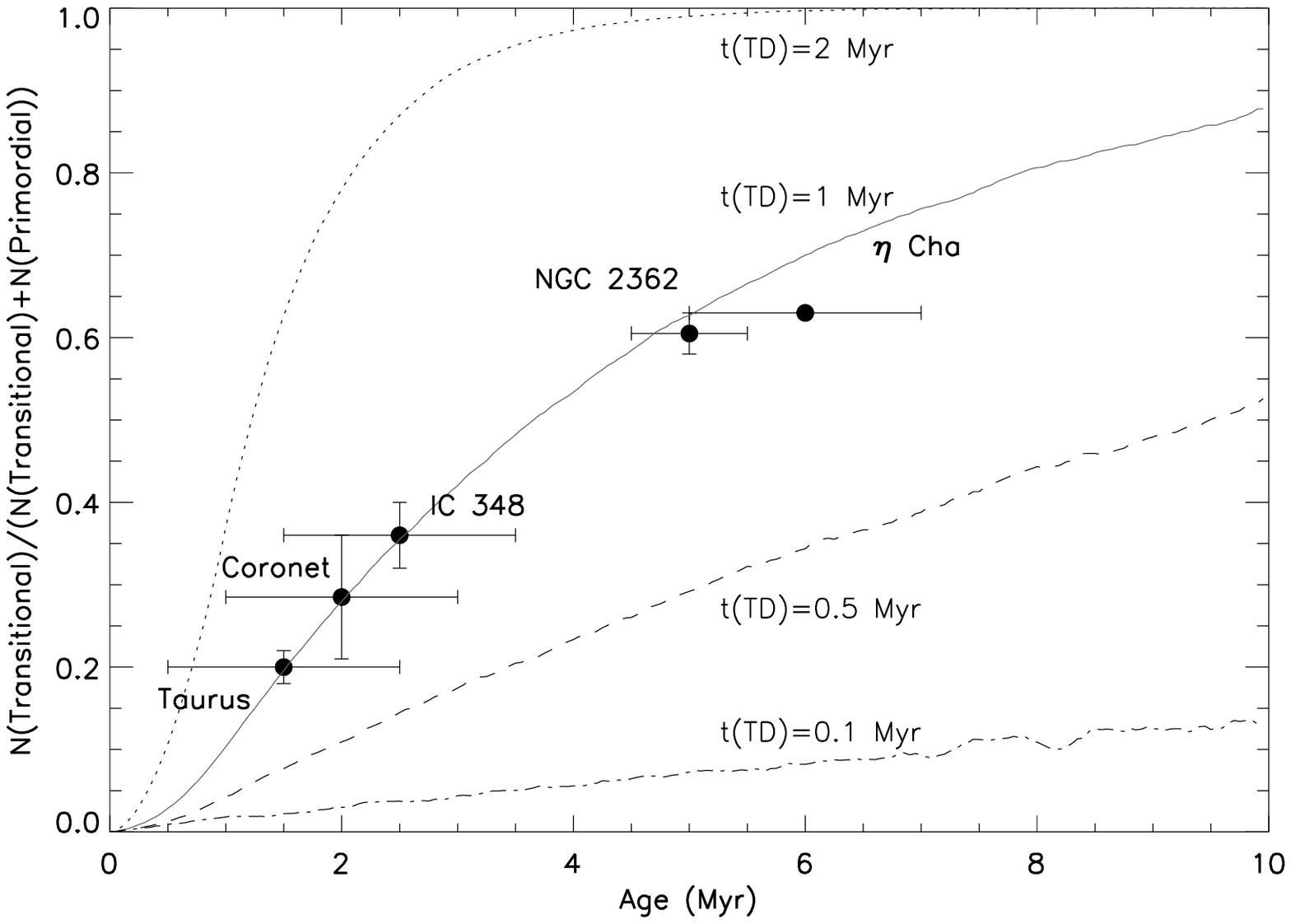}
\caption{(Top) The frequency of protoplanetary disks vs. time as probed by Spitzer observations.  For NGC 1333, IC 348, and 25 Ori the 
frequencies are taken from the literature \citep{Gutermuth2008,Lada2006,Hernandez2007b}.  The frequency for Taurus is taken from the 
number of "Class II" objects in Taurus reported by \citet{Luhman2009} and the number of "Class III" objects showing evidence for 
weak IR excess (e.g. FW Tau).  The frequency for NGC 2362 is taken from data presented in \citet{CurrieLada2009}, which is 
in agreement with estimates from \citet{DahmHillenbrand2007}.  
(Bottom) The transitional disk frequency vs. time for different clusters in our study.  Overplotted are the predicted 
transitional disk frequencies from our parameteric model, assuming a typical protoplanetary disk lifetime
of 3 Myr and varying amounts of that time spent as transitional disks (0.1--2 Myr).
}
\label{timetranall}
\end{figure}

\begin{figure}
\plotone{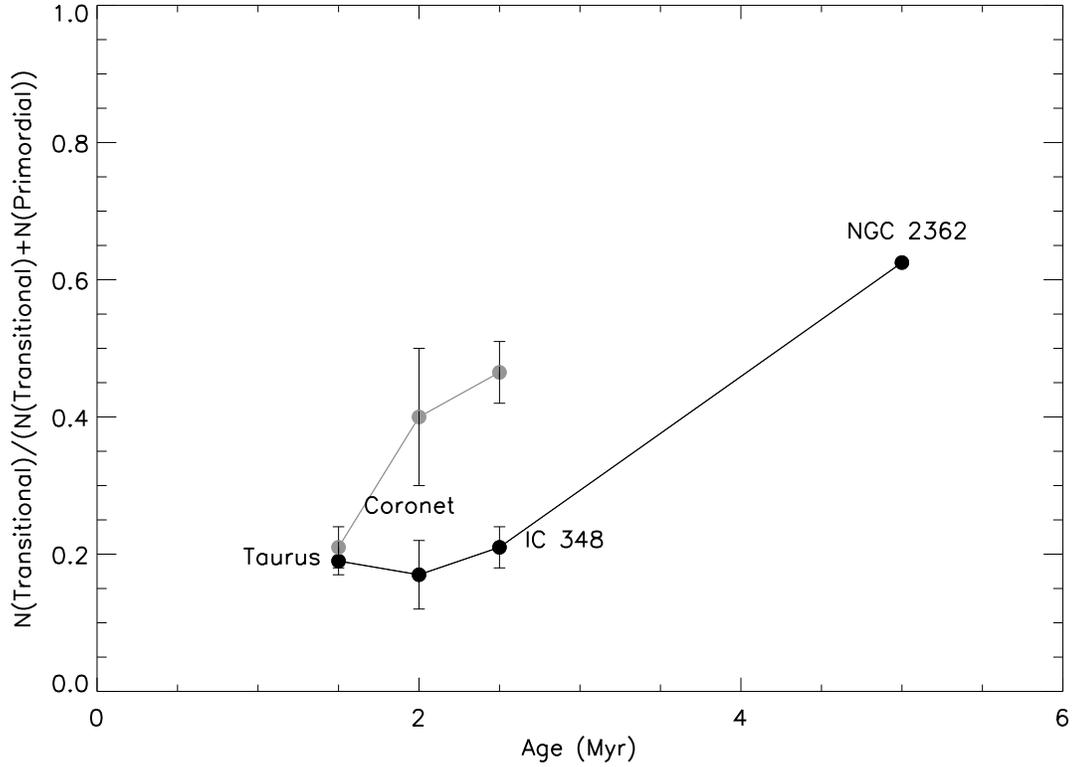}
\caption{ The transitional disk frequency vs. time for different clusters separated by spectral type: K5--M2 stars (lower locus) 
and later stars (upper locus).   The error bars simply define the range in values from Table \ref{freqtranall}, taking into account 
disks with uncertain states.}
\label{freqtranstype}
\end{figure}
%\begin{figure}
%\epsscale{1}
%\plottwo{cra-205.eps}{cra-466.eps}
%\caption{HiHi.}
%\label{twoexamples}
%\end{figure}
%\begin{figure}
%\plottwo{cra-205_innerhole.eps}{cra-466_innerhole.eps}
%\caption{HiHi.}
%\label{innerholeexamples}
%\end{figure}
%\begin{figure}
%\plottwo{cra-205_mass.eps}{cra-466_mass.eps}
%\caption{HiHi}
%\label{massexamples}
%\end{figure}

\begin{figure}
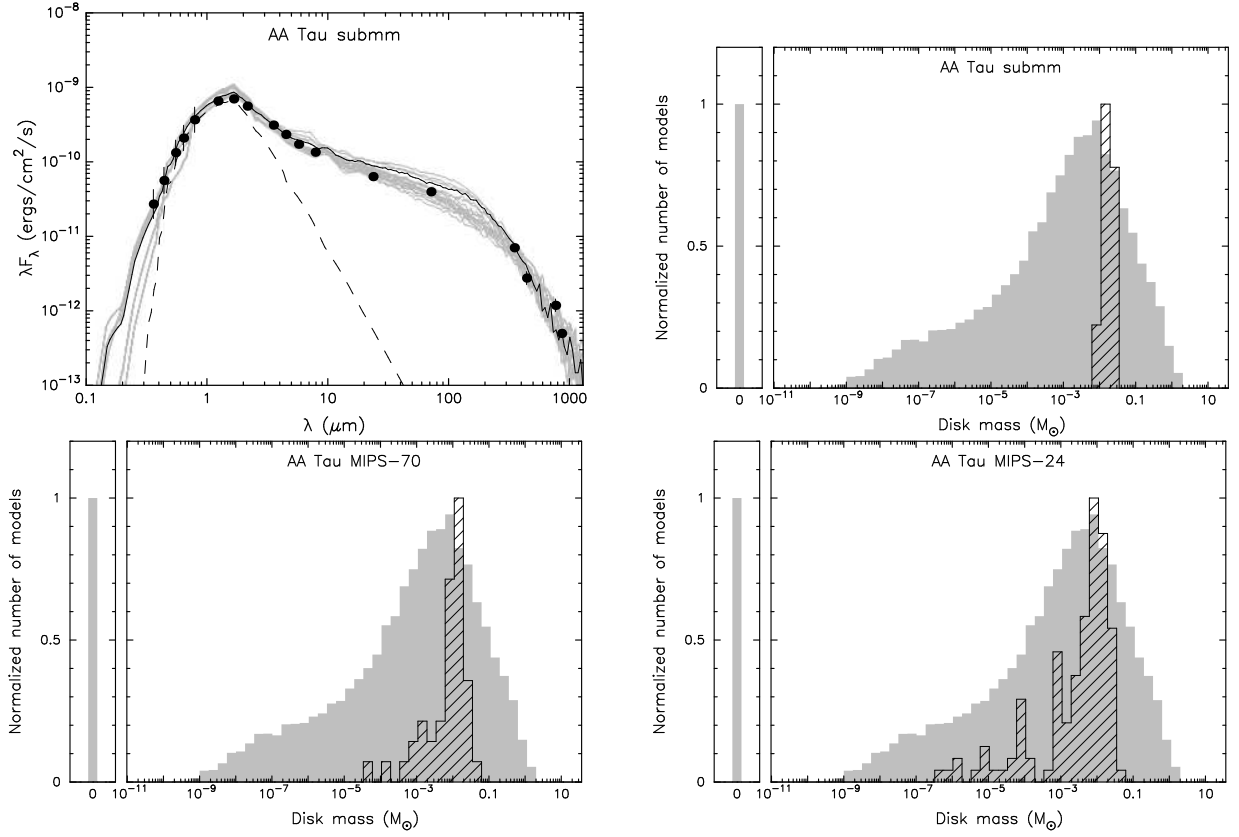

\centering
\plottwo{aatau_sed.eps}{aatau_submm.eps}
\plottwo{aatau_mips70.eps}{aatau_mips24.eps}
\caption{Analysis of AA Tau following \citet{Robitaille2007}.  The top-left panel 
shows the best-fit SED models to the AA Tau data from the optical through submillimeter; 
the top-right panel shows the histogram of disk masses from these best-fit 
models.  The bottom two panels show the histogram of disk masses if we restricted 
our data to 70 $\mu m$ or less (bottom-left panel) and 24 $\mu m$ or less (bottom-right panel).
}
\label{aatau}
\end{figure}

\begin{figure}
%\plotone{cdfaw05_tab2.ps}
\plotone{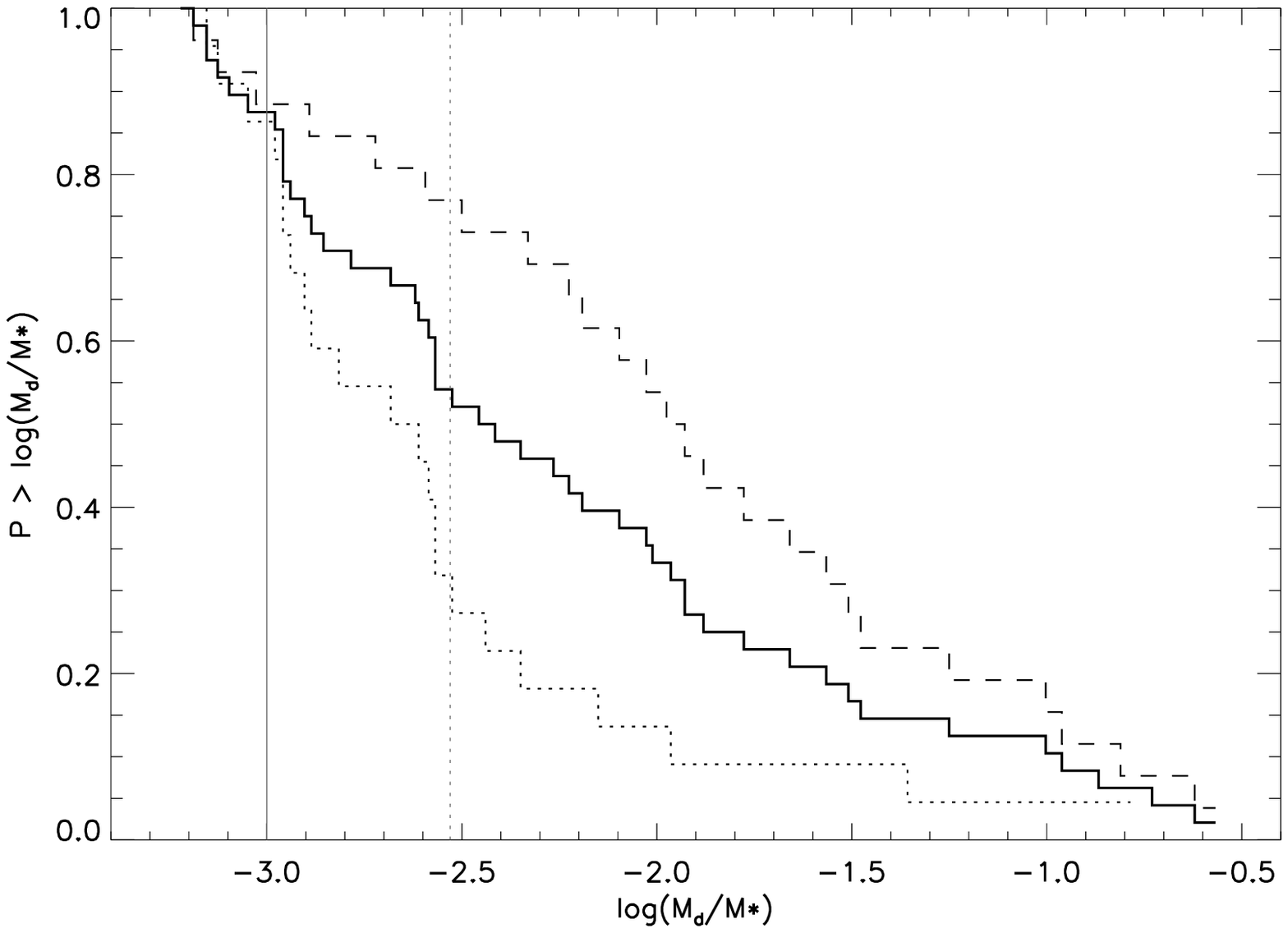}
\caption{Cumulative distribution function of primordial disk masses for 48 K5--M6 Taurus members listed in 
\citet{Andrews2005} based on \citeauthor{Andrews2005}'s disk mass estimates.  
Also shown are distribution functions for the 'Table 1-only' sample and 'Table 2' sample 
from \citet[][see Appendix for definitions]{Andrews2005} identified by dotted and dashed distribution 
functions, respectively.
For reference, our adopted limits for primordial disk masses -- 0.001 M$_{\star}$ and 0.003 M$_{\star}$ -- 
are shown as vertical solid and dotted grey lines, respectively.  
}
\label{cdfinitial}
\end{figure}

\begin{figure}
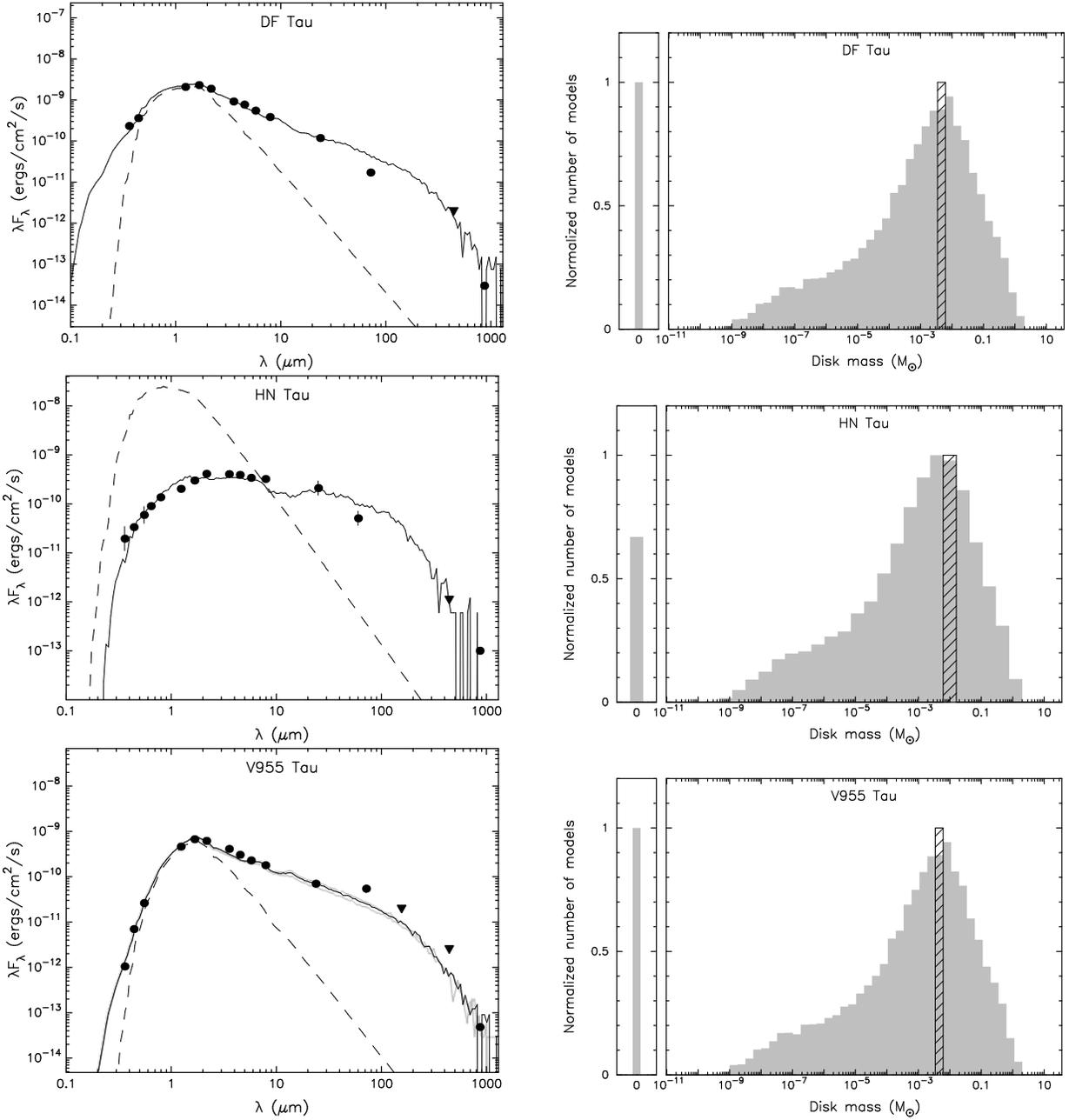

\centering
\plottwo{dftau_sed.eps}{dftau_mass.eps}
\plottwo{hntau_sed.eps}{hntau_mass.eps}
\plottwo{v955tau_sed.eps}{v955tau_mass.eps}
\caption{SED modeling results for Class II objects with 
optically-thick near-to-mid IR emission (primordial disks) 
in Taurus studied in \citet{Andrews2005} that have 
M$_{disk}$ $<$ 0.001 M$_{\star}$ based 
on the \citet{Andrews2005} SED modeling: DF Tau, HN Tau, and V955 Tau.  
The division corresponding to M$_{disk}$/M$_{\star}$ = 10$^(-3)$ for DF Tau, 
HN Tau, V955 Tau occurs at M$_{disk}$ = 1$\times$10$^{-3}$, 8.5 $\times$ 10$^{-4}$, 
and 5.5 $\times$ 10$^{-4}$, respectively.  Using 
the Robitaille models, we find that all of these sources have disk masses larger than 0.001 M$_{\star}$.}
\label{tauruspd1}
\end{figure}

\begin{figure}
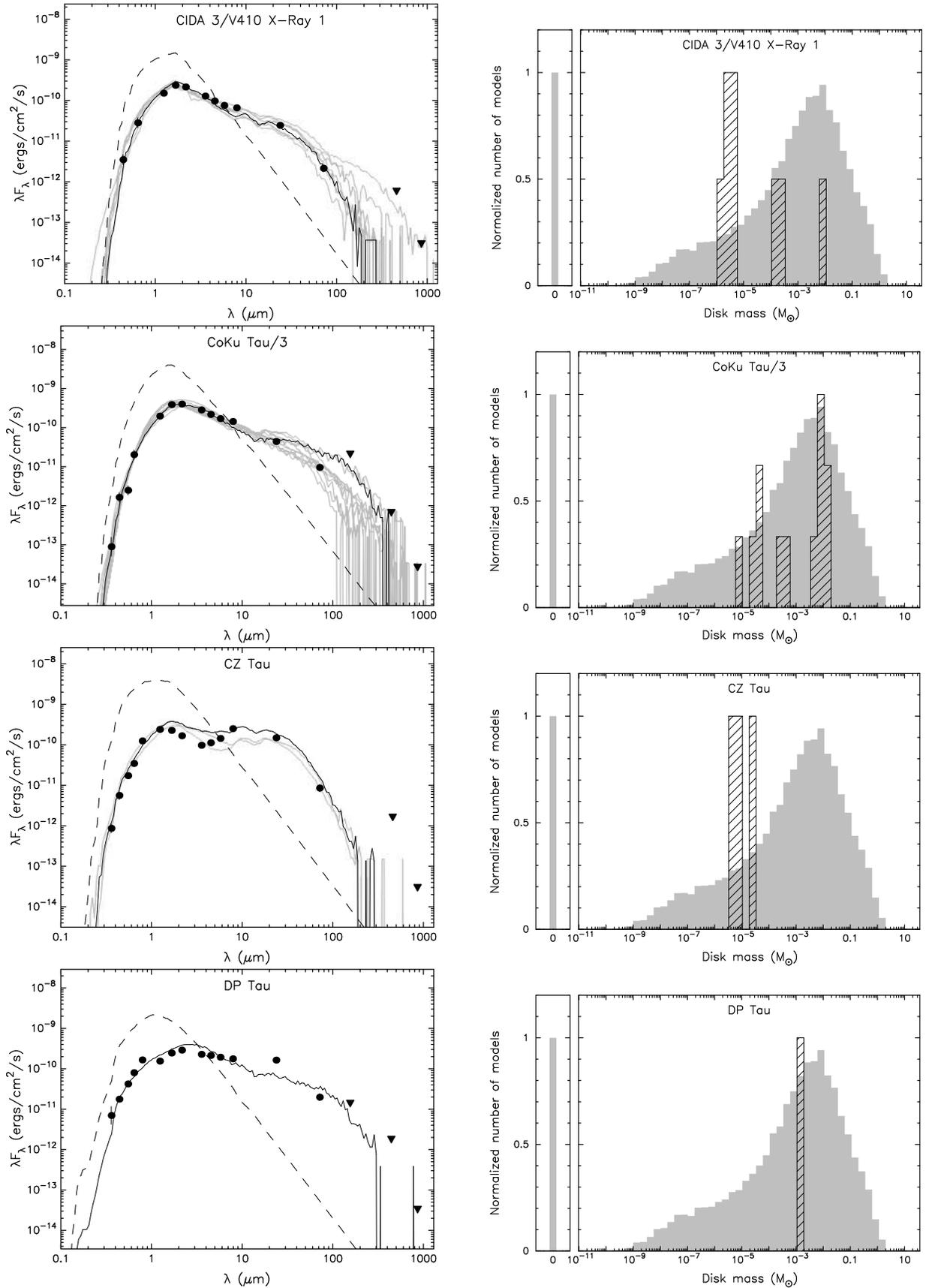

\centering
\plottwo{cida3_sed.eps}{cida3_mass.eps}
\plottwo{cokutau3_sed.eps}{cokutau3_mass.eps}
\plottwo{cztau_sed.eps}{cztau_mass.eps}
\plottwo{dptau_sed.eps}{dptau_mass.eps}
\caption{SED fits and distribution of masses from best-fitting 
models for other primordial disks
in Taurus with data from \citet{Andrews2005}.  
The division corresponding to M$_{disk}$/M$_{\star}$ = 10$^(-3)$ for CIDA 3/V410 X-Ray 1, 
CoKu Tau/3, CZ Tau, and DP Tau occurs at M$_{disk}$ = 6 $\times$10$^{-4}$, 6 $\times$ 10$^{-4}$, 
5.5 $\times$ 10$^{-4}$, and 4 $\times$ 10$^{-4}$, respectively.
Based on the \citet{Andrews2005} submillimeter estimates, all of these 
have disk masses less than 0.001 M$_{\star}$.  Our analysis finds that 
only two of the four have such low masses.}
\label{tauruspd2}
\end{figure}

\begin{figure}
%\plotone{cdfaw05_tab2.ps}
\plotone{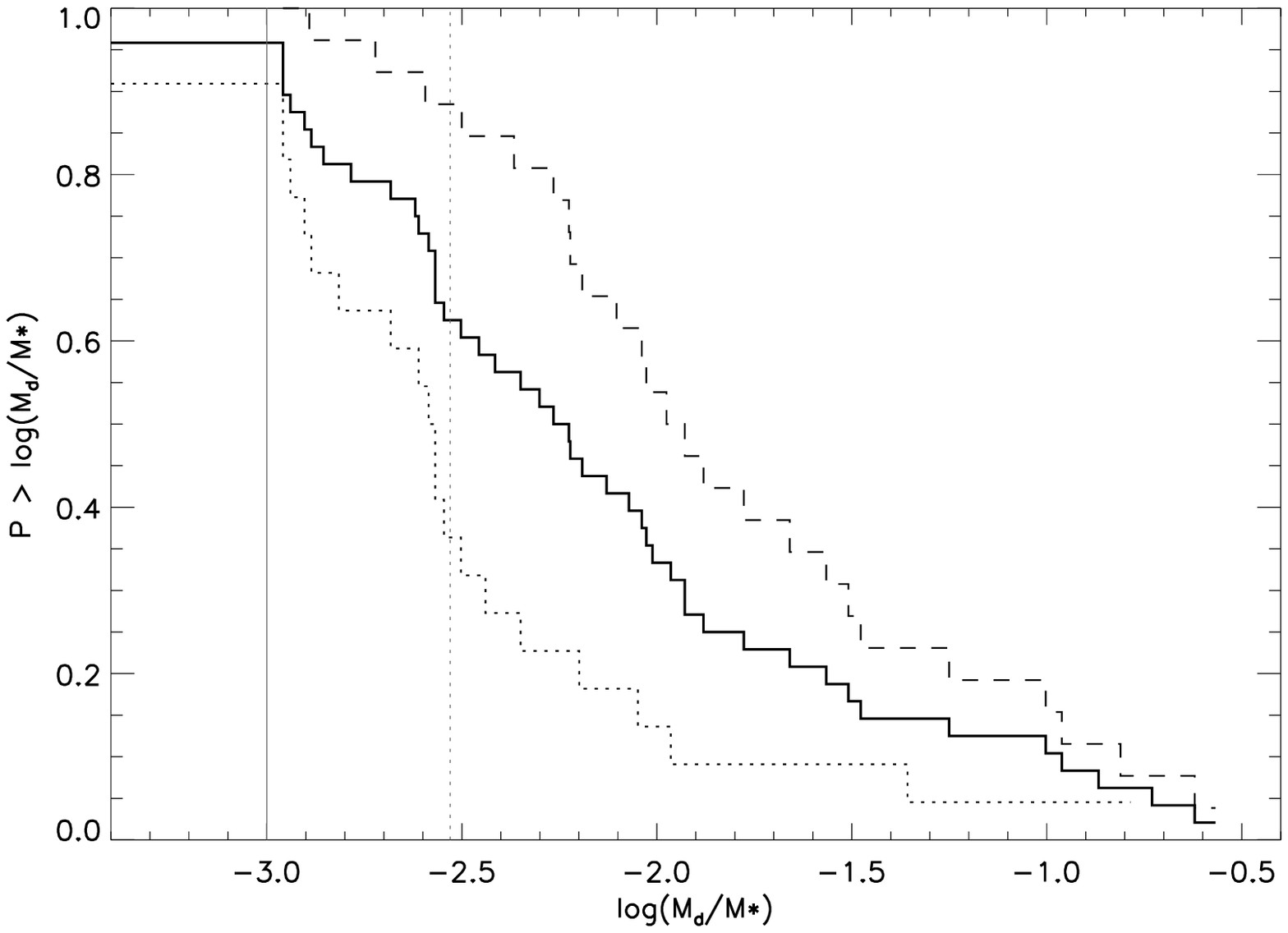}
\caption{Cumulative distribution function of primordial disk masses for 48 K5--M6 Taurus members listed in 
\citet{Andrews2005} taking into account revised mass estimates for Taurus members discussed in the Appendix.
Our results show that the disk mass range of 0.001--0.003 M$_{\star}$ defines a meaningful lower limit 
for the masses of optically-thick primordial disks. 
}
\label{cdffinal}
\end{figure}

\appendix
\section{Identifying Transitional Disks}

\subsection{Wavelength Range Required to Meaningfully Constrain Disk Masses}
To determine the wavelength range necessary to meaningfully constrain disk 
masses, we follow \citet{Robitaille2007} and display SED modeling results for 
AA Tau, a "typical" optically-thick primordial disk (Figure \ref{aatau}).  
For modeling, we assume an extinction range of A$_{V}$ = 1.45 to 2.45 -- the average between 
the \citet{Furlan2006} and \citet{Rebull2010} estimates $\pm$ 0.5 mags -- and 
use the optical, 2MASS, Spitzer-IRAC/MIPS, and submillimeter data as listed 
by \citet{Robitaille2007}.  We adopt the same photometric errors as well.  

The best-fit SED including all these data (top-left panel) 
 reproduces the \citeauthor{Robitaille2007} results. 
Our distribution of masses (top-right) is sharply peaked 
about M$_{disk}$ = 0.015 M$_{\odot}$ with a median 
value of $\sim$ 0.02 M$_{\odot}$, perfectly overlapping
 with their results .  Now 
we remove longer wavelength data to see if the predicted 
range of disk masses changes.  Restricting ourselves to 
data equal to or shortward of 70 $\mu m$, the dispersion in 
disk masses is larger, but clearly and sharply peaked about 
M$_{disk}$ $\sim$ 0.015 M$_{\odot}$ (bottom-left) with a 
median value of $\sim$ 0.01 M$_{\odot}$.  Modeling optical through 
MIPS-24 data yields a peak noticeably shifted towards slightly 
lower disk masses ($\sim$ 8$\times$10$^{-3}$ M$_{\odot}$).  More generally, 
the distribution of disk masses traces the shape of the distribution for 
all possible models (light grey shaded region).

While submillimeter data is required to yield the most precise 
disk mass estimates, this exercise indicates that far-IR wavelengths 
effectively probe the bulk disk mass.  Disk mass estimates derived 
from modeling optical through far-IR data (e.g. 70 $\mu m$) show good agreement 
with modeling that includes submillimeter data.  Therefore, we 
conclude that a submillimeter detection is preferred but not 
required to estimate disk masses with a precision useful 
for our study.  On the other hand, if far-IR/submm data is not 
available, disk mass estimates are very imprecise and likely not useful 
for our study.

\subsection{Masses of Primordial Disks and Transitional Disks}
SED modeling of optical-to-submm data for stars in 1--2 Myr-old 
clusters like Taurus provides useful benchmark values 
for disk properties that can be utilized to assess disk evolution.  
For instance, comparing near-to-mid IR emission from disks to
 the median and upper/lower-quartile Taurus SED is often used to 
identify sources showing evidence for warm dust depletion and/or settling 
in the inner disk \citep[cf.][]{Hartmann2005,Furlan2006,CurrieLada2009}.  Similarly, the 
median and upper/lower quartile of masses for optically-thick primordial disks compared to 
stellar masses as inferred from submillimeter measurements ("fractional 
disk masses") can be used to assess dust depletion in the outer disk, which 
traces the total mass of dust \citep{Beckwith1991,Wood2002,Andrews2005}.
Submillimeter data is particularly important in this paper since identifying 
homologously depleted transitional disks -- those with low disk masses 
compared to primordial disks-- presupposes 
that we can identify appropriate lower limits for primordial disk masses.  In this section, 
we revisit and expand upon the study of \citet{Andrews2005} to identify 
the range of fractional disk masses for primordial disks, showing that the 
appropriate lower limit is M$_{disk}$ = 0.001--0.003 M$_{\star}$.

The \citet{Andrews2005} study includes data for 153 Taurus members (their Table 1).  
\citet{Andrews2005} estimated disk masses for 44 of the members from SED modeling (their Table 2), 
since they have submillimeter detections and high-quality near-to-far IR 
photometry: we hereafter refer to this sample as the 'Table 2 sample'.
  Disk masses for the other 109 members were estimated solely from their submillimeter 
fluxes/upper limits: we hereafter refer to this sample as the 'Table 1-only sample'. 

To determine fractional disk masses for disks in both samples, we adopt 
spectral types from \citet{Luhman2009}, use our stellar mass estimates, and 
restrict the samples to K5--M6 stars.  Because our goal is identifying the masses of 
optically-thick primordial disks, we further trim the sample of 
Class I protostars, Class III diskless stars, known transitional disks with inner holes (e.g. GM Aur, UX Tau), 
and other disks with optically-thin near-to-mid IR emission listed in Table 
\ref{taurusstate} (e.g. V836 Tau, JH 223).  Finally, we remove foreground nonmembers (St 34) and other 
stars lacking MIPS data reported in \citet{Luhman2009} and \citet{Rebull2010} 
that is necessary for SED modeling (e.g. CIDA-11, CIDA-12, and FV Tau/c).  

Our selection criteria leaves 22 stars in the Table 1-only 
sample and 26 in the Table 2 sample.  The sample of stars that were not modeled 
by \citet{Andrews2005}, and thus listed in Table 2, is dominated by Class III diskless 
members.  This characteristic explains the precipitous drop in the number of 
Table 1-only stars analyzed (22) compared to the nominal sample (109).

Figure \ref{cdfinitial} shows the cumulative distribution functions for 
both samples, adopting \citet{Andrews2005} disk mass estimates 
and our stellar mass estimates.  For the Table 2, the 
median and mean fractional disk masses are 
 M$_{disk}$ $\sim$ 0.011 M$_{\star}$ and M$_{disk}$ $\sim$ 0.039 M$_{\star}$.  
The interquartile range covers 
0.003 M$_{\star}$ to 0.033 M$_{\star}$.  About 88\% (23/26) have masses 
 greater than or equal to 0.001 M$_{\star}$: the low-mass outliers are DF Tau, HN Tau, 
and V955 Tau.  For the Table 1-only sample, about 82\% (18/22) have masses 
greater than or equal to 0.001 M$_{\star}$: the low-mass outliers are 
CIDA-3, CoKu Tau/3, CZ Tau, and DP Tau.  To more definitively determine whether 
any member in either sample has a low-mass disk, we fit their 
SEDs using the \citet{Robitaille2007} model grid and data reported from 
Spitzer \citep[][]{Luhman2009,Rebull2010} and ground-based studies, following 
the methods described in \S 4 of the main text.

As listed in Table \ref{taurusextra} and shown by Figure \ref{tauruspd1}, inferred disk masses 
for DF Tau, HN Tau, and V955 Tau using the Robitaille models
are substantially higher than values listed by \citet{Andrews2005}.  The median masses of the 
best-fit models are 5--9 $\times$10$^{-3}$ M$_{\star}$, exceeding both the 
0.001 M$_{\star}$ limit and the previously quoted lower-quartile fractional disk mass. 
Thus, we conclude that all of the optically-thick primordial disks modeled by 
\citet{Andrews2005} have inferred masses greater than M$_{disk}$ = 0.001 M$_{\star}$.

We obtain similar results for the low-mass outliers in our second sample.
Figure \ref{tauruspd2} shows the SEDs and distribution of masses from best-fit Robitaille 
models for CIDA-3, CoKu Tau/3, CZ Tau, and DP Tau.  CoKu Tau/3 and DP Tau have inferred masses 
greater than 0.001 M$_{\star}$.  CZ Tau and CIDA-3 have much lower masses and represent the only optically-thick 
primordial disks from the \citet{Andrews2005} sample with inferred
 masses lower than 0.001 M$_{\star}$.

Figure \ref{cdffinal} shows the revised cumulative distribution functions for the fractional 
 masses of optically-thick primordial disks surrounding K5--M6 stars in Taurus.
  Except for the sources individually modeled in this section (listed in 
Table \ref{taurusextra}), we adopt disk mass values from \citet{Andrews2005} for simplicity.  
Of these members, $\sim$ 96\% (46/48) have masses greater than or equal to 0.001 M$_{\star}$. 
The median and interquartile range for the Table 2 sample is 0.011 M$_{\star}$ and 0.006--0.033 M$_{\star}$.
For the combined sample of 48 stars, these values are 5$\times$10$^{-3}$ M$_{\star}$ and 0.0024--0.015 M$_{\star}$; however, 
most of the masses are derived from submillimeter fluxes alone, so individual mass estimates accounting for the entire disk 
should be larger and the median and quartile disk masses should likewise be larger.

Since nearly all primordial disks in the \citet{Andrews2005} sample have masses 
greater than 0.001 M$_{\star}$ and the lower quartile disk 
mass for the modeled sample is $>$ 0.003 M$_{star}$, we conclude that disks with masses lying 
below 0.001--0.003 M$_{\star}$ lie below the range expected for 
primordial disks.  Thus, based on modeling current optical through submillimeter data, 
defining the lower mass limit for optically-thick primordial disks 
at 0.001--0.003 M$_{\star}$ is justified.

\end{document}

%% file: tab_list.tex
\begin{deluxetable}{lllllllllllllllllllll}
 \tiny
%\rotate
%\documentstyle[10pt]
%SPMquot"(0pt)
\setlength{\tabcolsep}{0.02in}
%\tablewidth{0pc}
%\linewidth{0.1 in}
%\tabletypesize{\scriptsize}
%\tabletypesize{\scriptsize}
\tabletypesize{\tiny}
\tablecolumns{20}
\tablecaption{Members Listed in Forbrich and Preibisch (2007) And Included in This Work}
\tiny
\tablehead{{Name}& {RA (2000)}&{DEC (2000)}}
\startdata
S-CrA 	& 19:01:08.60 	&  -36:57:21.3\\
F-3 	& 19:01:15.86 	&  -37:03:44.3\\
FP-6 	& 19:01:19.39 	&  -37:01:42.0\\
FP-8 	& 19:01:22.40 	&  -37:00:55.4\\
CrA-134 & 19:01:25.75 	&  -36:59:19.3\\
CrA-135 & 19:01:27.15 	&  -36:59:8.6\\
V709 	& 19:01:34.84 	&  -37:00:56.7\\
HD176386B & 19:01:39.15 &  -36:53:29.6\\
FP-18 	& 19:01:39.34 	&  -37:02:07.8\\
TY-CrAabcd & 19:01:40.81 &  -36:52:34.0\\
IRS2 	& 19:01:41.55 	&  -36:58:31.6\\
HBC677 	& 19:01:41.62 	&  -36:59:53.1\\
FP-23 	& 19:01:43.12 	&  -36:50:20.9\\
IRS5ab 	& 19:01:48.02 	&  -36:57:22.4\\
FP-25 	& 19:01:48.46 	&  -36:57:14.5\\
IRS6A 	& 19:01:50.45 	&  -36:56:38.1\\
V710 	& 19:01:50.66 	&  -36:58:09.9\\
IRS8 	& 19:01:51.11 	&  -36:54:12.5\\
IRS9 	& 19:01:52.63 	&  -36:57:00.2\\
RCrA 	& 19:01:53.67 	&  -36:57:08.3\\
IRS7w 	& 19:01:55.31 	&  -36:57:22.0\\
FP-33 	& 19:01:55.61 	&  -36:56:51.1\\
FP-34 	& 19:01:55.76 	&  -36:57:27.7\\
FP-35 	& 19:01:55.85 	&  -36:52:04.3\\
IRS-7e 	& 19:01:56.39 	&  -36:57:28.4\\
FP-37 	& 19:01:57.46 	&  -37:03:11.9\\
FP-38 	& 19:01:58.32 	&  -37:00:27.5\\
T-CrA 	& 19:01:58.79 	&  -36:57:50.1\\
V702 	& 19:02:01.92 	&  -37:07:43.0\\
B1858 	& 19:02:01.94 	&  -36:54:00.1\\
HBC-679 & 19:02:22.13 	&  -36:55:41.0\\
CrA-159 & 19:02:33.07 	&  -36:58:21.1
\enddata
\tablecomments{Sources with an ``FP" prefix followed by a number identify Coronet Cluster members 
uniquely identified by \citet{Forbrich2007}.
The number following ``FP" identifies the row number of the source in Table 2 of that paper.}
\label{newsources}

\end{deluxetable}

%% file: tab_new.tex
\begin{deluxetable}{lllllllllllllllllllll}
 \tiny
%\rotate
%\documentstyle[10pt]
%SPMquot"(0pt)
\setlength{\tabcolsep}{0.02in}
%\tablewidth{0pc}
%\linewidth{0.1 in}
%\tabletypesize{\scriptsize}
%\tabletypesize{\scriptsize}
\tabletypesize{\tiny}
\tablecolumns{20}
%\tablecaption{New Spitzer Data for Coronet Cluster Members}
\tablecaption{Photometry for Members Listed in Forbrich and Preibisch (2007)}
\tiny
\tablehead{{Name}&{J}&{$\sigma$(J)}&{H}&{$\sigma$(H)}&{K$_{s}$}&{$\sigma$(K$_{s}$)}&{[3.6]}&
{$\sigma$([3.6])}&{[4.5]}&{$\sigma$([4.5])}&{[5.8]}&{$\sigma$([5.8])}&[8]&{$\sigma$([8])}&
{[24]}&{$\sigma$([24])}
}
\startdata
S-CrA&8.19& 0.02 & 7.05 & 0.024 & 6.10 & 0.02 & 5.060 & 0.005 & 4.386 & 0.005 & 3.699 & 0.005 & 2.969 & 0.004 & - & - \\
FP-3&-& -& -& -& -& -& 13.971 & 0.014 & 13.371 & 0.012 & 12.092 & 0.020 & 11.616 & 0.029 & -& -\\
FP-6&-& -& -& -& -& -& -& -& -& -& 12.299 & 0.017 & 11.458 & 0.033 & 6.106 & 0.005 \\
FP-8&-& -& -& -& -& -& -& -& -& -& -& -& 12.697 & 0.092 & -& -\\
CrA-134&11.61& 0.02 & 9.78 & 0.03 & 8.77 & 0.023 & 7.876 & 0.002 & 7.367 & 0.002 & 6.862 & 0.003 & 6.537 & 0.002 & -& -\\
 CrA-135&10.81& 0.02 & 10.11 & 0.02 & 9.89 & 0.02 & 9.599 & 0.005 & 9.601 & 0.004 & 9.551 & 0.007 & 9.611 & 0.007 & -& -\\
 V709&8.67&0.03 & 7.97 & 0.04 & 7.71 & 0.02 & 7.863 & 0.002 & 7.717 & 0.003 & 7.533 & 0.004 & 7.512 & 0.004 & 6.794 & 0.010 \\
$^{1}$HD&7.60& -& 7.49 & 0.09 & 7.14 & 0.08 & 7.837 & 0.003 & 7.781 & 0.003 & 6.243 & 0.003 & 4.706 & 0.003 & -& -\\
176386 &\\
FP-18&11.34&0.03 & 10.66 & 0.03 & 10.37 & 0.03 & 10.110 & 0.005 & 10.097 & 0.005 & 9.979 & 0.007 & 9.927 & 0.010 & 9.364 & 0.104 \\
$^{1}$TY CrA&7.49&0.02 & 6.97 & 0.03 & 6.67 & 0.02 & 6.217 & 0.005 & 6.085 & 0.005 & 4.478 & 0.004 & -& -& -& -\\
abcd&\\
IRS-2&13.75&0.02 & 9.74 & 0.02 & 7.07 & 0.02 & 5.181 & 0.004 & 4.281 & 0.003 & -& -& 2.230 & 0.003 & -& -\\
 HBC 677&10.47& 0.02 & 9.04 & 0.02& 8.09 & 0.03 & 7.725& 0.003 & 7.201 & 0.003 & 6.528 & 0.004 & 5.859 & 0.004 & -& -\\
FP-23&16.92&0.18 & 15.23 & 0.11 & 14.77 & 0.10 & -& -& -& -& -& -& -& -& -& -\\
IRS-5ab&16.9&--& 13.73 & 0.01 & 10.59 & 0.01 & 7.456 & 0.002 & 6.293 & 0.001 & -& -& 3.937 & 0.003 & -& -\\
FP-25&15.56&-& 13.14 & -& 12.49 & 0.15 & -& -& 9.115 & 0.003 & -& -& 7.017 & 0.003 & -& -\\
IRS-6A&16.39&0.20 & 12.14 & -& 10.31 & -& 8.480 & 0.003 & 7.687 & 0.004 & -& -& 6.286 & 0.003 & -& -\\
V710&15.96&0.10 & 11.13 & 0.03 & 7.62 & 0.04 & -& -& 3.659 & 0.003& -& -& -& -& -& -\\
IRS-8&14.92&0.04 & 12.69 & 0.03 & 11.37 & 0.02 & 9.911 & 0.005 & 9.358 & 0.005 & 8.744 & 0.006 & 7.946 & 0.005 & -& -\\
IRS-9&-&-& -& -& -& -& 8.031& 0.003 & 7.313 & 0.003 & -& -& -& -& -& -\\
 R CrA&6.952&0.021 & 4.955 & 0.024 & 2.910 & 0.318 & -& -& -& -& -& -& -& -& -& -\\
IRS-7w&12.55&-& 13.25 & 0.32 & 10.76 & -& 8.635 & 0.004 & 7.073 & 0.003 & -& -& -& -& -& -\\
FP-33&-&-& -& -& -& -& 10.181 & 0.005 & 9.025 & 0.004 & -& -& 6.860 & 0.004 & -& -\\
FP-34&-&-& -& -& -& -& 9.730 & 0.004& 8.293 & 0.004 & -& -& -& -& -& -\\
FP-35&-&-& -& -& -& -& -& -& 13.387 & 0.016 & 11.660 & 0.018 & 10.921 & 0.020 & 7.532 & 0.021 \\
IRS-7e&-&-& -& -& -& -& 9.417 & 0.004 & -& -& -& -& -& -& -& -\\
FP-37&15.58&0.07 & 14.02 & 0.04 & 13.19 & 0.04 & 12.320 & 0.007 & 12.264 & 0.006 & 11.925 & 0.089 & 12.086 & 0.037 & -& -\\
FP-38&16.961&-& 15.275 & 0.139 & 13.704 & 0.060 & 12.426& 0.007 & 12.102 & 0.008 & 11.508 & 0.017 & 12.008 & 0.050 & -& -\\
T-CrA&8.93&0.03 & 7.70 & 0.04 & 6.60 & 0.02 & -& -& 4.819 & 0.005 & 4.217 & 0.005 & -& -& -& -\\
 V702&8.90&0.02 & 8.48 & 0.04 & 8.35 & 0.03 & 8.444 & 0.004 & 8.329 & 0.003 & 8.282 & 0.0048 & 8.280 & 0.003 & 8.122 & 0.034 \\
B185839.6&-& -& -& -& -& -&  14.099 & 0.019 & 14.766 & 0.054 & -&-& -& -& 8.617 & 0.048 \\
-3658&\\
 HBC-679&10.33&0.03 & 9.50 & 0.04 & 9.23 & 0.03 & 9.031 & 0.003 & 8.967 & 0.004 & 8.901 & 0.005 & 8.868 & 0.006 & 8.258 & 0.039 \\
% HBC-680&9.324&0.028 & 8.296 & 0.045& 7.952 & 0.019 & 7.707 & 0.003 & 7.256 & 0.002 & 6.674 & 0.004 & 6.107 & 0.003 & 4.195 & 0.001 \\
 CrA-159&10.59& 0.03 & 9.26 & 0.02& 8.45 & 0.020 & 7.565 & 0.002 & 7.016 & 0.003 & 6.372& 0.003 & 5.887 & 0.004 & 3.704 & 0.001 \\
\enddata
\tablecomments{Notes -- 1) Inspection of the IRAC mosaic shows that TY CrAabcd and HD 176386B are contaminated by nebular emission at 5.8 $\mu m$ 
and 8 $\mu m$. 2).} 
\label{newphotometry}

\end{deluxetable}

%% file: tab_diskstate.tex
\begin{deluxetable}{llllllllllllll}
 \tiny
%\rotate
%\documentstyle[10pt]
%SPMquot"(0pt)
%\setlength{\tabcolsep}{0.003in}
%\linewidth{0.1 in}
%\tabletypesize{\tiny}
%\tabletypesize{\scriptsize}
\tabletypesize{\small}
\tablecolumns{11}
\tablecaption{Provisional Evolutionary States for Coronet Cluster Disks with New Spitzer Photometry Based on 1--24 $\mu m$ Data}
\tiny
%\tablehead{{ID}&{ST}&{ST Ref.}&{EW(H$_{\alpha}$)}&{log(M$_{disk}$/M$_{\odot}$)}&{R$_{in, min}$ (AU)}&{$\beta$}&{Disk Type}&{Notes}}
%\tablehead{{ID}&{ST}&{Si08 iden.}&{Er09 iden.}&{EW(H$_{\alpha}$)}&{M$_{dust}$(M$_{\odot}$)}
%&{R$_{in, rsub}$}&{R$_{in, AU}$}&{Optically-thin?}&{Disk State}}
\tablehead{{ID}&{ST}&{ST Ref}&{ A$_{V}$(best) }&{$\tau_{mid-IR}$}&{Inner Hole? (R$_{sub}$)}&{Disk Class}\\{}&{}&{}&{}&{}&{}}
\startdata
S-CrA & G5/K5$^{1}$& 1 &2& thick& n & PD\\
CrA-159 & M2$^{2}$ & est.&3&thick&n & PD\\
IRS-2 & K2 &2&-&thick& n & PS\\
CrA-134 & K3 &2 & 11&thick& n & PD\\
CrA-135 & M4 & 2&0.5&-- &--& Star\\
V709 & K1 & 7&0.5&thin/--& -- & Star/TD(HD)/DD?\\
HD 176396B & K7 &2& 0.75&? & n & ?\\
FP-18 & M3 &est.& 0.4& --& -- & Star\\
TY-CrA & B8/K2 & 3,4& 3.5&? & n & ?\\
HBC-677 & M2 &2& 3.9&thick& n & PD \\
IRS-5ab & K5 & 2,5& --&thick& n & PS\\
FP-25 & M? & 7& --&thick& n & PS\\
IRS-6 & M1 & 2&10.5&thick& n & PD\\
V710 & K5--M0 &5& -&thick& n & PS\\
IRS-7 & ? & -- & - & thick & n &PS\\
IRS-8 & M2 &est.&13.5&thick& n & PD\\
R-CrA & A5 & 7&?&thick&n & PD\\
T-CrA & F0 & 6&0.3&thick&n & PD\\
V702 & G5 & 7&0.25&--&-- & Star\\
HBC-679 & K2 &7& 1.5&--& -- &Star\\
\enddata
\tablecomments{The disk states are identified as follows: PS=Protostar, PD = Primordial Disk, and TD = Transitional Disk, 
Star = Stellar Photosphere (no circumstellar material).  The transitional disks are further divided into those 
with inner holes (IH) and those that are homologously depleted (HD).  
The disk states here are considered to be provisional since these Coronet Cluster stars lack sensitive 
far-IR/submm data.
The spectral type references are the following: 1) \citet{Carmona2007}, 2) \citet{MeyerWilking2009}, 
3) \citet{Marraco1981}, 4) \citet{Casey1998}, 5) \citet{Nisini2005}, 6) \citet{Patten1998}, 
and 7) \citet{Forbrich2007}.  
"Est" means that the spectral type was estimated by
modeling the optical to near-IR SED using the \citet{Robitaille2006} grid and the \citet{Currie2010b} 
effective temperature scale.
Note (1) -- The fit to the photosphere for this star is generally poor; the best-fit 
A$_{V}$ corresponds to a K0 stellar photosphere model.  However, the exact value has no 
bearing on our classification.
Note (2) -- \citet{Patten1998} list the spectral type for this star as M3--M5.  However, 
we cannot reproduce the observed optical to near-IR SED with a M3--M5 photosphere using 
either the \citet{Currie2010} Te scale and dwarf colors or with the \citet{Robitaille2006} grid: 
all good-fitting models ($\chi^{2}-\chi^{2}_{best}$ $<$ 3) predict that 
its photosphere is hotter than $\sim$ 3800 K and thus its spectral type 
is earlier than M0--M2.  Since other stars classified 
as M3--M5 by \citet{Patten1998} are reclassified as M0--M2 by \citet{MeyerWilking2009} based on 
higher signal-to-noise data, we list this star as an M2 star.
}
%\label{dwarfcolors}
\label{diskstatenew}
\end{deluxetable}

%% file: tab_diskstateold.tex
\begin{deluxetable}{lllllllllllllll}
 \tiny
%\rotate
%\documentstyle[10pt]
%SPMquot"(0pt)
\setlength{\tabcolsep}{0.01in}
%\linewidth{0.1 in}
\tabletypesize{\tiny}
%\tabletypesize{\scriptsize}
\tablecolumns{12}
\tablecaption{Provisional Evolutionary States for Coronet Cluster Disks Studied in \citet{SiciliaAguilar2008} Based on 1--70 $\mu m$ Data}
\tiny
%\tablehead{{ID}&{ST}&{Disk Mass$^{1}$ (M$_{\odot}$)} &{R$_{in}$(R$_{sub})}&{Disk State(Si08)}&{Disk State(Er09)}&{Disk State}}
\tablehead{{ID}&{ST}&{ST Ref.}&{A$_{V}$(best)}&{$\tau_{mid-IR}$}&{Inner Hole?(R$_{sub}$)} & {Disk Class} &{Disk Class} & {Disk Class}\\
{}&{}&{}&{}&{}&{}&{(Si08)}&{(Er09)}&{(This Work)}}
\startdata
CrA-205 & M4--M6& 8,9&0.25&thin&y,2000&TD(IH)&PD&TD(IH)\\
CrA-432 & M5--M7& 8,9&0.5&thick&n&PD&--&PD\\
CrA-453 & M4.5 & 10&1.5&--&--&Star&--&Star\\
CrA-465 & M5--M7.5 & 8,10&0&thick&n&PD&--&PD\\
CrA-466 & M2 & 10&6.6&thick&n&TD(IH)&PD&PD\\
CrA-4107& $\ge$ M4.5& 8,9&1&thick&n&PD&--&PD\\
CrA-4109& $>$ M5&9, est&1&thin&y,300&TD(IH)&TD(IH)&TD(IH)\\
CrA-4110& M5 & 9&0.4&thick&n&PD&--&PD\\
CrA-4111& M4.5&9&0&thin&y,70&TD(IH)&TD(IH)&TD(IH)\\
G-1& M0 &9,est.&3&thin&2&PD&--&TD(HD)\\
G-14&M4.5&10&1.9&thin&n&TD(IH)&PD&TD(HD)\\
G-30&M3.5&10&0.09&thin&n&TD(IH)/DD&--&TD(HD)\\
G-32&$>$ M5&10,est.&8--15&thick&n&PD&--&PD\\
G-49&M4 &10&0.07&--&--&Star&--&Star\\
G-65&M1--M2&10&14--15&thick&n&TD(IH)&PD&PD\\
G-85&M0.5&9, est.&17&thick&n&PD&--&PD\\
G-87&M1.5&10&14&thin&1&TD(IH)&PD&TD(HD)\\
G-94&M3.5 &10&0.6&thin&n&TD(IH)/DD&--&TD(HD)/DD\\
G-95&M1 &10&5&thin&y,20&TD(IH)/DD&--&TD(IH)/DD\\
G-102&M5 &10&0.7&thin&n&TD(IH)/DD&--&TD(HD)/DD\\
\enddata
\tablecomments{
References for spectral types are the following: 8) \citet{LopezMarti2005}, 9) Sicilia-Aguilar et al. 2010, in prep., 
10) \citet{SiciliaAguilar2008}.  "Est" means that the spectral type was estimated by 
modeling the optical to near-IR SED using the \citet{Robitaille2006} grid and the \citet{Currie2010b} 
effective temperature scale.}
%\label{dwarfcolors}
\label{diskstateold}
\end{deluxetable}

%% file: tab_irs.tex
\begin{deluxetable}{llllllllllllll}
 \tiny
%\rotate
%\documentstyle[10pt]
%SPMquot"(0pt)
%\setlength{\tabcolsep}{0pt}
\tabletypesize{\tiny}
\tablecolumns{13}
\tablecaption{IRS Flux Density Measurements Used in Modeling Coronet Cluster Stars from \citet{SiciliaAguilar2008}}
\tiny
\tablehead{{ID}&{Wavelengths Sampled ($\mu m$)}&{Flux Density (mJy)}}
%}
\startdata
CrA-205 & 9,13,17,21,27 & 1.03 $\pm$ 0.06, 0.70 $\pm$ 0.08, 0.91 $\pm$ 0.30, 1.10 $\pm$ 0.30, 2.08 $\pm$ 0.26\\
G-14 & " & 4.95 $\pm$ 0.20, 4.70 $\pm$ 0.15, 5.53 $\pm$ 0.45, 4.75 $\pm$ 0.34, 3.62 $\pm$ 0.49\\
G-65 & 6,7,9, 10, 11, 12 & 53.06 $\pm$ 12.80, 48.30 $\pm$ 30.30, 49.96 $\pm$ 13.27, 47.9 $\pm$ 12.30, 48.10 $\pm$ 23.98, 43.1 $\pm$ 32.70\\
G-87 & 9, 13, 17, 21, 27 & 10.02 $\pm$ 0.31, 12.56 $\pm$ 0.29, 15.55 $\pm$ 4.84, 19.04 $\pm$ 4.88, 25.37 $\pm$ 6.19\\
\enddata
\tablecomments{The wavelength sampling for G-65 is different than the others because G-65 lacks data longwards of 
$\sim$ 15 $\mu m$.}
\label{irsdenspoint}
\end{deluxetable}

%% file: tab_taurus.tex
\begin{deluxetable}{llllllllllllll}
 \tiny
%\rotate
%\documentstyle[10pt]
%SPMquot"(0pt)
%\setlength{\tabcolsep}{0pt}
\tabletypesize{\tiny}
\tablecolumns{13}
\tablecaption{Properties of Selected Taurus Sources}
\tiny
\tablehead{{Name}&{ST}&{M$_{\star}$}&{A$_{V}$}&{A$_{V}$ range}&{K$_{s}$-[8]}&
{K$_{s}$-[24]}&{$\tau_{mid-IR}$}&{Inner} \\
{}&{}&{(M$_{\odot}$)}&{}&{}&{}&{}&{}&{Hole?}}
%}
\startdata
UX Tau & K5&1.0& 0.75&0.25-1.25&1.56 & 5.42 & thin&y\\
V807 Tau & K5&1.0&1.4&0.9-1.9&1.46&3.91&thin&n\\
LkCa 15 & K5&1.0&1.2&0.7-1.7&1.58&5.06&thin&y \\
FY Tau & K5&1.0&5&4--6&1.68&3.93&thin&n\\
JH 112 & K6&0.95&3.4&2.72--4.08&1.97&5.34&thick&n\\
V836 Tau & K7&0.85&1.05&0.55--1.55&1.76&4.77&thin&n\\
DN Tau & M0&0.75&0.7&0.2--1.2&1.99&5.00&thin&n\\
VY Tau & M0&0.75&1.1&0.6--1.6&1.53&4.35&thin&n\\
GO Tau & M0&0.75&2&1.5--2.5&1.85&4.93&thin&n\\
IP Tau & M0&0.75&0.6&0.1--1.1&1.68&4.74&thick&n\\
HO Tau & M0.5&0.65&1.3&0.8--1.8&1.90& 4.79& thick&n\\
DH Tau & M1&0.6&1.9&1.4--2.4&1.28&4.79&thin&?\\
FX Tau & M1.5&0.57&2&1.5--2.5&1.90&4.85&thick&n\\
CY Tau & M1.5&0.57&1.7&1.2--2.2&1.89&4.11&thick&n\\
04333905 & M1.75&0.57&7.9&2.4--3.6&1.49&5.81&thin&y\\
+2227207\\
JH-223 & M2&0.55&2.2&1.6--2.6&1.72&4.27&thin&n\\
GH Tau & M2&0.55&1&0.5--1.5&1.79&4.51&thick&n\\
\\
ZZ Tau & M3&0.4&1.2&0.7--1.7&1.39&3.91&thin&n\\
XEST13-10 &M3&0.4&6&4.8--7.2&1.52&4.40&thin&n\\
CIDA 8 & M3.5&0.3&2.25&1.75--2.75&1.49&4.30&thin&n\\
04231822&M3.5&0.3&8&6.4--9.6&1.53&5.11&thin&n\\
+2641156\\
%ITG-40 & M3.5&0.3&20.5&16.4--24.6&1.35&4.64&thin&n\\
04153916&M3.75&0.275&2&1.5--2.5&1.62&4.93&thin&n\\
+2818586\\
FP Tau & M4&0.25&0.5&0--1&1.52&4.53&thin&n\\
%MH0 6 & M4.75&0.2&1.56&4.27&thin&n&10$^{-4}$ & 5$\times$10$^{-4}$&--&--&--&PD & PD/TD(HD) ?\\
ITG-15 & M5&0.175&3.25&2.6--3.9&1.62&4.79&thin&n\\
04202555&M5.25&0.15&2.1&1.6-2.6&1.69&5.37&thin&?\\
+2700355\\
%04210934+2750368& M5.25&0.15& 0 & 0--0.5 & 1.18 & 3.23 & thin &n\\
%04213459& M5.5 & 0.12 & 2 & 1.5--2.5 & 1.39 & 3.23 & thin&n\\
%+2701388&\\
%04362151+2351165&M5.25&0.15&thin&?&5$\times$10$^{-4}$& 3.3$\times$10$^{-4}$&--&--&--&PD & PD/TD(IH)\\
%04284263 &M5.25&0.15&1&0.5--1.5&1.62&4.22&thick&n\\
%+2714039\\
%04163911 &M5.5&0.12&1&0.5--1.5&1.73&3.90&thick&n\\
%+2858491\\
\enddata
\label{taurusstate}
\tablecomments{The K$_{s}$-[8] and K$_{s}$-[24] colors listed are dereddened colors as determined by \citet{Luhman2009}.  
The stellar masses are computed using the \citet{Currie2010} effective temperature scale for K5--M2 stars and the 
\citet{Luhman2003} scale for later stars and the  \citet{Baraffe1998} isochrones. 
A$_{V}$ was determined individually by fitting the optical-to-near-IR SED for 
stellar photospheres using the \citet{Currie2010} intrinsic colors for a given 
effective temperature and using the \citet{Currie2010} and \citet{Luhman2003} 
effective temperature scales.  The range in A$_{V}$ assumes a nominal 20\% extinction 
uncertainty with a minimum uncertainty of 0.5 mags.}
\end{deluxetable}

%% file: tab_taurus2.tex
\begin{deluxetable}{llllllllllllll}
 \tiny
%\rotate
%\documentstyle[10pt]
%SPMquot"(0pt)
\setlength{\tabcolsep}{0pt}
\tabletypesize{\scriptsize}
\tablecolumns{1}
\tablecaption{Disk Evolutionary States for Selected Taurus Sources}
\scriptsize
\tablehead{{}&{}&{}&{}&{}&{}&{}&{}\\ 
{Name}&{M$_{disk}$}&{$\frac{M_{disk}}{M_{\star}}$}& {$\frac{M_{disk}}{M_{\star}}$}&{f($<$ 1,3 $\times$10$^{-3}$)  }&{Submm.}&
{M$_{disk}$} &{$\frac{M_{disk}}{M_{\star}}$}&{Disk Class}\\
{}&{}&{}&{(IQR)}&{}&{Ref.}&{(M$_{\odot,submm}$)}&{(submm)}}
%}
\startdata
UX Tau & 0.02 & 0.02 &[0.017,0.022] & 0,0&1 & 5.1$\times$10$^{-3}$& 5.1$\times$10$^{-3}$ &TD(IH)\\
V807 Tau & 3$\times$10$^{-3}$ &3$\times$10$^{-3}$ &[3$\times$10$^{-3}$,4$\times$10$^{-3}$]&0,0.5&1&10$^{-3}$&10$^{-3}$ & PD/TD(HD)\\
LkCa 15 & 0.01 & 0.01 &[0.01,0.025]&0,0&1& 0.05&0.05 & TD(IH)\\
FY Tau & 10$^{-4}$ & 10$^{-4}$ &[6$\times$10$^{-6}$,1$\times$10$^{-3}$]&0.76,0.82&1&7$\times$10$^{-4}$&7$\times$10$^{-4}$ & TD(HD)\\
JH 112 & 8$\times$10$^{-3}$ & 9.4$\times$10$^{-3}$ &[4$\times$10$^{-3}$,0.01]&0.06,0.12&1& 10$^{-3}$&1.05$\times$10$^{-3}$ & PD\\
V836 Tau & 0.01 & 0.012 &[0.012,0.014]&0,0&1&0.01&0.012 & PD\\
DN Tau & 0.01 & 0.013 &[0.013,0.03]&0,0&1,4&3$\times$10$^{-2}$&4$\times$10$^{-2}$ & PD\\
VY Tau & 2$\times$10$^{-4}$ & 2.7$\times$10$^{-4}$&[3.4$\times$10$^{-5}$,9$\times$10$^{-4}$]&0.76,0.91
&1&$<$ 5$\times$10$^{-4}$&$<$6.7$\times$10$^{-4}$ & TD(HD)\\
GO Tau & 6$\times$10$^{-3}$ & 8$\times$10$^{-3}$ &[5$\times$10$^{-3}$,0.023]&0,0&1&0.07 &0.09 & PD\\
IP Tau & 8$\times$10$^{-3}$ & 0.011 &[7$\times$10$^{-3}$,0.016]&0,0.05&1& 3$\times$10$^{-3}$&4$\times$10$^{-3}$ &PD\\
HO Tau & 8$\times$10$^{-3}$ & 0.012 &[0.013,0.013]& 0,0&1& 2$\times$10$^{-3}$&3.1$\times$10$^{-3}$ & PD\\
DH Tau & 5$\times$10$^{-3}$ & 0.01 &[0.01,0.018]&0,0.27&1&3$\times$10$^{-3}$&5$\times$10$^{-3}$ &PD/TD(IH)\\
FX Tau & 4$\times$10$^{-3}$ & 7$\times$10$^{-3}$ &[5$\times$10$^{-3}$,0.013]&0.09,0.22&1& 9$\times$10$^{-4}$&1.5$\times$10$^{-3}$ & PD\\
CY Tau & 8$\times$10$^{-3}$ & 0.014 &[0.014,0.025]&0,0&1,4& 6$\times$10$^{-3}$&0.015 & PD\\
04333905 &4$\times$10$^{-4}$ &7$\times$10$^{-4}$ &[1.7$\times$10$^{-4}$,2.7$\times$10$^{-3}$] & 0.63,0.68&--&--&-- & TD(IH)\\
+2227207\\
JH 223 & 7.9$\times$10$^{-5}$ & 1.4$\times$10$^{-4}$ &[8.5$\times$10$^{-5}$,1$\times$10$^{-3}$] & 
0.75,0.82&1&$<$3$\times$10$^{-4}$&$<$5.5$\times$10$^{-4}$ & TD(HD)\\
GH Tau & 4.1$\times$10$^{-3}$ & 7.5$\times$10$^{-3}$ &[4$\times$10$^{-3}$,0.01]&0.03,0.16&1& 
7$\times$10$^{-4}$&1.3$\times$10$^{-3}$ & PD\\
\\
ZZ Tau & 1$\times$10$^{-4}$ & 2.5$\times$10$^{-4}$ &[2$\times$10$^{-5}$,8$\times$10$^{-4}$]&
0.77,0.86&1,3&$<$4$\times$10$^{-4}$&$<$10$^{-3}$ & TD(HD)\\
XEST13-10 &4$\times$10$^{-4}$ & 1$\times$10$^{-3}$ &[2.5$\times$10$^{-4}$,3$\times$10$^{-3}$]&0.5,0.72&--&--&-- & TD(HD) \\
CIDA 8 & 8$\times$10$^{-3}$ & 0.03 &[0.03,0.04]&0,0&1&10$^{-3}$&3.3$\times$10$^{-3}$ & PD\\
04231822&5$\times$10$^{-4}$ & 1.7$\times$10$^{-3}$ &[3$\times$10$^{-4}$,5$\times$10$^{-3}$]&0.43,0.66&--&--&--& PD/TD(HD)\\
+2641156\\
%ITG-40 & M3.5&0.3&5$\times$10$^{-5}$& 1.7$\times$10$^{-4}$&--&--&-- & TD(HD)\\
04153916&4$\times$10$^{-4}$ & 1.5$\times$10$^{-3}$ &[4$\times$10$^{-4}$,6$\times$10$^{-3}$]&0.39,0.59&--&--&-- & PD/TD(HD)\\
+2818586\\
FP Tau & 2.5$\times$10$^{-4}$ & 1$\times$10$^{-3}$ &[4$\times$10$^{-4}$,4$\times$10$^{-3}$]&
0.53,0.68&2&$<$1.3$\times$10$^{-4}$&$<$5.2$\times$10$^{-4}$ & TD(HD)\\
%MH0 6 & M4.75&0.2&10$^{-4}$ & 5$\times$10$^{-4}$&--&--&-- & PD/TD(HD) ?\\
ITG-15 & 3$\times$10$^{-4}$ & 1.7$\times$10$^{-3}$ &[3.6$\times$10$^{-4}$,9$\times$10$^{-3}$] & 0.49,0.65&--&--&-- & PD/TD(HD)\\
04202555&4$\times$10$^{-4}$ & 2.7$\times$10$^{-3}$ &[1$\times$10$^{-3}$,3$\times$10$^{-3}$]&0.07,0.53&--&--&-- & PD/TD(IH)\\
+2700355\\
%04213459 & M5.5 & 0.12 & 7$\times$10$^{-6}$ & 5.8$\times$10$^{-5}$ & -- & -- & -- & TD & TD(HD)\\
%+2701388 &\\
%04362151+2351165&M5.25&0.15&thin&?&5$\times$10$^{-4}$& 3.3$\times$10$^{-4}$&--&--&--&PD & PD/TD(IH)\\
%04284263 &M5.25&0.15&2$\times$10$^{-4}$ & 1.3$\times$10$^{-3}$&--&--&--&PD & PD\\
%+2714039\\
%04163911 &M5.5&0.12& 2 $\times$10$^{-4}$ & 1.7$\times$10$^{-3}$&--&--&--&PD & PD/TD(HD)\\ 
%+2858491\\
\enddata
\label{taurusstate2}
\tablecomments{ IQR refers to the interquartile range of fractional disk masses from 
best-fit models.
The identification "Lu10" refers to \citet{Luhman2009}, where the classification listed is based 
on applying the \citet{Luhman2009} color criteria.  
The references for submillimeter data and (sub)millimeter-derived disk masses are the following: (1) \citet{Andrews2005}, (2) 
\citet{Jewitt1994}, (3) \citet{Jensen1994}, and (4) \citet{Beckwith1991}.  
The fractional disk masses are computed using the \citet{Currie2010} effective temperature scale for K5--M2 stars and the 
\citet{Luhman2003} scale for later stars and the  \citet{Baraffe1998} isochrones. Sources with disk 
evolutionary states listed as PD/TD(IH) or PD/TD(HD) have uncertain states: either SED modeling is inconclusive 
as to whether the disks likely have inner holes (PD/TD(IH)) or the classification depends on whether the 
full range or interquartile range of disk masses is used to identify primordial disks (PD/TD(HD)).  
\citet{Luhman2009} label all of these disks as primordial disks.}  
\end{deluxetable}

%% file: tab_taurusextra.tex
\begin{deluxetable}{llllllllllllll}
 \tiny
%\rotate
%\documentstyle[10pt]
%SPMquot"(0pt)
%\setlength{\tabcolsep}{0pt}
\tabletypesize{\scriptsize}
\tablecolumns{13}
\tablecaption{Optically-Thick Primordial Disks Around K5--M6 Stars 
in Taurus Listed As Having Low Masses In \citet{Andrews2005}}
\tiny
\tablehead{{Name}&{ST}&{M$_{\star}$}&{M$_{disk}$ (M$_{\odot}$)(AW05)}&{M$_{disk}$/M$_{\star}$ (AW05)$^{1}$}&
{M$_{disk}$ (M$_{\odot}$, this work)}&{M$_{disk}$/M$_{\star}$ (this work)}}
\startdata
HN Tau & K5 & 1.0 & 8$\times$10$^{-4}$ &8$\times$10$^{-4}$&6$\times$10$^{-3}$&6$\times$10$^{-3}$\\
V955 Tau & K7 & 0.85 &5$\times$10$^{-4}$&5.9$\times$10$^{-4}$&4$\times$10$^{-3}$&5$\times$10$^{-3}$\\
DF Tau & M2 & 0.55 & 4$\times$10$^{-4}$&7.3$\times$10$^{-4}$& 5$\times$10$^{-3}$&9$\times$10$^{-3}$\\
\\
CoKu Tau/3 & M1 &0.6 &$<$4$\times$10$^{-4}$&$<$6.7$\times$10$^{-4}$& 5$\times$10$^{-3}$ & 8$\times$10$^{-3}$\\
DP Tau & M1 & 0.6 &$<$5$\times$10$^{-4}$&$<$8.3$\times$10$^{-4}$&2$\times$10$^{-3}$&3$\times$10$^{-3}$\\
CIDA-3/V410 X-Ray 1 & M2 & 0.55 &$<$4$\times$10$^{-4}$&$<$7.3$\times$10$^{-4}$&2$\times$10$^{-5}$&
4$\times$10$^{-5}$\\
CZ Tau & M3 & 0.4 &$<$4$\times$10$^{-4}$&$<$10$^{-3}$&2$\times$10$^{-5}$&5$\times$10$^{-5}$\\

\enddata
\tablecomments{Notes (1) -- Fractional disk mass is derived using our estimates for 
stellar mass.}
\label{taurusextra}
\end{deluxetable}

%% file: tab_ic348.tex
\begin{deluxetable}{llllllllllll}
 \tiny
%\rotate
%\documentstyle[10pt]
%SPMquot"(0pt)
\setlength{\tabcolsep}{0.015in}
%\linewidth{0.1 in}
\tabletypesize{\tiny}
%\tabletypesize{\scriptsize}
\tablecolumns{9}
\tablecaption{Provisional Disk Evolutionary States for Selected Late-Type IC 348 Members based on 1--24 $\mu m$ Data}
\tiny
\tablehead{{}&{}&{}&{}&{}&{}&{}&{}&{Disk Class}&\\
%\tablehead{
{ID}&{ST}&{A$_{v}$(Best)}&{K$_{s}$-[8]}&{K$_{s}$-[24]}&{$\tau_{mid-IR}$}&{Inner Hole?}&{CK09}&{Lu10}&{This Work}}
\startdata
21 & K0 & 5.4 & 1.03 & 5.43 &thin & y & TD(IH) & TD & TD(IH)\\
26 & K7 & 8 & 2.03 & 4.90 &thick & n & PD & PD & PD\\
32 & K7 & 4.6 & 1.95 & 4.81 &thick & n &  PD & PD & PD\\
40 & K8 & 4 & 1.58 & 5.68 &thin & ? & PD & PD & TD(HD)/TD(IH)\\
55 & M0.5 & 11 & 1.93 & 4.86 &thick & n & PD & PD & PD\\
58 & M1.25 & 3.7 & 1.81 & 5.81 &thin & ? & PD & PD & TD(HD)/TD(IH)?\\
67 & M0.75 & 1.2 & 1.14 & 6.12 &thin & y & TD(IH) & TD & TD(IH)\\
68 & M3.5 & 2.25 & 1.43 & 4.14 &thin & ? & PD & PD & TD(HD)/TD(IH)?\\
72 & M2.5 & 1.7 & 0.75 & 5.65 &thin & y & TD(IH) & TD & TD(IH)\\
76 & M3.75 & 3.0 & 1.48 & 4.41 &thin & n & PD & TD & TD(HD)\\
97 & M2.25 & 5.5 & 1.30 & 4.85 &thin & y & PD & PD & TD(IH)\\
110& M2 & 5.0 & 1.62 & 5.90 &thin & y& PD & PD & TD(IH)\\
128 & M2 & 2.75 & 1.47 &3.54 &thin & n & PD & PD & TD(HD)\\
129 & M2 & 2.0 & 1.88 & 4.32 &thin & n & PD & PD & TD(HD)\\
135 & M4.5 & 0.9 & 1.41 & 3.48 &thin & n & PD & TD & TD(HD)\\
153 & M4.75 & 2.85 & 1.84 & 4.76 &thick& n & PD & PD & PD\\
194 & M4.75 & 3.1 & 1.46 & 4.88 &thin & n & PD & PD & TD(HD)\\
213 & M4.75 & 1.4 & 1.48 & 4.04 &thin & n & PD & PD & TD(HD)\\
214 & M4.75 & 1.1 & 1.48 & 4.69 &thin & n& PD & PD & TD(HD)\\
301 & M4.75 & 6.2 & 1.24 & 4.93 &thin & y &TD & TD & TD(IH)\\
308 & M4 & 11.8 & 1.31 & 4.92 &thin & y & PD & TD & TD(IH)\\
1928 & M5.5 & 1.4 & 1.74 & 3.68 &thick & n & PD & PD & PD\\
8078 & M0.5 & 5.8 & 1.82 & 5.22 &thick & n & PD & PD & PD\\
9024 & M0 & 3.75 & 1.94 & 4.72 &thick & n & PD & PD & PD\\
10352 & M1 & 3.5 & 1.88 & 5.24 &thick & n & PD & PD & PD
\enddata
\tablecomments{A$_{V}$ (Best) 
corresponds to the extinction that produces the best fit to the observed 
SED.  The range in A$_{V}$ used for SED modeling with the Robitaille models 
follows the method described for Taurus stars.  
The identification "Lu10" refers to \citet{Luhman2009}, where the classification listed is based
on applying the \citet{Luhman2009} color criteria.
The classification listed for CK09, \citet{CurrieKenyon2009}, is based on 
applying their color-criteria.  The disk states listed here are considered to 
be "provisional" since IC 348 lacks sensitive far-IR and submm data.}
%Question marks in the spectral type column denote spectral types assigned based on 
%fitting the optical and near-IR fluxes.  
%Disks whose evolutionary status is unclear are denoted 
%by question marks in the third column.  
%IDs 168 and 295 are labeled as 'homologously depleted' disks 
%instead of transition disks because their MIPS 24 $\mu m$ emission is significantly depleted relative to the flat disk and median 
%Taurus values, though changing the label only negligibly affects our conclusions.  
\label{ic348state}
\end{deluxetable}

%% file: tab_ngc2362.tex
\begin{deluxetable}{llllllllllll}
 \tiny
%\rotate
%\documentstyle[10pt]
%SPMquot"(0pt)
\setlength{\tabcolsep}{0.015in}
%\linewidth{0.1 in}
\tabletypesize{\tiny}
%\tabletypesize{\scriptsize}
\tablecolumns{11}
\tablecaption{Provisional Disk Evolutionary States for Late-Type NGC 2362 Members Based on 1--24 $\mu m$ Data}
\tiny
\tablehead{{}&{}&{}&{}&{}&{}&{}&{}&{Disk Class}&\\
%\tablehead{
{ID}&{ST}&{A$_{v}$(Best)}&{K$_{s}$-[8]}&{K$_{s}$-[24]}&{$\tau_{mid-IR}$}&{Inner Hole? (R$_{sub}$,AU)}&{CL09}&{Lu10}&{This Work}}
\startdata
   3& M2$^{1}$&0.6& 1.45& 5.89&thin& y,4.10 &TD(IH)&PD&TD(IH)\\
   36 &K5$^{1}$ &0.5&1.70&4.51&thin&n &TD(HD)&PD&TD(HD)\\
   41$^{2}$ &K1& 0.3&0.19&1.34&thin& n & TD(HD)&-&TD(HD)\\
   63$^{2}$ &M0$^{1}$& 0.7&0.52&2.76&thin&n & TD(HD)&-&TD(HD)\\
   85&M2$^{1}$ & 1.5&1.55&4.58&thin&n &TD(HD)&PD&TD(HD)\\
   111&K7& 0.75&2.5&5.34&thick& n & PD &PD&PD\\
   139&M2& 0.4&3.48&6.82&thick& n & PD &PD&PD\\ 
   168&K3& 0.5&0&3.48&thin& y,5.20 & TD(IH)&TD&TD(IH)\\
   177&M0$^{1}$& 0.8&1.82&5.33&thin & ?$^{3}$ & TD(HD)&PD&PD/TD(HD)/TD(IH)?\\
   187&M0.5& 1.5&2.64&4.81&thick& n & PD &PD& PD\\
   194&K5&1.2&1.06&5.46&thin & y,7.57 & TD(IH)& TD& TD(IH)\\
   196&K5$^{1}$& 0.1& 0.22&2.35&thin& y,3.63 & TD(IH)&TD &TD(IH)\\
   202&M2.5& 0.9&2.53&5.92&thick & n & PD & PD& PD\\
   204&M2.5& 2& 1.46&4.69&thin & n & TD(HD)& PD &TD(HD)\\
   213&M3$^{1}$& 0.8&1.28&3.84&thin & n & TD(HD)& TD&TD(HD)\\
   214&K7$^{1}$& 0.35&1.5&4.17&thin & n & TD(HD)&PD &TD(HD)\\
   219&M0& 0.8&1.65&3.87&thin & n & TD(HD)&PD &TD(HD)\\
   229&M1.5& 0.95&1.38&3.92&thin & n & TD(HD)&PD &TD(HD)\\
   251&M2$^{1}$& 1.25&1.7&4.68&thin& n & TD(HD)&PD &TD(HD)\\
   267&M2$^{1}$& 0.85&1.33&4.19&thin& n & TD(HD)&PD &TD(HD)\\
   295&M2$^{1}$& 0.4&0.34&4.20&thin& y,1.29 & TD(IH)&TD &TD(IH)\\
\enddata
\tablecomments{
The identification "Lu10" refers to \citet{Luhman2009}, where the classification listed is based
on applying the \citet{Luhman2009} color criteria.
The identification CL09 refers to the disk evolutionary states listed in \citet{CurrieLada2009}.
The disk states listed from this work are considered to be provisional since 
NGC 2362 lacks sensitive far-IR/submm data.  To compare our results with those presented in 
\citet{CurrieLada2009}, we list the inner hole sizes in AU.  (1) Denotes approximate spectral types assigned based on 
fitting the optical and near-IR data.  Likewise, A$_{V}$ (Best) is 
corresponds to the extinction that produces the best fit to the observed 
SED.  (2) The detections of these sources were confirmed by a rereduction of the MIPS-24 $\mu m$ 
data and PRF-fitting photometry.
%Question marks in the spectral type column denote spectral types assigned based on 
%fitting the optical and near-IR fluxes.  
%Disks whose evolutionary status is unclear are denoted 
%by question marks in the third column.  
%IDs 168 and 295 are labeled as 'homologously depleted' disks 
%instead of transition disks because their MIPS 24 $\mu m$ emission is significantly depleted relative to the flat disk and median 
%Taurus values, though changing the label only negligibly affects our conclusions.  
(3) The best-fit models are evenly divided between those that 
have an AU-scale inner hole and those that don't.  For members with disks that 
have an inner hole (marked "y"), we add the inner hole size in AU as determined in 
\citet{CurrieLada2009}.}
\label{n2362state}
\end{deluxetable}

%% file: tab_etacha.tex
\begin{deluxetable}{llllllllllllll}
 \tiny
%\rotate
%\documentstyle[10pt]
%SPMquot"(0pt)
\tabletypesize{scriptsize}
\setlength{\tabcolsep}{0pt}
%\tabletypesize{\tiny}
\tablecolumns{14}
\tablecaption{Disk Evolutionary States for Late-Type $\eta$ Cha Sources}
%\tiny
\tablehead{{}&{}&{}&{}&{}&{}&{}&{}&{}&{Disk Class}&{}&{}\\
{Name}&{ST$^{1}$}&{M$_{\star}$}&{K$_{s}$-[8]}&{K$_{s}$-[24]}&
%{$\tau_{mid-IR}$}&{Inner Hole?}&{M$_{disk}$}&{$\frac{M_{disk}}{M_{\star}}$}&{$\frac{M_{disk}}{M_{\star}}$}&
{$\tau_{mid-IR}$}&{Inner}&{M$_{disk}$}&{$\frac{M_{disk}}{M_{\star}}$}&
 {f($<$ 1,3$\times$10$^{-3}$)}&{Si09} &{Lu10}&{ This Work}\\
{}&{(M$_{\odot}$)}&{}&{}&{}&{}&{Hole?}&{}&{}&{}}
\startdata
RECX-1 & K7+M0,K6 & 0.95&0.23 &0.38&-&n & - & - &--&star & star & star\\
RECX-3 & M3,M3.25 & 0.35&0.26 & 0.92&thin&y & 3.5$\times$10$^{-8}$ &1$\times$10$^{-7}$&1,1&TD(IH) & TD & TD(IH)\\
RECX-4 & M1.75,M1.3 & 0.6&0.30 & 0.89&thin&y & 1.8$\times$10$^{-7}$ &3$\times$10$^{-7}$&1,1&TD(IH) & TD & TD(IH)\\
RECX-5 & M3.8,M4 & 0.25&0.97 & 4.81&thin &y & 7$\times$10$^{-4}$ &2.8$\times$10$^{-3}$&0.14,0.73&TD(IH)& TD & TD(IH)\\
RECX-6 & M3,M3 & 0.375&0.25 & 0.41&-&n & - & - &--&star & star & star\\
RECX-7 & K6.9+M1,K6 & 0.95&0.17 & 0.26&-&n & - & -&--&star & star &star\\
RECX-9 & M4.4+M4.7, & 0.225&1.37 & 3.96 &thin&n &1.4$\times$10$^{-4}$ & 6$\times$10$^{-4}$&0.53,0.75&
TD(HD) & PD & TD(HD)\\
&M4.5\\
RECX-10& M0.3,M1 & 0.6&0.24 & 0.28 &-&n & - & - &--&star & star &star\\
RECX-11& K6.5,K5.5 & 1&1.69 & 3.98 &thin&n & 3.3$\times$10$^{-3}$ & 3.3$\times$10$^{-3}$&0.13,0.50&PD & PD & PD\\
RECX-12& M3.2,M3.25 & 0.35&0.34 & 0.46&-&n & - & - &--&star & star & star\\
J0841& M5.5,M4.75 & 0.2&1.50 & 3.91 &thin&n & 3$\times$10$^{-5}$ & 2$\times$10$^{-4}$&1,1&TD(HD) & PD & TD(HD)\\
J0843& M5,M3.25 & 0.35&2.92 & 5.91&thick&n & 1$\times$10$^{-3}$ & 3$\times$10$^{-3}$&0.31,0.50&PD & PD & PD\\
J0844& M5.3, M5.75 & 0.1&1.86 & 4.51 &thin?&n & -- & --&--&PD & ? & PD/TD(HD)$^{2}$\\
J0838& M5, M5.25 & 0.15&0.53&-&- & n & -&-&--&star & star & star\\
J0836& M5.3, M5.5 &0.12&0.58 &-&- &n & -&-&--&star & star & star
\enddata
\tablecomments{
The identification "Lu10" refers to \citet{Luhman2009}, where the classification listed is based
on applying the \citet{Luhman2009} color criteria.
The identification "Si09" refers to \citet{SiciliaAguilar2009}'s disk classification translated 
into our nomenclature.
1) The first entry for spectral type comes from \citet{Lyo2004} and the 
second one comes from \citet{LuhmanSteeghs2004}.  We compute the stellar masses and fractional disk masses using the 
\citet{LuhmanSteeghs2004} spectral types and the stellar masses for a given effective temperature at 6 Myr from 
the \citet{Baraffe1998} isochrones.  Our results do not leverage on whose spectral types we adopt. 2) For our 
statistics, we classify this disk as a primordial disk since it lacks far-IR data to needed to derive a meaningful 
disk mass estimate.  We omit a column for optical extinction since 
it is negligible for this cluster \citep[e.g.][]{Lyo2004,LuhmanSteeghs2004}.}
\label{etachastate}
\end{deluxetable}

%% file: tab_all.tex
\begin{deluxetable}{llllllll}
 \tiny
%\rotate
%\documentstyle[10pt]
%SPMquot"(0pt)
\setlength{\tabcolsep}{0.003in}
%\linewidth{0.1 in}
%\tabletypesize{\tiny}
%\tabletypesize{\scriptsize}
\tabletypesize{\small}
\tablecolumns{7}
%\tablecaption{Evolutionary States for Coronet Cluster Disks with New Spitzer Photometry}
%\tablecaption{Disk Evolutionary States as a Function of Spectral Type}
\tablecaption{Frequency of Transitional Disks With Time (Corrected for Contamination)}
\tiny
\tablehead{{Cluster} & {Age (Myr)}&{Spectral Type Range} & {f(TD)$_{lower}$} & {f(TD)$_{upper}$}}
\startdata
Taurus & 1--2 & K5--M6 & 0.18 (25/143) & 0.22 (32/145) \\
       &      & K5--M2 & 0.17 (13/75) & 0.21 (16/77) \\
       &      & M2.5--M6 & 0.18 (12/68) & 0.24 (16/68)\\ 
\\
Coronet &1--3 & K5--M6 & 0.21 (4/19) & 0.36 (8/22)\\
        &     & K5--M2 & 0.12 (1/9) & 0.22 (2/9)\\
       &      & M2.5--M6& 0.30 (3/10) & 0.50 (6/12)\\
\\
IC 348 & 2.5 & K5--M6 & 0.32 (40/124) & 0.40 (49/124)\\
       &      & K5--M2 & 0.18 (6/33) & 0.24 (8/33)\\
       &      & M2.5--M6 & 0.42 (34/81) & 0.51 (41/81)\\
\\
NGC 2362 & 5  &K5--M3 & 0.58 (11/19) & 0.63 (12/19)\\
%         &    &K5--M2 & 0.56 (9/16) & 0.75 (12/16)\\
\\
$\eta$ Cha & 6--8 &K5--M6 & 0.63 (5/8) & "\\
\enddata
\tablecomments{The transitional disk frequency is defined as f(TD)/(f(TD)+f(PD)).  
The range in transitional disk frequencies (f(TD)) for Taurus accounts for 
disks with uncertain/indeterminable states.  The range in frequencies 
for the Coronet, IC 348, and NGC 2362 also account for the range of contamination 
estimates based on SED modeling of Taurus data.}
%\label{dwarfcolors}
\label{freqtranall}
\end{deluxetable}